\documentclass[11pt]{article}\usepackage{amsmath,amssymb,multirow,jheppub,graphicx}
\usepackage{centernot}
\usepackage{color}
\allowdisplaybreaks[1]
\usepackage{tikz}
 \tikzset{node distance=2cm, auto}
\usepackage{bbm} 
\usepackage{youngtab}

\renewcommand{\Re}{\text{Re }}
\renewcommand{\Im}{\text{Im }}

\def\Im{\text{Im}}
\def\Re{\text{Re}}

\def\tr{\text{tr}}

\def\hat{\widehat}
\def\bar{\overline}

\def\I{{\cal I}}

\def\Z{{\mathbb Z}}
\def\R{{\mathbb R}}
\def\coeff#1#2{{\textstyle {\frac {#1}{#2}}}}

\def\half{\coeff 12}
\def\N{{\cal N}}
\usepackage{verbatim}   
\usepackage[all]{xy}
\usepackage{bm}  
\def\Dslash{{\rlap{\raise 1pt \hbox{$\>/$}}D}}
\def\Pslash{{\rlap{\raise  1pt \hbox{$\>/$}}\,\partial}}

\newcommand{\diff}{\mathrm{d}}

\newcommand{\Diff}{{\mathcal{D}}}

\newcommand{\be}{\begin{equation}}      
\newcommand{\ee}{\end{equation}}      
\newcommand{\bea}{\begin{eqnarray}}      
\newcommand{\eea}{\end{eqnarray}}

\newcommand{\im}{\mathrm{i}}

\newcommand{\rmc}{\mathrm{c}}
\newcommand{\rme}{\mathrm{e}}

\newcommand{\rmL}{\mathrm{L}}
\newcommand{\rmR}{\mathrm{R}}
\newcommand{\rmV}{\mathrm{V}}

\newcommand{\z}{\mathsf{z}}

\title{Strongly coupled QFT dynamics via TQFT coupling}
\author[1]{Mithat  \"Unsal}
\affiliation[1]{Department of Physics, North Carolina State University, Raleigh, NC 27607, USA}
\emailAdd{unsal.mithat@gmail.com}
\abstract{
We consider a class of quantum field theories and quantum mechanics, which we couple to $\Z_N$ topological QFTs, in order to classify non-perturbative effects in the original theory.  
The $\Z_N$  TQFT structure arises naturally from turning on a classical background field for a  $\Z_N$  0- or 1-form  
global symmetry.
In  $SU(N)$ Yang-Mills theory  coupled to  $\Z_N$ TQFT,  
 the non-perturbative expansion parameter  is  
$\exp[-S_I/N]= \exp[-{8 \pi^2}/{g^2N}]$ both in the  semi-classical  weak coupling domain and  strong coupling domain, corresponding to a fractional topological charge configurations.   
To classify  the non-perturbative effects in original $SU(N)$ theory, we must use $PSU(N)$ bundle and lift configurations (critical points at infinity)  for which there is no obstruction back to $SU(N)$. These  provide a refinement of instanton sums:
     integer topological  charge, but   crucially  fractional action configurations  contribute, providing a TQFT protected generalization of resurgent semi-classical expansion to strong coupling.  
      Monopole-instantons (or fractional instantons) on $T^3 \times S^1_L$  can be interpreted as tunneling events in the 't Hooft flux background  in the $PSU(N)$ bundle.  
 The  construction provides a new perspective to the strong coupling regime of QFTs and resolves a number of old standing issues, especially, fixes the conflicts between  the large-$N$ and  instanton analysis.   
 We   derive the mass gap at $\theta=0$ and gaplessness at $\theta=\pi$ in $\mathbb{CP}^{1}$ model, and mass gap for arbitrary $\theta$ in $\mathbb{CP}^{N-1}, N \geq 3$ on $\R^2$.     }

\begin{document}
\maketitle

\section{Introduction}

The applications of coupling a topological quantum field theory (TQFT) to  quantum field theory (QFT)  received recent interest in the discussion  
of mixed anomalies \cite{Gukov:2013zka, Kapustin:2014gua, Gaiotto:2017yup}.  In this context, it is used to extract   kinematic constraints  imposed by symmetry on   a QFT.  
This is an exact, albeit non-dynamical information about quantum theory. 

Here, we would like to explore the implications of coupling a TQFT to QFT in the study of non-perturbative dynamics, a territory that we cannot expect to be exact, in search of reliable approximations.  We want to take advantage of  robustness of TQFT in parts of the story. 
One of our physical goals  is to understand  what controls the strength of non-perturbative effects in an asymptotically free  QFT on large $d$-dimensional manifold $M_d$, where 
$M_d$ serves as a regularization for $\R^d$.  More practically, we would like to understand the role of topological defects such as instantons, 
monopole instantons and fractional instantons in the dynamics much more precisely both in weak coupling semi-classical domain and in strong coupling domain by using the restrictions that follows from TQFT.

\begin{figure}[t]
\vspace{-1.5cm}
\begin{center}
\includegraphics[width = 0.9\textwidth]{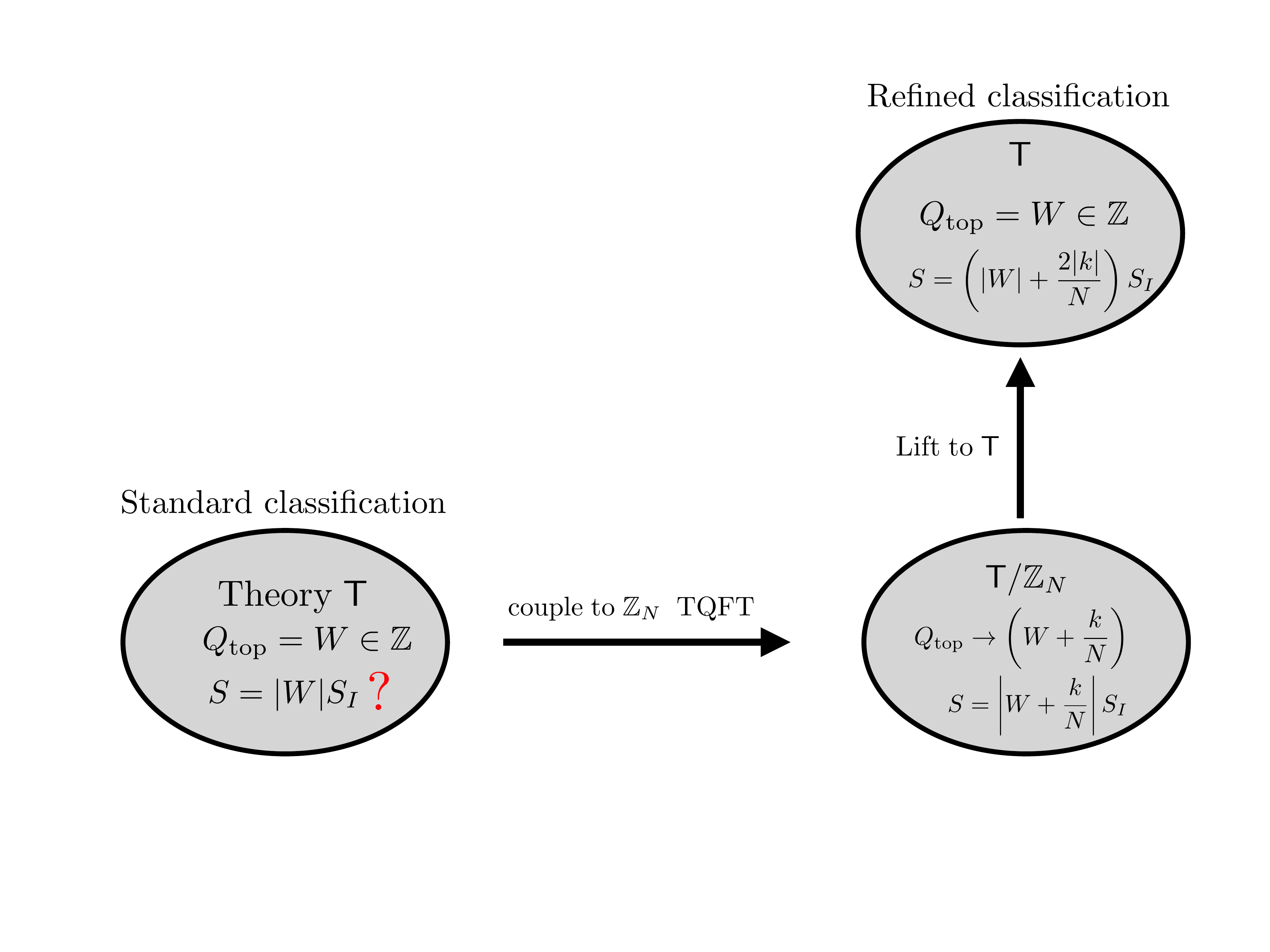}
\vspace{-1.5cm}
\caption{By coupling a QFT to a TQFT,   
the  topological charge   and action become fractional. Such configurations do not lift to 
the partition function of the theory ${\mathsf T}$, but  their {\it integer} topological charge and {\it fractional} action composites do. Even  the $W=\frac{1}{N}$ fractional instantons contribute to some observables in ${\mathsf T}$, but not to partition function.  
We  demonstrate   that the non-perturbative expansion parameter in theory is $\exp[-S_I/N]$ instead of 
  $\exp[-S_I]$. This statement was already known to be true in semi-classical domain. The TQFT coupling 
  tells us that it is also true in strong coupling domain.   
    }
\label{fig:bigpicture}
\end{center}
\end{figure}

Let us denote a theory by  $\mathsf T$ and its global symmetry by $G$. 
To be less abstract, we develop three  (extremely) parallel stories.  Quantum mechanics of a  particle on a circle in the presence of  a potential with $N$ degenerate minima, that we call $T_N$ model for short,   $2d$ $\mathbb {CP}^{N-1}$ model,   and $4d$  $SU(N)$  Yang-Mills theory.  $T_N$ quantum mechanics 
and Yang-Mills theory possess a $\Z_N$ global symmetry,  0-form or 1-form, respectively.   $\mathbb {CP}^{N-1}$ has a $SU(N)/\Z_N$ global symmetry.  One can turn on a classical background field for these symmetries (also called background gauging).\footnote{We use {\it background gauging} for turning on a classical background field for a global symmetry  and coupling it in a gauge invariant manner to the theory. Note that with a background field turned on, it is still meaningful to look at correlators that transform under background gauging. These are  non-trivial and transform covariantly. 
 There is no path integration at this stage. 
 If we choose to sum over all possible backgrounds, i.e, perform a path integral over the background field by making it dynamical,  we refer to it as {\it gauging}.
 In this case, the  correlators that transform under  gauge transformations  are identically zero by Elitzur's theorem \cite{Elitzur:1975im}. Both of these ideas are used throughout this work. 
 }    
In doing so,   coupling QFT $\mathsf T$  to a $\Z_N$ TQFT becomes strictly necessary, we call the resulting theory as  $\mathsf T/\Z_N$   theory.\footnote{Strictly speaking, $\mathsf T/\Z_N$  is appropriate  description  for QM and Yang-Mills. In $\mathbb {CP}^{N-1}$, it is slight abuse of language. To turn on  $SU(N)/\Z_N$ background, we first turn on an $SU(N)$ background field,  and then, couple the theory to 
a $\Z_N$ TQFT.  This is what we mean by  $\mathsf T/\Z_N$  model in  $\mathbb {CP}^{N-1}$.} As a result of this procedure,   we learn the existence of new non-perturbative effects and scales   relevant to the original theory, see Fig.~\ref{fig:bigpicture}.

\begin{figure}[t]
\vspace{-1.5cm}
\begin{center}
\includegraphics[width = 0.7\textwidth]{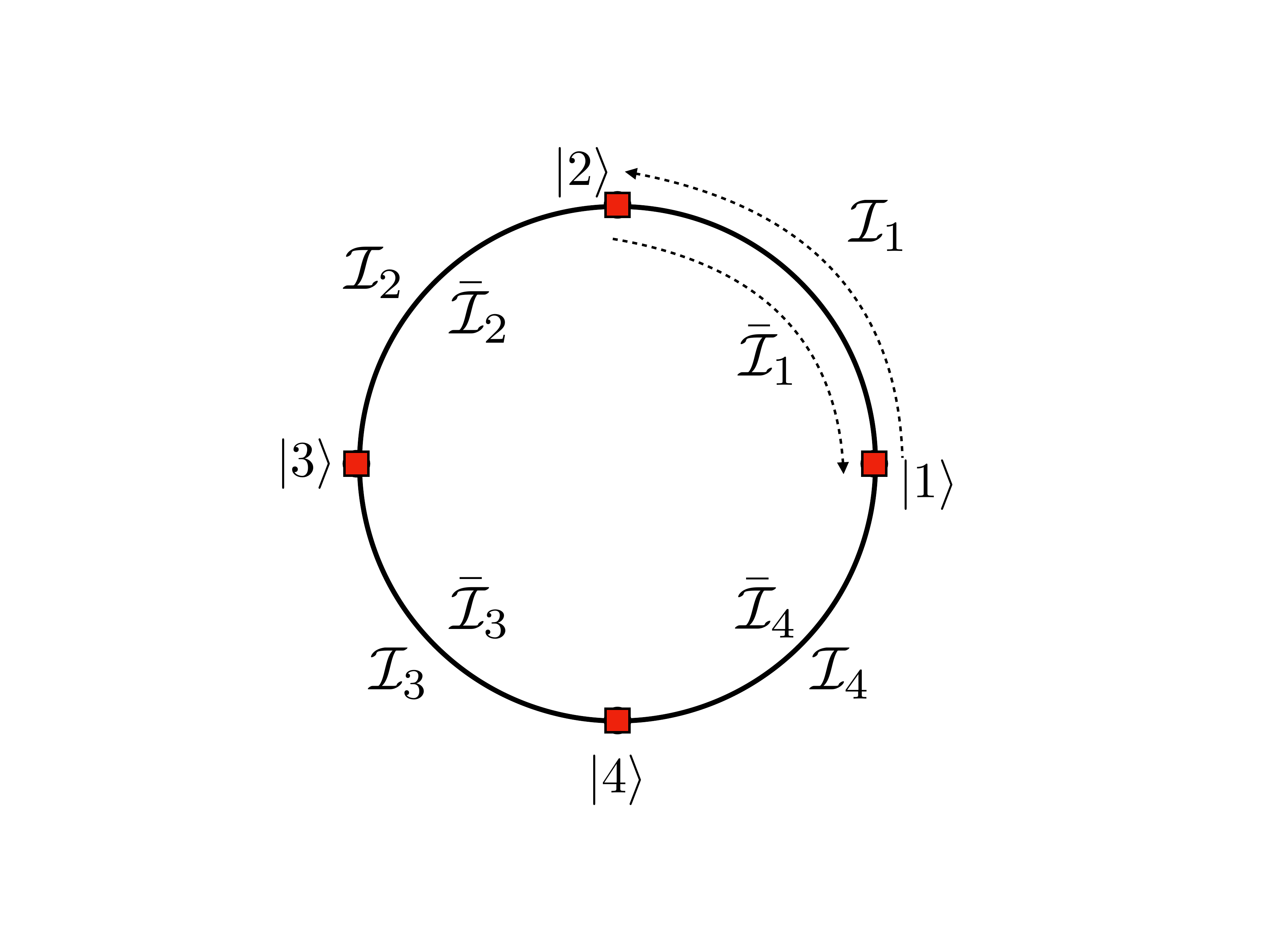}
\vspace{-1cm}
\caption{$T_N$ model is quantum mechanics with periodic potential $-\cos(Nq)$ on a circle with fundamental domain 
$ q \in [0, 2\pi]$. In the Born-Oppenheimer approximation, it is equivalent to $N$-site lattice model.  In partition function, we sum only over periodic paths, corresponding to  integer winding number topologies,  but obviously, there are other saddles with fractional winding number.   }
\label{fig:TN}
\end{center}
\end{figure}

A recurring theme  can be described  simply in quantum mechanics. 
Consider $T_N$ model, 
 with $\Z_N$ shift  symmetry, Fig.\ref{fig:TN}.   The partition function  is described as a sum over 
periodic paths on $S^1$, it receives contribution  only from  integer winding number $W \in \Z$ configurations. Refer to the action of $W=1$ instanton as $S_I$.   There are 
clearly fractional instantons  $I_j$  with  winding number $W \in  \frac{1}{N} \Z$  and action $S_I/N$ in the theory, which contribute at leading order to the spectrum,  but forbidden to contribute to  partition function, because they are not periodic paths in $S^1$.   
Now, if we couple the theory to $\Z_N$ TQFT and  gauge  $\Z_N$,  these configurations become  closed paths in  $S^1/\Z_N$, 
 but there is an obstruction to lift them  to closed paths in original  $S^1$. Yet, a pair of loops with winding numbers $+ 1 -1=0$  in  $S^1/\Z_N$ lift to  $S^1$  
as $\frac{1}{N} -  \frac{1}{N} =0$ 
 and contribute to the partition function with action $2S_I/N$.  The 
  precise semi-classical  description of this saddle here  is {\it critical point at infinity} \cite{Behtash:2018voa} which we explain in QM discussion. 
 So, there are zero winding, and fractional action  non-BPS configuration contributing to partition function in $T_N$ quantum mechanics.  

 We formalize this construction as follows in QM and QFT. 
\begin{itemize}
\item  In $d=1,2, 4$, we describe  a $\Z_N$  TQFT and couple it to the theory  $\mathsf T$:  quantum mechanics,   $\mathbb{CP}^{N-1}$  and Yang-Mills, 
respectively. 
\item By turning on a   classical $\Z_N$  background field or by gauging it (summing over all possible backgrounds, corresponding to 
$\mathsf T/\Z_N$), we find a set of saddles   which do {\it not} lift to the  original theory $\mathsf T$.  But fractional action  non-BPS composites of such objects lift to $\mathsf T$ and contribute to partition function. 
\item We can view  the  (fixed) classical $\Z_N$  TQFT backgrounds as (fixed)  't Hooft fluxes or   twisted boundary conditions  in $d=2, 4$
\cite{tHooft:1979rtg, tHooft:1977nqb}. 

\item In  Yang-Mills theory, 
 the non-trivial configurations in the $SU(N)/\Z_N$ bundle do not uplift to $SU(N)$ bundle, but  certain  integer topological charge  and fractional action configurations do. 
  The same statement is also true in  $\mathbb {CP}^{N-1}$, where $U(1)/\Z_N$ vs.  $U(1)$ are appropriate bundles. 
  Both of these statements are independent of strength of coupling, in particular, valid at both weak and strong coupling due to TQFT coupling.

\item Since in order to find the  saddles in $ T_N$ model, we need to directly look at   either      $ T_N$  with classical $\Z_N$  background $(A^{(1)}, A^{(0)})$  or equivalently, $\Z_N$  twisted boundary conditions  or      $( T_N/\Z_N)_p $ models where $\Z_N$ is gauged,  
we claim that in order to find  the relevant saddles of $SU(N)$ gauge  theories,  we should first consider either  $SU(N)$ with  background gauge  field $(B^{(2)},B^{(1)}) $  for the  $\Z_N^{[1]}$  1-form symmetry or  $(SU(N)/\Z_N)_p$ theory where $\Z_N^{[1]}$ is gauged. 
After finding these configurations, we can  patch  them up   to find the ones that can be lifted to $SU(N)$ theory.  These are  fractional action  (e.g. $2S_I/N$)   configurations that contribute to the partition function of  $SU(N)$ theory.  

\item   These configurations  are TQFT-protected, and exist both at weak and strong coupling.  
In compactified theories, they exist both at $T^{d-1}_{\rm large} \times S^1_{\rm small}$ as well as  
  $T^{d}_{\rm large}$ where the latter is a place holder for $\R^d$. 
 
\item This construction shows that, in $d=2$ $\mathbb{CP}^{N-1}$ and  
$d=4$ Yang-Mills theory, the non-perturbative effects are not controlled by the (BPST) instanton factor $\rme^{-S_I + \im \theta}$  \cite{Polyakov:1975yp, Belavin:1975fg},  but instead by exponentially larger   
  $\rme^{-S_I/N + \im \theta/N}$ 
 fractional instanton factor, due to TQFT-protection.   

\item This extends our earlier result in  flavor $\Omega_F$-twisted $\mathbb{CP}^{N-1}$ model on semi-classical small $\R \times S^1_L$  \cite{Dunne:2016nmc, Dunne:2012ae} to large $\R \times S^1_L$, to the strong coupling domain, by augmenting flavor twisted theory with a $\Z_N$ TQFT  background (i.e. with 't Hooft twist.) 
  It also extends some  results in semi-classical 
 deformed Yang-Mills  theory, $\N=1$ SYM,  QCD(adj) and some other QCD-like theories  (see \cite{Unsal:2007jx, Unsal:2007vu, Unsal:2008ch,Shifman:2008ja}) from small $\R^3  \times S^1_L$  to large  $\R^3  \times S^1_L$, to the strong coupling domain by coupling these systems to  $\Z_N$  TQFTs.  
 
 
\item  Based on matching  weak coupling semi-classical descriptions, matching of global symmetries and 
mixed 't Hooft anomalies, and improving an old work  of 
Fateev, Frolov, Schwarz, and   Berg,   L\"uscher 
\cite{Berg:1979uq, Fateev:1979dc} in crucial ways,  
we show  that low-energy spectrum of the $ \mathbb{CP}^{N-1} $ model is described by 
  $N$-flavor massive  Schwinger model or its non-abelian bosonization, the   mass deformation of $SU(N)_1$ WZW model with an extra scalar.  
  
  \item The mass parameter in Schwinger model, or mass deformed  WZW models, due to matchings of semi-classical descriptions, map  to a fractional instanton effect $\rme^{-S_I/N}$ in $ \mathbb{CP}^{N-1} $, which is nothing but the   strong scale of the theory.    By studying renormalization group for topological defect operators  \cite{Kosterlitz:1974sm} (derived  originally to explain  the Kosterlitz-Thouless phase transition \cite{Kosterlitz:1973xp}), we describe  the low energy spectrum of the theory. 
  
  \item  We determine the mass gap and low energy spectrum  for arbitrary $\theta$ in $\mathbb{CP}^{N-1}$  on $\R^2$   via this procedure. For $N \geq 3$, the spectrum is gapped for any $\theta$, and composed of an  adjoint and singlet.  For the $N=2$ theory, the theory is gapped at $\theta \neq \pi$.  
  We show the existence of a gapless triplet at $\theta=\pi$.   Our formalism give a prediction for the mass gap in the $\mathbb{CP}^{1}$ model 
  as $m_{\rm gap} (\theta)  =  \Lambda \left| \cos \frac{\theta}{2} \right|^{{2}/{3}}$, which we hope can be tested by other means. 


\end{itemize}

 Our  construction  resolves many puzzles concerning conflicts between  large-$N$ vs. instantons,  
instantons vs.   multi-branch structure of vacua,  incorrect  $\theta$-angle dependence of vacuum energy due to instantons,  
instantons vs.   $\eta'$ puzzle, and many others mentioned in  books and reviews \cite{ Coleman198802, Vainshtein:1981wh, Schafer:1996wv, Marino:2015yie}.  
In retrospect, our current construction proves why recent studies on $\R^{d-1} \times S^1$ in the context of adiabatic continuity and resurgent  semi-classical analysis  
 \cite{Dunne:2016nmc, Dunne:2012ae, Unsal:2008ch,Shifman:2008ja}
capture  interesting non-perturbative dynamics of QFT on $\R^d$. 
In the light of our construction, it may be worthwhile to    revisit the   meron idea of Gross et. al. \cite{Callan:1977gz, Gross:1977wu, Callan:1977qs} on $\R^d$.  This idea requires some fixing which can easily be done,  for the $N=2$ case,  see  Sec.\ref{sec:meron}. It also needs correct generalization to $N \geq 3$.  

\subsection{Earlier works on Yang-Mills on  $T^3 \times \R$: Not all QM reductions are  the same.}
We finalize the introduction by describing  historical  precedents to this work\footnote{We thank anonymous referee for encouraging us to write this part 
which clarifies the relation between the present  work and earlier discussions of Yang-Mills theory in a quantum mechanical reduction down to small $T^3 \times \R$.}, especially concerning the different class of constructions  in  a quantum mechanical reduction on    $T^3 \times \mathbb R$.  
This will also give us an  opportunity to highlight where our work resembles or differs from the earlier works. 

An exact solution to self-duality equation with   topological charge $Q=1/N$ and action 
$\frac{8 \pi^2}{g^2N}$ on $T^4$ is first found by   't Hooft \cite{tHooft:1981nnx}, and generalized by  van Baal  \cite{vanBaal:1982ag}.  't Hooft original physical motivation was to find configurations which 
 survive in the large-$N$ limit, and which are exponentially more important than the usual instantons even at finite-$N$.
 (The usual instantons are suppressed as $e^{-N}$ in the large-$N$ limit, and are irrelevant.) 
 These solutions  can  be interpreted as configurations in  $SU(N)$  with discrete flux background turned on,  or in 
 $PSU(N)$ bundle.  However, these are  {\it constant}   solutions, and  the geometry of $T^4$ must satisfy a certain aspect ratio to have a 
 $\frac{8 \pi^2}{g^2N}$  action. 
   Unlike on $\R^3 \times S^1$ where it is easy to find non-trivial (space-time dependent) fractional instanton solutions, the monopole-instantons,  where dynamics abelianize at long distances due to adjoint-Higgsing sourced by the vev of gauge holonomy,  
 it proved to be   extremely difficult to find non-trivial solutions in systems which do not abelianize at long distance. 
 
  Discrete  flux background and   twisted boundary conditions are  used  to reduce  $\N=1$ $SU(N)$ SYM to   quantum mechanics on $T^3 \times \R$ and employed to calculate supersymmetric index in Witten's work \cite{Witten:1982df}.     The work closest in spirit to our construction, working with non-supersymmetric theories, and aiming to preserve  the  vacuum structure and  fractional instantons  in a quantum mechanical limit, are the works of  Gonzalez-Arroyo et. al. In their construction, dynamics   do not abelianize,  and    they have found by numerical simulations on latticized $T^3 \times \R$ that  time-dependent fractional instanton solutions with action   $\frac{8 \pi^2}{g^2N}$ exist  in the presence of 't Hooft flux 
  \cite{GarciaPerez:1989gt, GarciaPerez:1992fj,  GarciaPerez:1993jw, GonzalezArroyo:1995zy,  
GonzalezArroyo:1995ex, GonzalezArroyo:1996jp, 
 Montero:2000mv, GonzalezArroyo:1997uj, Gonzalez-Arroyo:2019wpu}.  The analytic form of these solutions is still an open and important problem. 
 The difference between the works of Witten and Gonzalez-Arroyo,  and our work is following. 
  We reduce deformed YM on $\R^3 \times S^1_\beta$ (which is adiabatically connected to pure YM on $\R^4$) down to  $T^3 \times  S^1_\beta$ by using {\it magnetic GNO flux} in the co-weight lattice of $SU(N)$. This is  more refined data than discrete 't Hooft flux.   
  Magnetic $N$-ality determines the discrete magnetic flux.    The fact that we can turn on $ \bm \mu_1$ GNO flux  and this is degenerate in energy with   $\bm {\mu_1 - \alpha_1},  \ldots, \bm   \mu_1 - \sum_{j=1}^{N-1} \bm \alpha_j $  (all with discrete magnetic flux one) helps us to identify the tunneling in QM with the monopole-instantons on $\R^3 \times S^1_\beta$, which  carry magnetic charges in the  extended simple root system  of $SU(N)$.  In both construction, 
  there are $N$-classical vacua (sourced by turning on a classical background) and $N$- fractional instantons 
  with topological charge $Q=1/N$ and action  $\frac{8 \pi^2}{g^2N}$, an extremely desirable aspect.

 A frequently  asked question is the relation between Luscher  \cite{Luscher:1982ma}  and van Baal's  works \cite{vanBaal:1986ag, vanBaal:2000zc}, which employs  $T^3 \times \R$, without incorporating discrete flux, and our present work.\footnote{van Baal also have important works on the discrete flux backgrounds and classification of the configurations in $PSU(N)$ bundle.}  These works   would like to benefit from the 
 absence of phase transitions in finite volume.  However, unlike our construction, in  these compactification, vacuum structure is not persistent.  In particular, in $SU(N)$ theory,  there are $N^3$ vacua related to perturbative  breaking of $(\Z_N^{[0]})^3$ center symmetry to all orders in perturbation theory, instead of  $N$ classical vacua that emerge from classical  flux insertion.     
 However, $0$-form symmetries cannot break spontaneously in QM unless enforced by a mixed anomaly.\footnote{In fact, $d-1$ form symmetries cannot break spontaneously in $d$ dimensions unless enforced by a mixed anomaly.}
  Indeed, there are instantons in the theory leading to a unique vacuum. 
 These  instantons between   $N^3$ perturbative vacua  are {\it not}  classical fractional instantons with action $ \frac{8\pi^2}{g^2N}$ or instantons with action   $ \frac{8\pi^2}{g^2}$. Rather, these instantons arise from balancing of the classical action with one-loop potential for the gauge holonomy, i.e, they are ``quantum instantons", with action of order $O(1/g)$, i.e, they are not related to  fractional instantons that we are talking about, that exist in thermo-dynamic limit on $\R^3 \times S^1$ or the ones that appear in $PSU(N)$ bundle classification.  
 
Furthermore,  if we consider the large-$N$ limit of the  Luscher  \cite{Luscher:1982ma}  and van Baal's compactification  \cite{vanBaal:1986ag, vanBaal:2000zc}, the tunneling amplitude between $N^3$ perturbative vacua  has an $N$ dependence of the form $e^{-\sqrt N}$, and     
 the tunneling becomes forbidden in the $N \rightarrow \infty$ limit. That is, even in finite volume, there is a phase transition in the large-$N$  limit (which is also a thermo-dynamic limit) of their construction, between the small  $T^3 \times \R$ theory and large  $T^3 \times \R$ theory according to 
 $(\Z_N^{[0]})^3$ center-symmetry realization. 
    (This phase transition is  actually  the reason why Eguchi-Kawai reduction \cite{Eguchi:1982nm} in its original form without discrete flux fails \cite{Bhanot:1982sh, Kiskis:2003rd}.) 
  In sharp  contrast, in our  and  Gonzalez-Arroyo's constructions, there are $N$-classical vacua, induced by classical background,  and the tunneling between them is a classical fractional instanton effect with action $ \frac{8\pi^2}{g^2N}$ and does not disappear in the large-$N$ limit.   (This is  actually  the reason why Twisted-Eguchi-Kawai reduction  with judiciously chosen discrete flux satisfies large-$N$ volume independence  \cite{GonzalezArroyo:2010ss}.)  
   Very importantly,  in our set-up, the fractional 
 instantons induce effects such as  $\exp [ - \frac{8\pi^2}{g^2N} + i \frac{\theta}{N}]$.  One of the most interesting non-perturbative phenomena in infinite volume gauge theory is that observables are  $N$-branched functions, where each branch is $2 \pi N$ periodic, yet, the observables are $2 \pi$ periodic, with cusps at $\theta=\pi$  \cite{Witten:1980sp}. 
 The  set of fractional instantons (including critical points at infinity)  are  compatible with a complete Fourier basis,   which can  
 describe such non-trivial observables with multi-branched structure, and this compatibility of saddles with a complete basis for Fourier expansion of observables is  far from trivial.  Indeed, we have the same intentions with 
  Luscher  and van Baal,  but thanks to the GNO and discrete  flux backgrounds we use, our quantum mechanical constructions have a  remembrance of fractional  topological configurations  on $\R^3 \times S^1$ and  the configurations in $PSU(N)$ bundle even on large-$T^4$. Therefore, the constructions with flux are more powerful  strategies   to understand non-perturbative Yang-Mills dynamics across different length scales. 
  
  Finally, perhaps, a point that is not sufficiently understood (or emphasized)   in the past, is the fact that the flux backgrounds have a re-interpretation 
  as coupling the QFTs to TQFT (backgrounds) \cite{Kapustin:2014gua}. These theories are globally different, but  locally the same. For example, mass spectrum are completely identical on large $T^4$.  The TQFT coupling instructs us that the existence of fractional topological charge configurations is independent of $T^4$ size, and is valid at  both weak and strong coupling.  This is explored in detail in present work and in a companion paper \cite{Unsal:2021cch}.

\section{Quantum mechanical $T_N$ and $(T_N/\Z_N)_p$ models }
Consider quantum mechanics of a particle on a circle 
with a potential of the form 
\begin{align} 
V(q) &= - \cos(Nq),  \qquad q \sim q+ 2\pi 
\end{align}
$q$ is physically identified with $ q+ 2\pi$.  The theory has a $\Z_N$ global shift symmetry, 
 \begin{align} 
\Z_N:  q  \mapsto  q+  \frac{2\pi}{N} 
\label{sym}
\end{align}
The system has $N$-degenerate minima to all orders in perturbation theory.  
 The  partition function, $Z(\beta)=\tr[\exp(-\beta H)]$,  in the $\beta \rightarrow \infty$ limit,   is dominated by the lowest-$N$ states. Within the  Born-Oppenheimer approximation, one can forget about higher states in the spectrum, as non-perturbative instanton induced splittings are exponentially smaller than splittings between harmonic oscillator levels. In this limit, the system may be  viewed as  $N$-site lattice model with nearest neighbor hopping interaction, described by a tight-binding approximation.  For convenience, we call this system $T_N$ model. See Fig.\ref{fig:TN}.  Some semi-classical and anomaly  aspects of this model are  studied in  \cite{Unsal:2012zj, Kikuchi:2017pcp,Gaiotto:2017yup}.
 
This  simple system provides a playground for some  ideas  that we will  employ in diverse  non-trivial QFTs. Hence, we explore in detail 
the connection between various ways of looking to this system and interpret them in simple physical terms. The connection between the 
$\Z_N$  twisted boundary conditions, and turning on  $\Z_N$  background field is  well known. These constructions have a precise interpretation as coupling QM to $\Z_N$ TQFT, which is fairly useful in QFT. 
Later, we may also choose to gauge $\Z_N$  symmetry. 
  This is equivalent to summing over all possible background fields or twisted boundary conditions, or performing a path integral over the TQFT.   Gauging, of course, changes the theory, from 
$T_N$ to $T_N/\Z_N$. In the latter, $q$ and  $q+  \frac{2\pi}{N}$ are physically identified. i.e, we end up with the particle on a circle with 1-minimum on the circle.  However, there are different versions of this 1-site theory that we call 
$(T_N/\Z_N)_p$, depending how we gauge  $\Z_N$.  The $p$ in the $(T_N/\Z_N)_p$ has three equivalent, but somehow different sounding interpretations, as  discrete theta angle $\theta_p$, or  level $p$ Chern-Simons  term that is present in TQFT, or  picking  Bloch state with quasi-momentum $p$ from the Hilbert space of the original theory, which we describe in detail, as it translates verbatim to QFT.

 \vspace{0.5cm}
\noindent
{\bf Basics:}
The path-integral representation of the partition function  is 
\begin{align}
 Z(\beta) & = \tr[\rme^{-\beta H}]= \int_{q(\beta)= q(0)} \Diff q \;  \exp(-S[q]),    \qquad {\rm where} \cr  
 S[q] &= \frac{1}{g} \int d \tau \left( \half \dot q^2 + V(q) \right)  + \frac{ \im \theta}{2 \pi} \int \diff q 
 \label{action}
\end{align}
The trace   $\tr[ \cdot ]$ in the operator formalism translates in the path integral formalism to integration over  all  paths obeying the {\it periodic} boundary condition, $q(\beta)= q(0)$.   The periodic paths are classified according to the homotopy group of the maps    
\begin{align}
q(\tau): S^1_\beta \rightarrow S^1,   \;\; \pi_1(S^1)=\mathbb{Z}
\end{align}
 which is nothing but winding number (corresponding to  topological charge):
\begin{align}
W= \frac{1}{2\pi} \int \diff q \in {\mathbb Z}.  
\end{align}
 which is integer valued.

Despite the fact that the partition function is given by summing  over integer winding number  
$W \in \mathbb Z$
configurations,   it is  clear  that there are  classical solutions with fractional topological charge,  called fractional instantons, see  Fig.\ref{fig:TN}.\footnote{These are 
obvious in quantum mechanics, but not  always so obvious in QFT. This  is the reason why  we lay out  a somewhat abstract formalism  in QM, which carries over to QFT.  Obvious facts in QM will map to non-obvious statements in QFT  both at weak and strong coupling on $d$-manifolds, $M_d$ including $\R^d$ limits.  } The configuration with $W=1$ is  an instanton, 
and its action is $S_{W=1}\equiv S_I$.  This  is our yard-stick throughout  the problem. The  configurations with    $W=1/N$ are referred to as  fractional instantons, and their action is $S_I/N$. These are tunneling configurations connecting neighboring minima $j\rightarrow j+1$. 
 Clearly, a single fractional instanton ${\cal I}_j$ cannot contribute to the partition function, as it does not obey the periodic boundary condition in \eqref{action}.  Equivalently, these are periodic paths  in $S^1/\Z_N$ which cannot be lifted to   $S^1$.  
 However, an   ${\cal I}_j$,   $\bar {\cal I}_j$ pair, which has $W=0$,  but action $2 S_I/N$,  {\bf must} contribute to the partition function  since it is a periodic path in $S^1$.   
 
  \vspace{0.5cm}
\noindent
{\footnotesize
{\bf Digression on critical points at infinity and resurgent cancellations:} 
  The  ${\cal I}_j$,   $\bar {\cal I}_j$ configurations can be viewed 
 as critical points at infinity.  Here, we will go over this concept fairly quickly.  See  \cite{Behtash:2018voa} for details and \cite{Dunne:2016nmc} for broad review. 
  The fact that  the combination of 
 ${\cal I}_j$ and  $\bar {\cal I}_j$ configuration  
 is often  called ``unstable" in old literature is a red-herring.  
 In semi-classics,  the critical point at infinity has a non-Gaussian Lefschetz thimble  (steepest descent cycle)  that one must  integrate  over. This contribution  maps to second order terms  in cluster expansion as
  \begin{align}  
  \frac{\beta^2}{2!} [{\cal I}_j] [\bar {\cal I}_j]  + \frac{\beta}{1!} [{\cal I}_j \bar {\cal I}_j]_{\pm}   
  \label{cluster}
\end{align}  
 where the first, maximally extensive term in $\beta$ is dilute (non-interacting) instanton gas contribution and sub-extensive term 
 capture the correlated  instanton-anti-instanton event $ [{\cal I}_a \bar {\cal I}_a]_{\pm}$  which is two-fold ambiguous.  Remarkable fact that 
  comes from resurgent analysis is that this ambiguity cancels against  the ambiguity of  lateral Borel  resummation of  perturbation theory  for vacuum energy,  $\mathbb B E_{0,\pm} $. 
  At second order in semi-classics, we find the cancellation, and 
   \begin{align}  
 \Im \mathbb BE_{0,\pm} + \Im [ {\cal I}_j \bar {\cal I}_j]_{\pm} =0
\end{align}   
The combination is unambiguous and meaningful within resurgent semi-classical analysis framework \cite{Dunne:2016nmc}.  
}
 
   \vspace{0.5cm}

Strictly speaking,   action $2 S_I/N$ and $W=0$  configurations can equivalently be viewed as uncorrelated events  
$ [{\cal I}_j]$  and $[\bar {\cal I}_j] $ and correlated bion events   $[{\cal I}_j \bar {\cal I}_j]_{\pm}$, for which   its real unambiguous part 
 $\Re [{\cal I}_j \bar {\cal I}_j]$   contribute to the partition function, as both correspond to  periodic paths in $S^1$.\footnote{Throughout the paper, we will not bother much with the resurgent structure, but it  is often  at the back of our minds. It is important that most of the critical points that contribute to actions in QFTs and QM are ``critical points at infinity", and is decomposable to BPS and non-BPS configurations.  We encourage the reader to go over 
   \cite{Behtash:2018voa} if necessary.   This background knowledge allows us to take some steps  swiftly in the course of this work. }
  Similarly,   configuration such as an ordered product  $ \prod_{j=1}^{N}   {\cal I}_j =  {\cal I}_1 {\cal I}_2 \ldots  {\cal I}_N$ with  integer topological charge $W=1$,  and integer action $ 1 \times S_I$ must contribute to the partition function  since it is also a  periodic path.  
  More generally, generically non-BPS configurations with  action $(W + \frac{2 \bar n}{N})  S_I$ and topological charge 
  $W \in \Z$ contribute to partition function.

 Let us now  derive the  fractional instanton decomposition of the path integral.  Since we already know the energy spectrum, we can take a short cut and   use it to derive the fractional instanton sum and see explicitly    the paths that contribute at leading semi-classics.

Within the Born-Oppenheimer approximation, we can write a tight-binding Hamiltonian to describe low energy physics 
(We shifted zero point energy  $\hbar \omega/2$ to zero, 
and $ \xi =  K   \rme^{-S/N} $ is the non-perturbatively small hopping parameter.) 
\begin{align}
H= - \sum_{j=1}^{N} \xi \rme^{i \theta/N} | j+1\rangle \langle j| + {\rm h.c.}  
\end{align} 
The  energy spectrum is given by diagonalizing the Hamiltonian: 
\begin{align} 
E_k(\theta)= -   2 K   \rme^{-S/N}   \cos  \frac{ \theta+ 2 \pi k} {N}  
\label{spectrum}
\end{align}
 The ground state depends on what the theta angle is.  The ground state  energy is  $ E_{\mathrm{G.S.}}(\theta)= \min_{k}  E_k(\theta)$, 
 exhibiting a two-fold degeneracy at odd-integer multiples of  $\theta=\pi$. 
 
Let us now turn the  partition function into a fractional instanton sum:
\begin{align} 
 Z(\beta, \theta)&=  \sum_{k =0}^{N-1}   \rme^{  2 \beta  K  \rme^{-\frac{S}{N}} \cos  \frac{ \theta+ 2 \pi k} {N} } \cr
&=  \sum_{k =0}^{N-1}   \sum_{n=0}^{\infty}   \sum_{\bar n =0}^{\infty}   \frac{1}{n!} \frac{1}{\bar n !}  \left( \beta K  \rme^{-\frac{S}{N} + \im  \frac{ \theta+ 
2 \pi k} {N}}  \right)^{n }     \left( \beta K  \rme^{- \frac{S}{N} -  \im \frac{\theta+ 2 \pi k} {N}}  \right)^{\bar n }    
\label{form2}
\end{align}
We can perform the summation over $k$  by using  the identity 
\begin{align}  
 \sum_{k=0}^{N-1}  
  \rme^{ \im 2 \pi   k (n-\bar n)/N}  = N \sum_{W \in \Z} \delta_{n-\bar n  - WN} 
  \end{align}
  converting the sum over $k$, which takes place in   the space of representations of $\Z_{N}$ translation symmetry    
  into   a sum over $W \in \Z$ which we will interpret as winding number. This gives the path integral in the leading order of  dilute gas approximation. 
  \begin{align} 
 Z(\beta, \theta) 
&= N  
 \sum_{W \in \Z}  \sum_{n=0}^{\infty}   \sum_{\bar n =0}^{\infty}   \frac{1}{n!} \frac{1}{\bar n !}  \left( \beta K  \rme^{-S/N + \im \theta/N}  \right)^{n}  \left( \beta K  \rme^{-S/N - \im \theta/N}  \right)^{\bar n}    \delta_{n-\bar n  - W N, 0} 
\label{constrained}
\end{align}
 where delta function gives a constraint on the fractional instanton sum. 
\eqref{constrained} can be interpreted as a grand canonical ensemble of fractional instantons. For a  term in the sum  with 
 $n$ fractional instanton, and  
$ {\bar n }$ fractional anti-instantons, the periodic orbits correspond to: 
\begin{align}
 n-\bar n  -W N=0,  \qquad  {\rm {i.e.,}} \;\;  n-\bar n=0 \;\;  {\rm mod} \;  N 
\end{align} 
These configurations which contribute to $ Z(\beta, \theta)$   comes with  fugacity 
\begin{align}
\rme^{-\frac{S_I}{N}( n+\bar n)} \;  \rme^{\im  \frac{\theta}{N}  (n- \bar n) }  = \rme^{- \left(W + \frac{2 \bar n}{N}\right)  S_I}  \;  \rme^{\im  W \theta } 
\label{form1}
\end{align} 
i.e, they possess fractional action, but  integer topological charge.

\begin{figure}[t]
\vspace{-.1cm}
\begin{center}
\includegraphics[width = 0.8\textwidth]{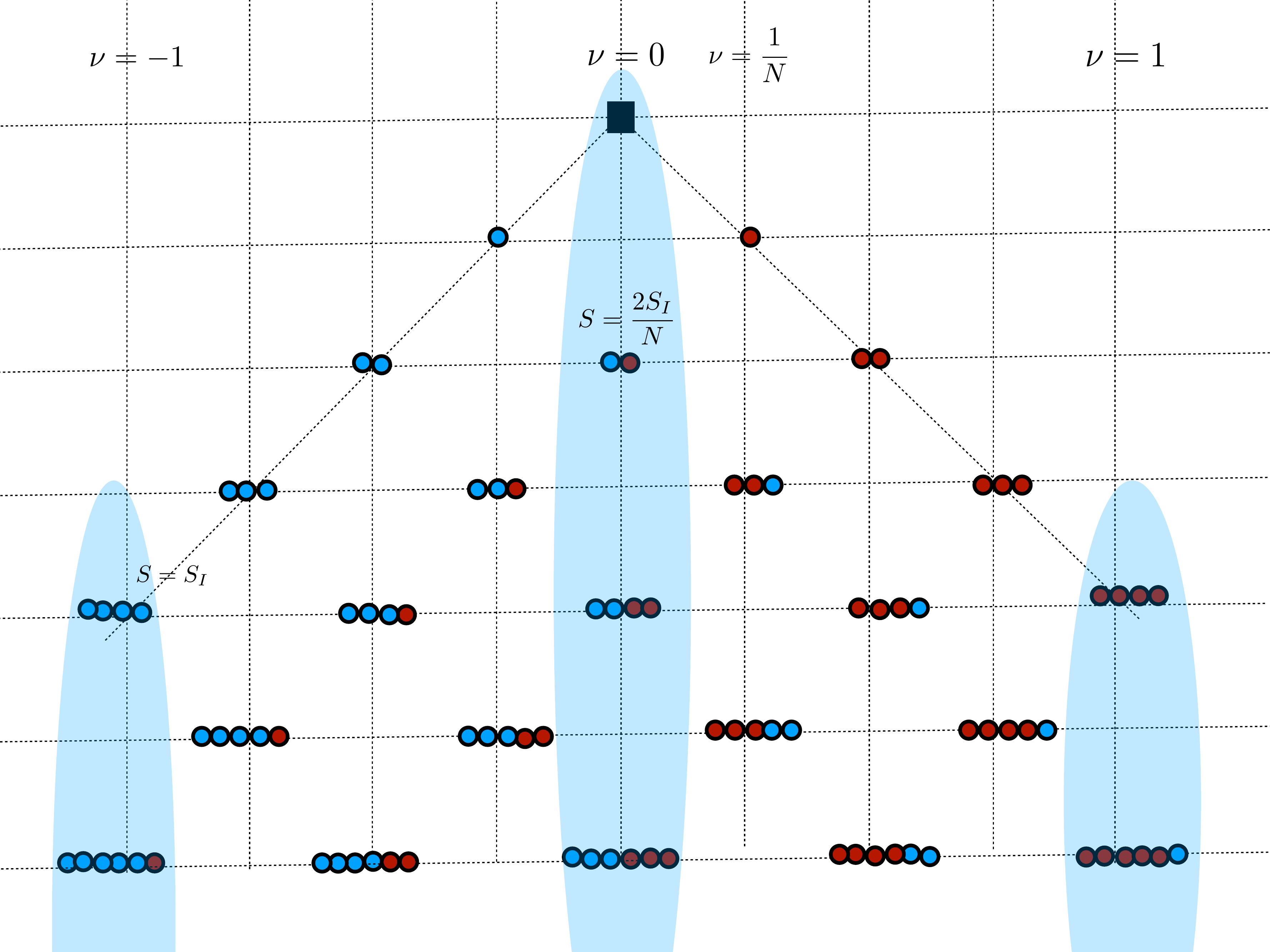}
\vspace{0.cm}
\caption{Only integer winding (topological charge) configurations corresponding to periodic paths contribute  to partition function. To detect fractional instantons in QM is visually trivial (See Fig.1).  Formally, it requires  either twisted boundary condition by using $\Z_N$ shift symmetry, 
or coupling to 
$\Z_N$ topological gauge theory.  The latter more abstract formalism is more useful in gauge theories in four-dimensions, to understand the true nature of saddles contributing to  Yang-Mills theory. 
Despite the fact that $W=\frac{1}{N}, S=S_I/N $ fractional topological charge and fractional  configurations are not contributing to partition function, $W=0, S=2 S_I/N $ fractional action configurations do. 
This is responsible for the emergence of multi-branch structure as a function of theta angle.  This construction tells us that trying to understand the saddles in $SU(N)$ gauge  theory by working directly with $SU(N)$ bundle is   not  correct. It neglects most  important contributions. 
 }
\label{fig:saddles}
\end{center}
\end{figure}

The simple observation, that saddles with fractional action, but integer topological charge contributes  to the path integral (see Fig.~\ref{fig:saddles}),  is at the heart of  all reliable semi-classical analysis of gauge theories, and sigma models, such as ${\cal N}=1$ SYM, Yang-Mills with double-trace deformations, and $\mathbb CP^{N-1}$ models, and is responsible for the multi-branched structure of the vacuum energy.     This structure naturally arise in the context of 
resurgence and sometimes called graded  resurgence triangle \cite{ Dunne:2012ae, Cherman:2014ofa}. 

In \eqref{constrained}, $K$ is the instanton prefactor 
that is not important for this particular discussion, and we will  ignore  it throughout the paper. 
$\beta$ is there due to position zero mode of a single  fractional instanton.  $\frac{\beta^n}{n!}$ arises due to the  integration  over  the positions  of $n$ fractional instantons, see e.g. \cite{Coleman198802}.    The overall factor $N$ is due to $N$ distinct classical minima in the $q \in [0,2\pi)$ fundamental domain.

The constrained sum can be simplified further. Solving the constraint, 
  \begin{align} 
 Z(\beta, \theta) 
 &= N \sum_{W \in \Z}   \left[ \sum_{\bar n=0}^{\infty}  \underbrace{ \frac{  (\beta  K  \rme^{-\frac{S_I}{N}})^{\bar n}}{{\bar n}!}  }_{ \bar n-\rm fractional \;  anti-inst.}
\underbrace{\frac{ (\beta  K  \rme^{-\frac{S_I}{N}})^{\bar n+  WN}     }{ (\bar n+ WN)!} }_{(\bar n+NW)- \rm fractional \;  inst.} \right] \rme^{\im  W \theta }  \cr
 &= N \sum_{W \in \Z}   \left[ I_{NW} (2 \beta  K  \rme^{-\frac{S_I}{N}} ) \right]  \rme^{\im  W \theta }  
\label{constrained-2}
\end{align}
where $I_{NW}$  is the modified Bessel function of order $NW$. We learn that  the series expansion of the modified Bessel function is actually an instanton sum. Since 
$Z(\theta+ 2 \pi)=  Z(\theta)$, \eqref{constrained-2} is nothing but the Fourier series expansion of the partition function:
\begin{align}
Z(\theta) = \sum_{W \in \Z} Z_W \rme^{\im  W \theta }, \qquad  \qquad  Z_W = N I_{NW} (2 \beta  K  \rme^{-\frac{S_I}{N}} )
\label{fourier}
\end{align}
Despite the fact that only  periodic orbits  with the  theta angle dependence  $e^{\im W \theta}$ contribute, the  non-trivial theta-angle dependence  which leads to multi-branched structure and $\theta$ dependence in the observables in the form $ \frac{ \theta+ 2 \pi k} {N}$ naturally arises! 
This fact will also help us in gauge theories on $\R^4$ and   $\R^3 \times S^1$. 

\subsection{Coupling to  $\Z_N$ TQFT background}
The quantum mechanical model we are studying  has a $\Z_N$ global  0-form shift symmetry.    First, we describe coupling it to a $\Z_N$ background gauge field. This stage  will help us to reveal the existence of fractional instantons that actually exist in the $T_N$ model, in a  more abstract 
way. But this somehow abstract formalism help us greatly in the QFT applications.   
 We can later choose to sum over    all possible backgrounds, which  is equivalent to gauging   $\Z_N$ completely, reducing the model to $T_N/\Z_N$, which is a 1-site model. 
  In our discussion, we aim to make the details fairly explicit and conceptual.  This will provide important insights that we will use in QFT on $d$-manifolds $M_d$. 

\vspace{0.5cm}
\noindent
{\bf  Coupling QM to TQFT:} We first couple our quantum mechanical model  to a  classical $\Z_N$ topological gauge theory  background \cite{Gukov:2013zka, Kapustin:2014gua}. 
The quantum topological theory is given  by the partition function
 \begin{align}
 Z_{{\rm top}, p} &= \int  \Diff   A^{(1)} \Diff A^{(0)}    \Diff  F^{(0)} \;  \rme^{ \im \int F^{(0)}  \wedge ( N A^{(1)}-\diff A^{(0)}) + \im p \int A^{(1)}}  
 \label{top}
 \end{align}
 Here, $F^{(0)}$ is a Lagrange multiplier zero-form field that we integrate over, and $i p \int A^{(1)}$ term is Chern-Simons term in $1d$, and $p$ must be an integer 
 for the theory to be gauge invariant.\footnote{This term will provide different possibilities in gauging. It will turn out to be related to discrete theta angle, which in turn, determines which Bloch state is picked from the lowest $N$ state after gauging. For now, $p$ term is not very important  and can be set to zero if desired.}

 Integrating $F^{(0)}$, we end up with a   pair, $(A^{(1)}, A^{(0)})$ which obeys 
\begin{align}
N A^{(1)}= \diff A^{(0)}, \qquad N \int A^{(1)}=  \int \diff A^{(0)} = 2 \pi \Z
\label{ZN}
\end{align}
and is  properly quantized.   The path integration of TQFT is fairly trivial, and yields: 
 \begin{align}
 Z_{{\rm top}, p}  &=  \int  \Diff   A^{(1)} \Diff A^{(0)}     \delta(  ( N A^{(1)}-\diff A^{(0)})    \;\; \rme^{ \im p \int A^{(1)}}  \cr
 &= \frac{1}{N}  \int \Diff A^{(0)}        \;\; \rme^{ \im  \frac{p}{N} \int dA^{(0)}}  \cr
 &= \frac{1}{N}   \sum_{k=0}^{N-1}  \rme^{  \im \frac{2 \pi p k}{N} }=  \delta_{p=0  \; {\rm mod} \; N}
 \label{top-2}
 \end{align}
$(A^{(1)}, A^{(0)})$   pair describe  a $\Z_N$  gauge field that can be turned on in quantum mechanical $T_N$ model  to probe  saddles, in particular, to probe the fractional instantons. 

The 0-form gauge invariance  of the action \eqref{top} is given by
\begin{align}
 A^{(1)} \mapsto   A^{(1)}  + \diff \lambda^{(0)},  \qquad  A^{(0)} \mapsto   A^{(0)} + N \lambda^{(0)}, \qquad  F^{(0)} \mapsto   F^{(0)} 
\end{align} 
To couple  a  classical $\Z_N$  background  field to the $q$-field, we declare 
 \begin{align}
q \mapsto  q - \lambda^{(0)}, 
\end{align}
Gauge invariant combinations of our quantum field $a$ and classical fields $(A^{(1)}, A^{(0)})$  are:  
\begin{align}
Nq+   A^{(0)}, \qquad   dq  + A^{(1)}= (\dot q + A_\tau)\diff \tau 
\end{align}
Hence, the partition function  in the classical $(A^{(1)}, A^{(0)})$ background can be found by promoting  $dq$ and $q$ to these gauge invariant combinations: 
\begin{align}
 Z &[ (A^{(1)}, A^{(0)}), p]   =\int    \Diff  F^{(0)} \;   \int_{q(\beta)=q(0)}  \Diff q \;  
  \rme^{ \im \int F^{(0)}  \wedge ( N A^{(1)}-\diff A^{(0)})   +  \im p \int A^{(1)}}   \cr
&  \times  \exp\left(  -\frac{1}{g} \int d \tau \left( \half (\dot q + A_\tau)^2 - \cos(Nq + A^{(0)}) \right)  + \frac{ \im \theta}{2 \pi} \int (\diff q   + A^{(1)} ) \right) \qquad  \
 \label{background2}
\end{align}
Note that at this stage,  we do not perform a path integral over $(A^{(1)}, A^{(0)})$. These are truly classical fields that we turn on according to what we want to do. Moreover, the path integral  $\int \Diff q$ is over periodic paths.

\vspace{0.5cm}
\noindent
{\bf Equivalence of symmetry twisted boundary condition to coupling QM to TQFT:}
The  $\Z_N$ background gauge field is equivalent to the twisted boundary condition in path integral.  Let us show this explicitly. Let $\mathsf {U}$ denote the translation operator  for $\Z_N$ shift symmetry  in the $T_N$ model.   Then,  the sum over  transition amplitude in quantum mechanics for sites $\ell$ units apart is given by: 
\begin{align}
 Z_\ell  & =   \tr [\rme^{-\beta H}  \mathsf {U}^\ell  ]  = \sum_{j=1}^{N} \langle j+ \ell | \rme^{-\beta H} |j \rangle  = 
  \int_{y(\beta)= y(0) + \frac{2 \pi}{N} \ell } \;  \Diff y \exp(-S[y]),   
  \label{tpf}
 \end{align}
 where $y(\tau)$ are paths obeying twisted boundary condition.  
Now, let us undo the twisted boundary condition in favor of a periodic boundary condition, and insert  its effects into the action. 
 Define 
 \begin{align}
 q(\tau)= y(\tau) - \frac{2 \pi \ell}{N \beta} \tau, \qquad   {\rm hence}   \;\; q(\beta)= q(0) \;\; {\rm mod} \; 2\pi.
 \end{align} 
 Plugging this into original action $S[y]$,   we find 
 \begin{align}
 S[q, \ell]= \frac{1}{g} \int d \tau \left[ \half \Big(\dot q+ \frac{2 \pi \ell}{N \beta} \Big)^2   - \cos \Big( Nq   + \frac{2 \pi \ell}{ \beta} \tau \Big)  \right]
 + \frac{ \im \theta}{2 \pi} \int \Big(\diff q +  \frac{2 \pi \ell}{N \beta} \diff \tau \Big)
 \label{bgrd}
 \end{align}
 which agrees  exactly  with  \eqref{background2}, with the  identification: 
 \begin{align}
A^{(1)}=  \frac{2 \pi \ell}{N \beta} \diff \tau,  \qquad  A^{(0)}=  \frac{2 \pi \ell}{ \beta} \tau
\end{align}
Happily, these obey  the conditions for a background $\Z_N$ gauge field  \eqref{ZN}. 
 
So, the twisted boundary conditions in \eqref{tpf}  are traded for a background field, and the partition function can equivalently be written as: 
\begin{align}
 Z_\ell  & =
  \int_{q(\beta)= q(0) } \Diff q  \;  \exp(-S[q,  \ell]),   
  \label{tpf-2}
 \end{align}
Note the explicit time dependence in the action \eqref{bgrd}. This is not a fluke,  it is in fact exactly as it should. This is the classical background that we are dealing with.  

One final remark concerns instanton equations in the $\Z_N$ background field. Using  Bogomolny factorization  with action \eqref{background2}, we find 
\begin{align}
(\diff q   + A^{(1)} ) = \pm [2V (Nq + A^{(0)})]^{1/2}, \qquad   \;\; q(\beta)= q(0) \;\; {\rm mod} \; 2\pi.
\label{ins1} 
\end{align} 
which is nothing but 
\begin{align}
\dot y   =   \pm [2V (Ny) ]^{1/2}, \qquad   \;\; y(\beta)= y(0) + \frac{2 \pi}{N} \;\; {\rm mod} \; 2\pi.
\label{ins2} 
\end{align} 
in the $y$ parametrization. So, to find the fractional instanton solution, we can  work with either parametrization.  Either request periodic path solutions  with background fields or paths solving \eqref{ins2} and satisfying twisted boundary conditions. These two simply related interpretation will help us to figure out a non-trivial solution in $PSU(N)$ bundle in  gauge theories on $\R^3 \times S^1$. 

In the  formulation   where $q(\tau)$ is a periodic path in the $(A^{(1)},  A^{(0)})$ background \eqref{background2},  the topological term is 
\begin{align}
 W= \frac{ 1}{2 \pi} \int (\diff q   + A^{(1)} ) )    
  =  \underbrace{\frac{ 1}{2 \pi} \int \diff q}_{\in \Z}   +   \underbrace{\frac{ 1}{2 \pi N} \int \diff A^{(0)} }_ {\in \frac{1}{N} \Z}        \in \frac{1}{N} \Z
  \label{pbc-1}
\end{align} 
where the first term is integer valued as it is over periodic paths, and the winding number is  fractional due to background.  
In the  formulation  where $y$ is a path satisfying twisted boundary condition   \eqref{tpf}, 
\begin{align}
 W= \frac{ 1}{2 \pi} \int \diff y     \in \frac{1}{N} \Z
 \label{tbc-1}
\end{align} 
This  is  now fractional due to the twisted boundary condition of  $y(\tau)$ field. 
The action corresponding to these fractional winding configurations are  also fractional $S= \frac{1}{N} S_I$  due to BPS  bound.

\vspace{0.5cm}
\noindent
{\bf Revealing fractional instantons formally:}
The existence of fractional instantons with fractional winding number is obvious from Fig.\ref{fig:TN}.  
Expressing the    partition function  $ Z[(A^{(1)},  A^{(0)}) ] \equiv Z_\ell (\beta, \theta)$  as an  fractional instanton sum.   
\begin{align} 
 Z_\ell (\beta, \theta) &=  \tr [  \rme^{-\beta H}   \mathsf {U}^\ell ]   \cr
&= \sum_{k =0}^{N-1}   \rme^{ \im \frac{2 \pi \ell k}{N}}   \rme^{  \xi \cos  \frac{ \theta+ 2 \pi k} {N} } \cr
&=  \left[ N \sum_{W \in \Z}   \sum_{\bar n =0}^{\infty}   \frac{    \xi^{ 2\bar n  + \ell+ W N   }   }{  (\bar n  + \ell+ W N)! \; \bar n !} 
     \rme^{ \im   W  \theta}  \right]   \rme^{ \im  \frac{\ell}{N} \theta }  \cr
     &=   \left[ N \sum_{W \in \Z}  I_{NW + \ell} (2 \xi)  \rme^{ \im   W  \theta}  \right]   \rme^{ \im  \frac{\ell}{N} \theta }  
\label{form4} 
\end{align}
The second line immediately emerges from our knowledge of energy  spectrum of the theory  \eqref{spectrum}  and the fact that eigenstates of the Hamiltonian are just 
Bloch states, and  $[ \mathsf {U}, H]=0$. 
To obtain  the third line, we can  expand the exponentials and sum over $k$ explicitly.  Not surprisingly, this is equivalent to the dilute gas of fractional instantons in the path integral picture. The fourth line is a nice mathematical identity which tells us that series expansion of {\it all} modified Bessel functions are actually instanton sums.   
The sum is over $\bar n$, and configurations that contribute are actually  $\bar n$ fractional anti-instantons,  and $(\bar n  + \ell+ W N)$ fractional instantons. 
The winding number of such a configuration is $ W + \frac{\ell}{N}$, and the sum has an overall  $\rme^{ \im   \frac{\ell}{N} \theta }$ dependence that arise from the fractional part of winding number.  These fractional instantons contribute to transition amplitudes in quantum mechanics.

Note that by turning on $(A^{(1)}, A^{(0)}) \equiv \ell$ background, we are not changing the theory or its Hilbert space, 
which, within Born-Oppenheimer approximation is just $N$ dimensional, that we denote as ${\cal H}_N$.  In simple words, we are summing over just transition amplitudes in quantum mechanics for states which are $\ell$ units apart.   The construction merely reveals the topological configurations which  play important role in the full theory.

Probably, one of the counter-intuitive sounding aspects  of this construction (yet widely familiar 
from the symmetric double-well potential, see  \cite{ZinnJustin:2002ru})  is following. We keep repeating that single fractional instanton $\I_j$  is not contributing to the partition function, as it is not a periodic path.  Yet, the energy gap in the theory  is sourced by single fractional instanton effect. The reason is  clear. Single fractional instanton gives rise to the transition matrix element between nearest neighbor sites,  which in turn determine the splittings  in the spectrum of the theory. 
 Say, at $\theta=0$, the gap in the spectrum is given by 
\begin{align}
 E_{\rm gap}= E_{k=1} -  E_{k=0}    = - 2 K   \rme^{-S_I/N}  \Big( \cos \ \frac{  2 \pi } {N}   -1\Big), 
   \label{gap0}
 \end{align}
a single fractional instanton effect, similar to the gap in the symmetric double-well.  

\subsection{Uses of gauging $\Z_N$} 
{\bf Discrete theta angle $\theta_p$ = level $p$ Chern-Simons = picking  Bloch state with momentum $p$}
Now, we can gauge  $\Z_N$ symmetry. Physically, gauging declares that instead of   $q$  being  physically equivalent to 
$q + 2 \pi$, it is identified with $q + \frac{2\pi}{N}$. With this procedure, the target-space  circle size is  effectively reduced to   $\frac{2\pi}{N}$.  In other words, $N$-site model is reduced to $1$-site,  and out of $N$ states in the lowest band, only one state survives.  Gauging, unlike turning a background as described above, changes the Hilbert space, by diluting it. It  reduces  ${\cal H}_N$ down to one state Hilbert space  ${\cal H}_1$.

Gauging $\Z_N$ symmetry amounts to summing over $\ell$,  $\sum_{\ell=0}^{N-1}  Z_\ell$. However, this is not the only choice in summation.  In fact, 
the sum over  $Z_\ell (\beta) $ can be done in $N$-different ways, corresponding to $N$ discrete $\theta$ angles, $\theta_p=  \frac{2 \pi  p }{N}$.   
This $\theta_p$ is nothing but the 1d  Chern-Simons term in the original topological gauge theory. 
As a result, 
\begin{align}
  Z_{( T_N/\Z_N)_p} & =  \int  \Diff A^{(1)}  \Diff A^{(0)}  \;  Z[ (A^{(1)}, A^{(0)}), p]   \;  \delta( N A^{(1)}-\diff A^{(0)}) \cr
 & \equiv \frac{1}{N}  \sum_{\ell=0}^{N-1} \rme^{- \im \frac{2 \pi \ell p }{N}}     \; Z_\ell   
  \label{tpf2}
 \end{align}
As already stated, as a result of gauging, the $N$-site model reduce to 1-site model, and  Hilbert space  gets diluted by a factor of $N$. Which state survives gauging  is 
dictated by  the choice of   $\theta_p$,  or equivalently, Chern-Simons level $p$.  
Using \eqref{form4} 
\begin{align}
  Z_{( T_N/\Z_N)_p}  &=  \frac{1}{N}   \sum_{\ell=0}^{N-1}  \rme^{- \im \frac{2 \pi \ell p }{N}}    Z_\ell (\beta) \cr
  &=   \frac{1}{N}   \sum_{\ell=0}^{N-1}  \rme^{- \im \frac{2 \pi \ell p }{N}}     \sum_{k =0}^{N-1}   \rme^{ \im \frac{2 \pi \ell k}{N}}   \rme^{  \xi \cos  \frac{ \theta+ 2 \pi k} {N} }  \cr 
  & =  \sum_{k =0}^{N-1}   \delta_{pk}  \rme^{  \xi \cos  \frac{ \theta+ 2 \pi k} {N} }   \cr
&=   \;  \rme^{  \xi  \cos  \frac{ \theta+ 2 \pi p} {N} } 
 \end{align}
 Only  the Bloch state with momentum $\frac{2 \pi p} {N}$ in the original Brillouin zone of the $N$-site model survives in the Born-Oppenheimer approximation. 

If we relax Born-Oppenheimer approximation and consider the whole Hilbert space,  then, the gauging procedure will pick just one-state with Bloch momentum $p$ from each band.   In this sense, the whole Hilbert space gets diluted by a factor of $N$. 

Finally, one can formalize the connection between   the topological configurations in the $( T_N/\Z_N)_p $ model, 
 the $ T_N$ model with background  $(A^{(1)}, A^{(0)}) \equiv \ell$ or the original theory without background fields.
  This is one of the crux of the matter that is most important in gauge theory where things are a bit more involved. 
  The object that we would  identify as instanton   with winding number $1$  and action $S_0$ in 
   the $( T_N/\Z_N) $ model is  the fractional instanton  with winding number $\frac{1}{N}$ and action $S_0= \frac{S_{I}}{N}$  of the $T_N$ model.   Therefore,   $( T_N/\Z_N)_p $  theory has non-perturbative data that is relevant to the original  $T_N $ theory.  
   
  Let us make this a bit more precise. 
 When we  gauge $\Z_N$, and sum  over all  $(A^{(1)}, A^{(0)})$,  (instead of just turning on  a  fixed classical  $\Z_N$ background), 
 the distinction between  isolated minima and  fractional instantons ${\cal I}_j$ disappear. We can in fact call
\begin{align}
{\cal I}_j  \mapsto  {\cal I}_1,  \; \forall j  \qquad {\rm by \;  gauging}\;   \Z_N
\end{align}
Then, we can make an obstruction vs. admission list: 
 \begin{itemize}
\item Instantons of the form   $[{\cal I}_1]^k, k\neq 0  \; {\rm mod} \;  N $,  which are   closed paths in $S^1/\Z_N$  cannot be lifted to $S^1$. 
\item  Instanton anti-instanton pairs  of the form $ [{\cal I}_1]^k [\bar {\cal I}_1]^k $ are closed paths   in  $S^1/\Z_N$,  and they can be   lifted to $S^1$ with $W=0$. 
These configurations possess fractional  quantized action in the original (ungauged) formulation. 
\item   Instantons in multiples of $N$   $[{\cal I}_1]^{Nq}, \; q \in \Z $  can also be   lifted to $S^1$, with $W=q$. These configurations possess integer quantized  action. 
\end{itemize}
   These facts  are by no means surprising, in fact, rather simple.   But  this leads us to the following intriguing situation. 
   \begin{itemize}
\item Since in order to find the  saddles in $ T_N$ model, we need to directly look at   either  
\begin{itemize}
\item   $ T_N$  with classical $\Z_N$  background $(A^{(1)}, A^{(0)})$  or equivalently, $\Z_N$  twisted boundary conditions 
\item     $( T_N/\Z_N)_p $ models where $\Z_N$ is gauged
\end{itemize} 
we claim that in order to find  the relevant saddles of $SU(N)$ gauge  theories,  we should first consider either  $SU(N)$ with  background gauge  field $(B^{(2)},B^{(1)}) $  for the  $\Z_N^{[1]}$  1-form symmetry or  $(SU(N)/\Z_N)_p$ theory where $\Z_N^{[1]}$ is gauged.\footnote{This is one reason we parallel the notation of quantum mechanics with gauge theory, and emphasize the correspondence: 
$T_N \Leftrightarrow SU(N)$ and  $(T_N/\Z_N)_p \Leftrightarrow  (SU(N)/\Z_N)_p$. The latter systems has      non-perturbative data that is relevant to the original  systems,  where it is interpreted as fractional.}
After finding these configurations, we can  patch  them up   to find the ones that can be lifted to $SU(N)$ theory.  These are  fractional action  (e.g. $2S_I/N$)   configurations that contribute to the partition function of  $SU(N)$ theory.  
\end{itemize}

\section{$ \mathbb{CP}^{N-1}$  in 2d and  $\Z_N$ TQFT} 

The $\mathbb {CP}^{N-1}$  model  may be expressed as  a $U(1)$ gauge theory with an    $N$-component  elementary field $z_i(x)$, obeying the constraint $\sum_{i=1}^{N}|z_i(x)|^2=1$. The theory is invariant under the action of a $SU(N)$  rotation on the  $z_i(x)$ field.   However,  $SU(N)$ does not act faithfully, since 
$z_i(x)$ is not a gauge invariant operator. The  global
symmetry that acts faithfully on states in  Hilbert space is 
\begin{align}
G = SU(N)/\mathbb Z_N = PSU(N)
\label{symsym}
\end{align}
  In other words, the  $\mathbb Z_N $ center of the  $SU(N)$ is also part of the  $U(1)$ gauge structure, and should not be counted in the  global symmetry. 
 All gauge invariant local operators of the $\mathbb {CP}^{N-1}$  model are in  $PSU(N)$ representations.

We would like to turn on an  ${SU(N)/\Z_N}$ background. To do this, we first introduce 
  a background $SU(N)$ one-form gauge field $A^{(1)}$ \cite{Gaiotto:2017yup, Tanizaki:2017qhf}.  Further, to gauge the   $\mathbb{Z}_N$  part, we turn on  a pair of 
  $U(1)$ 2-form and 1-form  gauge fields $(B^{(2)}, B^{(1)}) $, and a Lagrange multiplier   $ F^{(0)}$: 
     \begin{align}
 Z_{{\rm top}, p} &= \int  \Diff   B^{(2)} \Diff B^{(1)}    \Diff  F^{(0)} \;  \rme^{ \im \int F^{(0)}  \wedge ( N B^{(2)}-\diff B^{(1)}) + \im p \int B^{(2)}}  
 \label{top-cp}
  \end{align}
Therefore, 
  \begin{align}
N B^{(2)}= \diff B^{(1)},  \qquad N \int B^{(2)}=  \int \diff B^{(1)} = 2 \pi \Z
\label{ZN-cp}
\end{align}
and  also promote   the $SU(N)$ background  gauge field $A^{(1)}$   to a  $U(N)$ background  gauge field  
$\tilde A^{(1)}$:  
\begin{align}
\tilde A^{(1)} = A^{(1)} + \frac{ {\mathbbm 1} }{N} B^{(1)}  \qquad ({\rm locally})
\label{PSUN3}
\end{align}
The covariant derivative in the  $\tilde A^{(1)}$  background   takes the form 
\begin{align}
D(a,\widetilde{A}^{(1)} )z \equiv \diff z -\im  a z   + \im \widetilde{A}^{(1)}  z 
\end{align}
 where 
$a$ is the original dynamical  $U(1)$ gauge field and $\tilde A^{(1)}$  is classical. This couples the $\mathbb {CP}^{N-1}$  model to an  ${SU(N)/\Z_N}$  background field, and the 
theory now has a  1-form (background) gauge invariance 
\begin{align}
B^{(2)}\mapsto B^{(2)}+ \diff \lambda^{(1)},\;\; B^{(1)}\mapsto B^{(1)}+ N \lambda^{(1)}, \;\;
 \widetilde{A}^{(1)}\mapsto \widetilde{A}^{(1)}+\lambda^{(1)},   \;\;
 a\mapsto a+\lambda^{(1)}. 
\end{align}
which couples the  dynamical field with  the classical background. 
The gauge invariant combination of dynamical and classical fields are  given by 
\begin{align}
(a \mathbbm 1 - \tilde A^{(1)}), \qquad  (\diff a- B^{(2)})
\end{align}
Therefore, the action of the theory minimally coupled to the  background  field is given by 
\be
S_{\mathrm{bgrd}}={2\over g^2}\int  [D(a,\widetilde{A}^{(1)})z^{\dagger}\wedge *D(a,\widetilde{A}^{(1)} )z]-\im {\theta\over 2\pi}\int [\diff a- B^{(2)}],
\ee
BPS instantons  can be found by using  Bogomolnyi  factorization:
\begin{align}
D(a,\widetilde{A}^{(1)} )z=\pm \im * D(a,\widetilde{A}^{(1)} ) z
\label{ins3} 
\end{align}
As a result of  classical $PSU(N)$ background, the topological charge and action are no longer quantized in integer units, just like our simple QM problem. 
Rather, we have, as a perfect counterpart of \eqref{pbc-1},  
\begin{align}
Q=  {1 \over 2\pi}\int  (\diff a-B^{(2)}) = \underbrace{ \int \frac{\diff a}{2\pi}}_{\in \mathbb Z} +
 \underbrace{{1 \over  N} \int   \frac{\diff B^{(1)}}{2\pi}}_{\in \frac{1}{N} \mathbb Z}  \in \frac{1}{N} \mathbb Z
\label{non-integer}
\end{align} 
  The first term is integer valued because $a$ is in $U(1)$ bundle.  The second term arise from proper quantization   of  $\frac{\diff B^{(1)}}{2\pi}$. 
 Due to BPS nature of these configurations,   their action is fractional $S= \frac{S_I}{N}= \frac{4 \pi}{g^2N}$. 

Despite the fact that these saddles do not contribute to the partition function of the 
$\mathbb {CP}^{N-1} $ model without background fields, 
certain pairs of such configurations, such as   
\begin{align}
({\cal I}_a,  \bar {\cal I}_a) \qquad  {\rm for \;  which}  \;    Q= \frac{1}{N} - \frac{1}{N} =0,  \qquad  \qquad S=  \Big( \frac{1}{N} +  \frac{1}{N} \Big)S_I = \frac{2}{N} S_I 
\end{align}
do. There is no obstruction for them to contribute.  Similarly, 
\begin{align}
\prod_{a=1}^{N} {\cal I}_a \qquad {\rm for \;  which}  \;  Q= N \times \frac{1}{N} =1 \in \Z, \qquad  \qquad S=  N \times  \frac{1}{N}  S_I =  S_I 
\end{align}
also  contributes. 

More technically, $Q= \frac{1}{N}$ configurations  live  in a $PSU(N)$ bundle which cannot be uplifted to an $SU(N)$ bundle. Yet, composite configurations such as 
$({\cal I}_a,  \bar {\cal I}_a)$ pair and $\prod_{a=1}^{N} {\cal I}_a$ cary integer topological charge and  can be lifted to  an $SU(N)$ bundle.  

This argument proves the existence of non-BPS  fractional  action configurations in the vacuum of the $\mathbb {CP}^{N-1} $ model on arbitrarily large-$T^2$,  whose non-perturbative significance  is controlled by 
\begin{align}
S_a= \frac{S_{\cal I}}{N} &= \frac{  4 \pi }{g^2 N}, \qquad {\cal I}_{a, 2d} \sim \rme^{- \frac{  4 \pi }{g^2(\mu)N } + i \frac{\theta}{N}}, 
\qquad  \Lambda  = \mu  {\cal I}_{a, 2d},  
\label{frac-instanton}
\end{align}
where $ \Lambda$  is  the strong scale of the theory. 

The main point is following: Even when we consider $\mathbb {CP}^{N-1} $ sigma model without any insertion of $  B^{(2)} $  flux, we should still consider the field space in the $U(1)/\Z_N$ bundle, where both 
 topological charge and action are fractional. Then, we should sum over configurations that can be lifted to $U(1)$ bundle.  This is a well-defined mathematical prescription.  This, inevitably, leads us to the statement that the non-perturbative expansion parameter in the theory is $  \rme^{-  S_I/N} = \rme^{- \frac{  4 \pi }{g^2(\mu)N }}$ and it is  exponentially more important than BPST instanton: 
 \begin{align}
\rme^{- \frac{  S_I}{N} } \gg   \rme^{- 2 \frac{  S_I}{N} }\gg   \rme^{- 3 \frac{  S_I}{N} } \gg \ldots \gg   \underbrace{\rme^{-  {  S_I} }}_{\rm instanton}  
\label{frac-instanton-4}
\end{align}
Note that  $\gg$ indicates here  exponential hierarchies. In particular, at large-$N$ limit, 2d instantons scale as  $ \rme^{- \frac{  4 \pi }{g^2(\mu) }}  \sim  \rme^{- O(N^1)}$ and is completely suppressed,  while these configurations persist  as  $ \rme^{- \frac{  4 \pi }{g^2(\mu) N }}  \sim  \rme^{- O(N^0)}$.

\subsection{'t Hooft flux   using  $\frac{(SU(N) \times U(1) ) }{\mathbb Z_N}$ }
\label{flux-cp}
In $\mathbb {CP}^{N-1}$ model, by just using $U(1)$ gauge structure,  it is impossible to impose 't Hooft's twisted  boundary conditions. If we turn on an 
$SU(N)$ background field,   this also allows us to  turn on  a non-trivial  't Hooft flux  corresponding to twisted boundary conditions, 
 associated with  
\begin{align}
& \frac{SU(N) \times U(1)  }{\mathbb Z_N} 
\end{align}

Let $\Omega_{\mu}, \omega_\mu$ denote the transition functions for the  non-abelian and abelian parts. 
    $\Omega_1(x_2)$ is the transition function between $ (x_1 + L_{1}, x_2) \sim (x_1,  x_2)$. $\Omega_{\mu}$ function is independent of  $x_{\mu}$, but depends on other coordinate. 
For $\mathbb {CP}^{N-1}$   fields, we impose
\begin{align}
z(x_1 +L_1, x_2)&= \Omega_1(x_2)  z(x_1,  x_2)   \omega_1^{-1}(x_2), \qquad  \cr
z(x_1 , x_2 + L_2)&= \Omega_2(x_1)  z(x_1,  x_2)  \omega_2^{-1}(x_1)
\label{flux}
\end{align}
where $ \omega_1(x_2) \equiv  \rme^{\im \alpha_1 (x_2) }$ is pure phase. 
We can connect $z(x_1 +L_1, x_2+L_2)$ with  $z(x_1,  x_2)$ in two different ways. 
For consistency at the corners, the transition functions must satisfy 
\begin{align}
\Omega_1(L_2) \Omega_2(0)&= \Omega_2(L_1) \Omega_1(0)  \rme^{\im  \frac {2 \pi \ell }{N} } \cr
\omega_1(L_2) \omega_2(0)&= \omega_2(L_1) \omega_1(0)  \rme^{\im  \frac {2 \pi \ell }{N} } 
\label{flux2}
\end{align}
corresponding to $\ell=0,1, \ldots, N-1$-units of 't Hooft flux.   This is the  gauge invariant data in \eqref{flux}. 
The gauge covariance of the flux condition tells us that under a gauge  transformation,  e.g., $z(0, x_2) \rightarrow g(0, x_2) z(0, x_2) h^{-1} (0, x_2)$,
the transition matrices transform as     $\Omega_1(x_2) \rightarrow g(L_1, x_2) \Omega_1(x_2)  g^{-1}(0, x_2)$ etc. So, there is some  gauge  freedom in the choice of transition matrices.

As an example, we can satisfy  relations  \eqref{flux2} with the choices $\Omega_{1}({x_2})= \rme^{\im  \frac{2 \pi T \ell } {N} \frac{x_2}{L_2} }$, 
$\Omega_{2}=1$, $ \omega_{1}({x_2}) =  \rme^{\im  \frac{2 \pi \ell} {N} \frac{x_2}{L_2} }$, $\omega_{2}=1$ where 
$T= {\rm diag}(1,\ldots, 1, -(N-1))$. 
As a result,  
  \begin{align}
&\Omega_1(L_2)  = \Omega_1(0)  \rme^{\im  \frac {2 \pi \ell }{N} } \\
&\omega_1(L_2)  = \omega_1(0)  \rme^{\im  \frac {2 \pi \ell }{N} }
\end{align}

In the $U(1)$ gauge theory, only $\ell=0$ is allowed. If we turn on just $SU(N)$ background without 
turning on a 
background $\Z_N$ gauge field,  $\ell$  is still zero.  (This does not mean it is uninteresting, see Section \ref{sec:du}.)
 However, turning on a $PSU(N)$ background field, we can absorb the aperiodicity of the transition function of one part into the other. 


\subsection{Fractional instantons in $\mathbb {CP}^{N-1}$  on large $M_2$} 
Below, we review the standard textbook construction of instantons  and demonstrate how it is modified once a $PSU(N)$ background is turned on. 

Instantons in 2d are topologically non-trivial configurations with integer winding number.  To see this, note that finite action demands  
$D_{\mu} z =  0$ as  $|x| \rightarrow \infty$, a covariantly constant configuration. 
\begin{align}
z_i(x) = n_i e^{i \alpha (x)},   \;   \; n_i  \bar n_i=1 
\label{cov-const}
\end{align}
where  $n_i$ is a constant vector with unit norm.  Therefore,  in the perturbative vacuum,  
\begin{align}
a= \diff \alpha 
\end{align}
 which are  pure gauge configurations.   Let  $ \phi $ parametrize  $S^1_{ \infty}$ boundary of $\R^2$.  Pure gauge configurations are of the form  $\alpha(\phi) = W  \phi, W\in \Z$. 
 This gives a mapping from the boundary of the physical space to the  disjoint 
 classes of gauge transformations, which correspond to  homotopy classes $\pi_1(S^1) = \Z$.   This is the instanton number  or  
 topological charge
\begin{eqnarray}
Q=\frac{1}{2\pi} \int \diff a= \frac{1}{2\pi} \oint   a = \frac{1}{2\pi} \oint   \diff \alpha      \in \Z \;.
\label{theta}
\end{eqnarray}

Assume now we turn on a $PSU(N)$ background  field.
 Locally, the  $U(N)$ gauge field can be written as: 
\begin{eqnarray}
{{\cal A}^{(1)}} = ``\tilde A^{(1)} + a  {\mathbbm 1}"  \qquad \rm locally 
\end{eqnarray}   
where the gauge connection ${{\cal A}^{(1)}}$  is in $U(N) \cong (U(1) \times SU(N))/\Z_N$.

\vspace{0.5cm}
\noindent 
{\footnotesize
{\bf Example:} The simplest example of a $U(N)$  field of this form is following. Consider $F_{12}=\frac{2\pi \ell_{12}}{N L_1L_2}T$, with $T=\text{diag} (1,1,\dots,1,-(N-1))$ is a Cartan generator, and   $f_{12}= - \frac{2\pi \ell_{12}}{N L_1L_2} \mathbbm 1 $.  Hence,  $F_{{{\cal A}^{(1)}}}= -\frac{2\pi \ell_{12}}{ L_1L_2} 
 \text{diag} (0,0,\dots,0,1)$ is a $U(N)$ gauge field strength which cannot  be decomposed into 
 $SU(N)$ and   $U(1)$ parts, but can be decomposed to   $PSU(N)$ and   $U(1)/\Z_N$ parts. 
 The corresponding gauge fields are (setting $A_1=0, a_1=0$),  
$ A_2(x_1) = \frac{2\pi \ell_{12}}{N L_1L_2}T x_1,  a_2(x_1) = - \frac{2\pi \ell_{12}}{N L_1L_2} \mathbbm 1   
 x_1$ and the transition functions are given by 
 $\Omega_1 (x_2)= \rme^{- \im \frac{2\pi \ell_{12}}{N L_2}T x_2}$,   $\omega_1 (x_2)= \rme^{- \im \frac{2\pi \ell_{12}}{N L_2} {\mathbbm 1} x_2}$,  $\Omega_2=1, \omega_2=1$. 
 Hence, the transition functions obey \eqref{flux2}, corresponding to the insertion of $\ell_{12}$ units of 
't Hooft flux. 
 }
%
%
 
\vspace{0.5cm} 
\noindent
Now, the topological term can be written as 
\begin{eqnarray}
\frac{1}{2\pi N} \int   \tr  [F_{{{\cal A}^{(1)}}}]
\label{theta-2}
\end{eqnarray}
This  reduces to \eqref{theta} if  ${{\tilde A}^{(1)}}$   is  in $SU(N)$. But if ${{\tilde A}^{(1)}}$  is a connection in a  $PSU(N)$  bundle 
which  cannot be lifted to  an $SU(N)$ bundle,  
the topological charge in \eqref{theta-2} is no longer quantized in integer units, rather  it can assume fractional values, multiples of  $1/N$.  In this case, the dynamical field $a$ is not a simple $U(1)$ gauge field, but a $U(1)/\Z_N$ field.

Let us make this abstract argument concrete.  
For $U(N) $ gauge structure,  
the covariantly constant   configurations as  $|x| \rightarrow \infty$ takes the form: 
\begin{align}
 z_i( |x| \rightarrow \infty ) = e^{i T(\phi)} e^{i \beta (\phi)} n_i ,    
 \end{align} 
 instead of \eqref{cov-const}.  We  choose  
\begin{eqnarray}
T(\phi) + \beta(\phi)  = \frac{\phi}{N} \left( \begin{array} {cccc}
-1 & && \\
&-1&& \\
&& \ddots& \\ 
&&& (N-1) \\
\end{array} \right)  + \frac{\phi}{N}  {\mathbbm 1}  =  \left( \begin{array} {cccc}
0 & && \\
&0&& \\
&& \ddots& \\ 
&&& \phi \\
\end{array} \right) 
\end{eqnarray}
such that  $e^{i T(\phi)} e^{i \beta(\phi)} $ is a  proper $U(N)$ gauge transformation.   However,  $ \rme^{\im T(\phi)}$  does not live in an 
 $SU(n)$ bundle, rather   it is in the $PSU(n)$ bundle.  Similarly, $ e^{i\beta(\phi)} $  can no  longer be characterized by a proper  $U(1)$ gauge transformation, instead lives in  $U(1)/\Z_N$.  
\begin{align}
\rme^{\im T(\phi+ 2\pi)}&= \rme^{\im T(\phi)} \rme^{-\im \frac{2\pi}{N}} \cr
\rme^{\im \beta(\phi+ 2\pi)}&= \rme^{\im \beta(\phi)} \rme^{+\im \frac{2\pi}{N}}
\end{align}
This is an example of $\tilde A^{(1)}$ being a connection in   $PSU(n)$  bundle which cannot be  lifted to  $SU(n)$ bundle.

As a result, the dynamical gauge connection $a$ can no longer be characterized by a $U(1)$ gauge field, and $ \int da$ is not a multiple of $2\pi$.  Instead, the dynamical  $a$ and background  $\widetilde A^{(1)} $  are now intertwined in an inseparable way.  
Denoting  $g(\phi) = \rme^{i T(\phi)} \rme^{i \beta (\phi)} = 
{\rm Diag} \left(1, \ldots, 1, \rme^{ \im  N \beta (\phi)}    \right) $, the topological charge  \eqref{theta-2} can  now be written as
 \begin{align}
 \frac{1}{ N} \frac{1}{2\pi} \int    \tr [ -ig^{-1} \diff g ]  = \frac{1}{2\pi }   \int   \diff  \beta  = \frac{1}{N} \Z
 \end{align}
and it is   quantized in units of $\frac{1}{N}$, same as \eqref{non-integer}. Note that this   is the counterpart of \eqref{tbc-1} in quantum mechanics.  
It is clear that  $Q= \frac{1}{N}$ is in a $U(1)/\Z_N$ bundle,  and  cannot be lifted to an  $U(1)$ bundle. Yet, composite configurations such as 
$({\cal I}_a,  \bar {\cal I}_a)$ pair and $\prod_{a=1}^{N} {\cal I}_a$  can be lifted to $U(1)$ bundle. As in our quantum mechanics example,  these composites  are the non-trivial fractional    action configurations contributing 
to the partition function of  original $\mathbb {CP}^{N-1}$  even in the absence of  $PSU(N)$ background.

\begin{figure}[t]
\vspace{-1cm}
\begin{center}
\includegraphics[width = 0.7\textwidth]{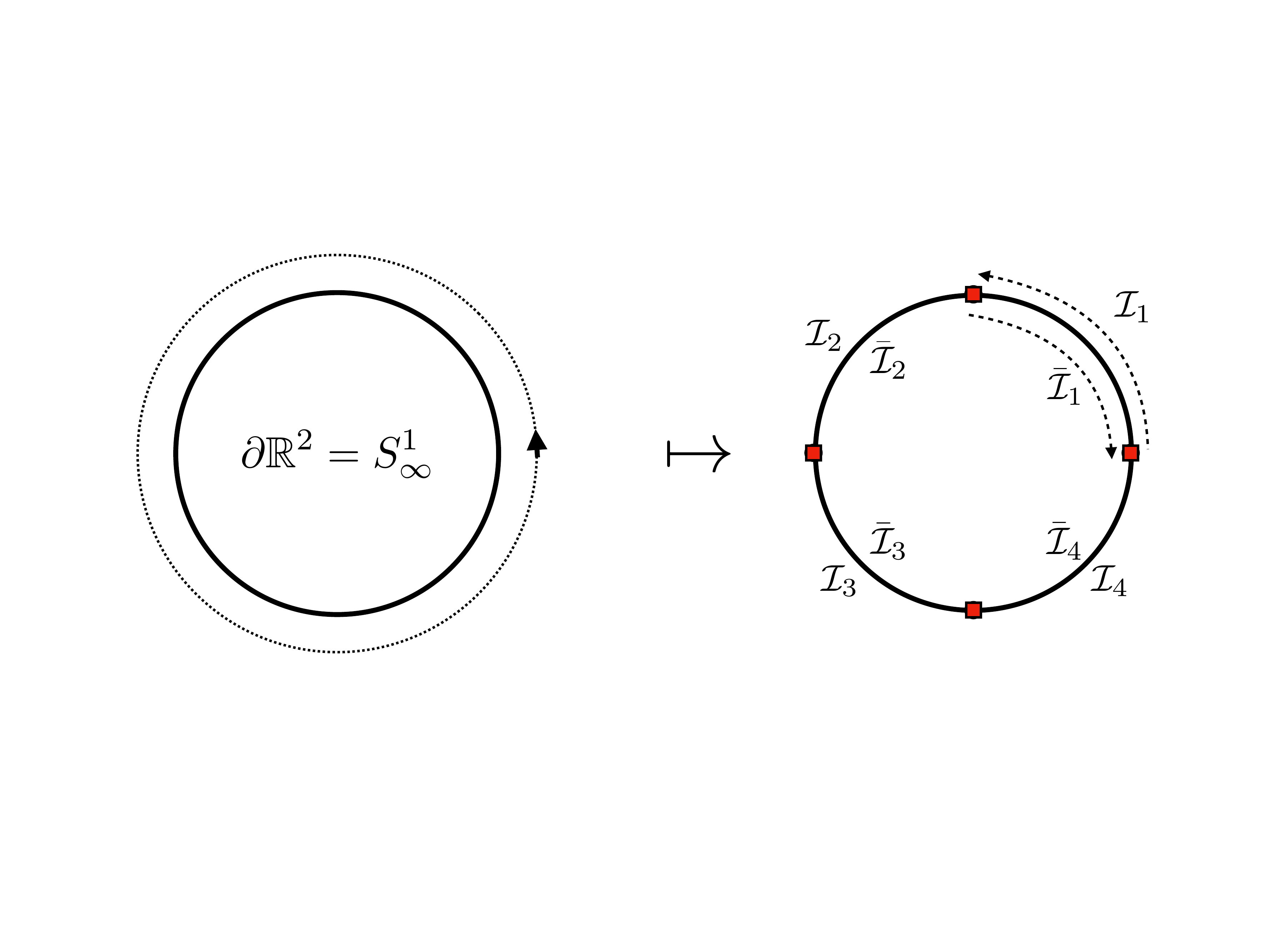}
\vspace{-2.cm}
\caption{Mapping from $S^1_{\infty}$ to field space. $Q= \frac{1}{2\pi} \oint   a = \frac{1}{2\pi} \oint   \diff \beta     \in  \frac{1}{N} \Z $, 
because $ \rme^{\im \beta(\phi)} $ is in 
$U(1)/\Z_N$ bundle. 
Figure is for $\mathbb {CP}^3$ model.  
 }
 \vspace{0cm}
\label{fig:mapping}
\end{center}
\end{figure}

\subsection{Flavor twist  {\it and}  't Hooft flux,    and relation to $\R \times S^1$}
\label{sec:du}
In a joint work with Dunne \cite{Dunne:2012ae}, we proved that if $\mathbb {CP}^{N-1}$ model is compactified on a cylinder with an $SU(N)$ flavor twisted boundary condition,   and if the $SU(N)$
 background is symmetric under $\Z_N$ shifts up to permutations,  i.e, 
 \begin{align}
&\tilde z(x_1 +L_1, x_2)= \Omega_F   \tilde z(x_1,  x_2), \qquad  \cr 
&   \Omega_F \sim   {\rm Diag} \left( \rme^{ \im  2 \pi \mu_1} \ldots,  \rme^{ \im  2 \pi \mu_N}   \right) =  {\rm Diag} \left(1,  \rme^{ \im \frac{2 \pi}{N}}, \ldots,     
   \rme^{ \im \frac{2 \pi (N-1)}{N}} \right)  
\label{flavor-twist}
\end{align}
 then there exists instantons with fractional topological charge and action, 
 \begin{align}
 W=\frac{1}{N}, \qquad S = \frac{1}{N} \frac{4 \pi}{g^2} \qquad ( {\rm weak\; coupling, \;no\;}   B^{(2)} {\; \rm  flux})
 \label{action-frac}
 \end{align}
  in the  semi-classical domain on   $\R \times S^1_L$.  
  See also  \cite{Fujimori:2017oab, Fujimori:2018kqp,  Ishikawa:2019tnw, Misumi:2014jua, Misumi:2019upg}. Fractional instantons are also described in earlier work on $\mathbb{CP}^{N-1}$  in \cite{Eto:2004rz, Eto:2006pg,Eto:2006mz, Brendel:2009mp, Bruckmann:2007zh}.

\begin{figure}[t]
\vspace{-.1cm}
\begin{center}
\includegraphics[width = 0.7\textwidth]{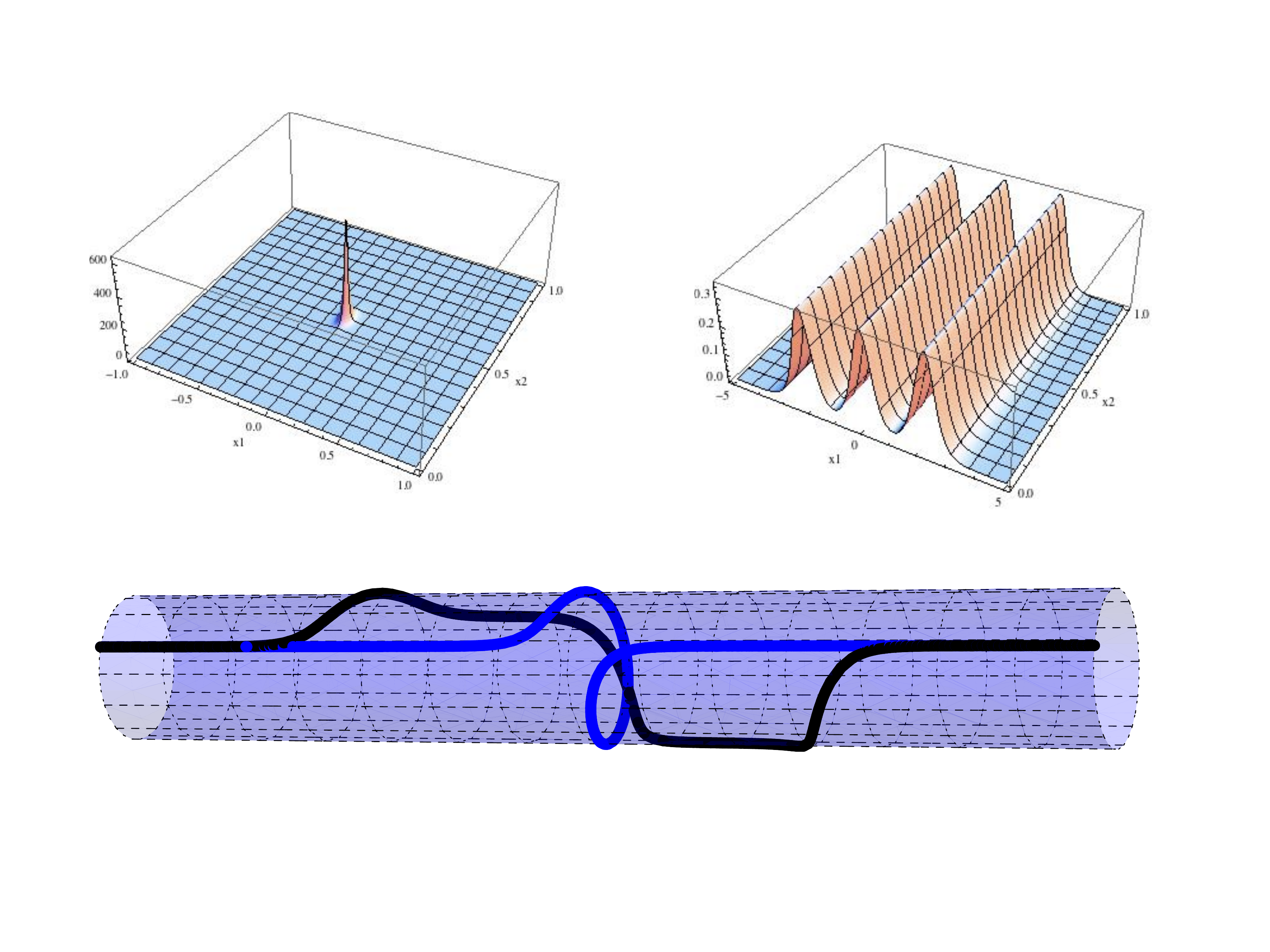}
\vspace{-1cm}
\caption{When the moduli parameter of instanton  is smaller than the scale set by the $\Omega_F$ background, instanton looks like a single lump (first figure), and when the moduli parameter is large, it fractionates to $N$-pieces (second figure).  Corresponding gauge holonomies for small and large instantons are also depicted, demonstrating fractionalization. For these solutions, 't Hooft flux is set to zero.  TQFT argument tells us that even when the instanton does not physically split to parts, 
it can  still be viewed as a composite  of $N$ $W=1/N$ fractional instantons. 
Figure borrowed from \cite{Dunne:2012ae}.  
 }
\label{fig:fraction}
\end{center}
\end{figure}

  It is    conjectured that this regime is continuously connected to the theory at large $\R \times S^1_L$ and $\R^2$.  The rationale  behind this is following.  The above  construction, in the operator formalism translates to 
\begin{align}
 Z_{\Omega_F }  & =   \tr [  \rme^{- L_1 H}  \hat  \Omega_F  ]   
  \label{tpf-7}
 \end{align} 
 where $ \hat  \Omega_F =    \prod_{k=1}^{N} \rme^{\im \frac{2 \pi k}{N} \hat Q_k}   =  \rme^{\im \sum_{k=1}^{N} \frac{2 \pi k}{N} \hat Q_k}$   where $\hat Q_k$ is the number operator associated with $z_k$ quanta. 
The Hilbert space of $\mathbb {CP}^{N-1} $ model fills in representation of $PSU(N)$, singlet, adjoint etc. 
For example, 
$\rme^{- L_1 H}  \hat  \Omega_F |(\bar z_i z^j) \rangle  = \rme^{- L_1 M_{\rm adj} } \rme^{\im (i-j) \frac{2 \pi}{N}} |(\bar z_i z^j) \rangle$ and the state sum of these states leads to $ (-1)  \rme^{- L_1 M_{\rm adj} }$ 
instead of    $ (N^2-1)  \rme^{- L_1 M_{\rm adj} }$ that would occur in thermal state sum.  There is also 
 a singlet  $\lim_{x \rightarrow y} \bar z_i(x) \rme^{\im \int_x^y a }  z^i (y) $   which is almost degenerate with the adjoint. (The difference is $O(1/N)$ \cite{Witten:1978bc}). 
 As a result, there are 
extremely powerful spectral cancellation in the Hilbert space $\cal H$ of $\mathbb {CP}^{N-1}  $. In particular, quite remarkably, 
\be
\lim_{N \rightarrow \infty}   Z_{\Omega_F}(L)  = \mathrm{e}^{-L E_{\rm ground}}
\label{simple2}
\ee
which means no state apart from the ground state contributes to the twisted partition function. 
This is the idea of quantum distillation, which tells us that the $Z_{\Omega_F}(L)$ must be an  analytic function of $L$ \cite{Dunne:2018hog}, similar to  cancellation that takes place in index calculations in supersymmetric theories \cite{Witten:1982df}.  
Note that the cancellation in the index calculation is a result of supersymmetry, in our case, the theory is bosonic and we engineered 
the spectral cancellation by using its global symmetry. The fact that this is possible is a remarkable aspect of the theory.

  Let us now see if we can learn something deeper from this construction concerning strongly coupled QFT,
  and non-perturbative configurations that contribute to path integrals. 
  Since anomalies are independent of the strength of the  coupling, we can try to take advantage of that. 
   The first hint in this direction is that  a mixed anomaly between $PSU(N)$ and $\mathsf C$ on $\R^2$   survives upon  
compactification on $\R \times S^1$ \cite{Tanizaki:2017qhf}  
if and only if   $\Omega_F$- twisted boundary conditions  \eqref{flavor-twist}   are  used,  and appropriate $B^{(2)}  =A^{(1)}_{\rm 1d}  \wedge L^{-1} \diff x_2 $ background is turned on.

Now, in the presence of $B^{(2)}$ on a general $\mathbf T^2$, we showed that  fractional topological charge and  fractional action configurations exist. Importantly, this statement is not restricted to only 
$S^1_{\rm large} \times S^1_{\rm small}$, but also valid at for $S^1_{\rm large} \times S^1_{\rm large}$, where the long distance theory is necessarily {\it strongly} coupled.

To connect weak and strong coupling regimes,   we reconsider $B^{(2)}$ flux along with $\Omega_F$ background.
  We use 
\begin{align}
\tilde z(x_1 +L_1, x_2)&= \Omega_1^F(x_2)  \tilde z(x_1,  x_2)   \omega_1^{-1}(x_2), \qquad  \cr
\tilde z(x_1 , x_2 + L_2)&= \Omega_2(x_1)  \tilde z(x_1,  x_2)  \omega_2^{-1}(x_1)
\label{flux-f}
\end{align}
where   we  embedded  both $B^{(2)}$ flux  and $\Omega_F$ background into transition matrix: 
\begin{align}
 \Omega_{1}^F({x_2})&= \rme^{\im  \frac{2 \pi T \ell } {N} \frac{x_2}{L_2} }  \Omega_F,  \qquad \Omega_{2}(x_1)=1, \cr 
  \omega_{1}({x_2}) &=  \rme^{\im  \frac{2 \pi \ell} {N} \frac{x_2}{L_2} }, \qquad \qquad   \omega_{2}(x_1)=1 
\label{flux-f2}
\end{align}
This  construction, in the   operator formalism, amounts to 
\begin{align} 
 Z_{\Omega_F, \ell}  & =   \tr [  \rme^{- L_1 H}  \hat  \Omega_F   \hat {\mathsf U}^{\ell} ]   
    \label{tpf-44}
\end{align}
where we separated $\hat  \Omega_F$,  and $ \hat {\mathsf U}^{\ell}$  for  clarity. 
In the path integral formalism, we can express it  either as  twisted boundary conditions, 
or coupling to  a background $\Z_N$ TQFT: 
\begin{align} 
 Z_{\Omega_F, \ell}  &  =   \int_{ \tilde z(L_1, x_2) = \Omega_1^F(x_2)  \tilde z(0,  x_2)   \omega_1^{-1}(x_2) }    \Diff  \tilde z 
 \; \exp{- \left( {2\over g^2}    \int  [D(a) \tilde z^{\dagger}\wedge *D(a) )\tilde z]   -\im {\theta\over 2\pi}\int [\diff a ] 
  \right)}
   \cr \cr
&= \int    \Diff  F^{(0)} \;   \int_{z(\beta)=z(0)}  \Diff z \;  \;\; 
  \rme^{ \im \int F^{(0)}  \wedge ( N B^{(2)}-\diff B^{(1)})  } \cr 
 & \times \exp - \left({2\over g^2}\int  [D(a,\widetilde{A}^{(1)})z^{\dagger}\wedge *D(a,\widetilde{A}^{(1)} )z]-\im {\theta\over 2\pi}\int [\diff a- B^{(2)}] \right)
   \label{tpf-4}
 \end{align} 
where $\ell$ is 't Hooft flux, and we set discrete theta angle to $p=0$.

The topological charge in $\mathbb {CP}^{N-1}$ model with $\Omega_F$  twist  is fractionally quantized  $W= \frac{1}{N}\Z$ for two,
 at first sight, seemingly different reasons. These  are 
 \begin{align}
Q=\frac{1}{2\pi}\int_{\rm tbc}  da &=  \frac{\mu_{i+1}- \mu_i }{2 \pi}   \Big|_{ \mu_i = \frac{2 \pi} {N} (i-1)} = \frac{1}{N} \cr
& = \frac{1}{N} \ell \Big|_{ \ell  =1} =   \frac{1}{N} 
\label{frac8}
 \end{align}
 The first equality is the property of  $\Z_N$ symmetric (up to permutation) background. It does not need the insertion of 't Hooft flux, but without that insertion, it seems to be deformable from $ \frac{1}{N}\Z$ topological charge. In what sense, is it 
 robust then? The point is, if  we wish to preserve a mixed anomaly that exists on $\R^2$ between $PSU(N)$ and 
 $\mathsf C$,  this particular background is the {\it unique} one  which achieves that\cite{Tanizaki:2017qhf}. The boundary condition \eqref{flavor-twist} remains invariant under  the intertwined combination of a center transformation  along with  $\Z_N$ cyclic permutation of fields, $z_i \rightarrow z_{i+1} $ \cite{Cherman:2017tey}.

 The second equality is  telling us that in the 
 't Hooft flux background, topological charge  is $\frac{1}{N}$  quantized, \emph {regardless} of what the value of $\Omega_F$ is. 
 Both of these statements are independent of coupling, and also persist at strongly coupled regime  on arbitrary size $T^2$. 
Next, we exhibit that this is  the case in two examples, one is large-$N$ limit of $\mathbb {CP}^{N-1}$  and the other is supersymmetric version.

 \vspace{0.5cm}
 \noindent 
{\bf Reminder of tunnelings in  $\mathbb  {CP}^{N-1}$ model:}   Here, we would like to detail few more information about tunneling events in  $\mathbb  {CP}^{N-1}$ model with $\Omega_F$ twisted background   on small $\R^1 \times S^1 $   that we will use later.   
Ref.~\cite{Dunne:2012ae}   also has an obvious reinterpretation on $\R^2= \R^{+} \times S^1$ with $\Omega_F$  and $ B^{(2)}$ flux background. In these backgrounds, the classical minima of the model are 
\begin{align}
z( |x| \rightarrow \infty) = \bm e_a, \qquad  {\bm e}_a= \Big( \ldots, 0, \underbrace{1}_{a^{\rm th}}, 0, \ldots \Big), \qquad a=1, \ldots, N
\end{align}
Let $\tau$ denotes coordinate on $S^1_{L}$ where we take $L \rightarrow \infty$.  Then, the  minimal tunneling events correspond to 
\begin{align}
\Delta z = z(\tau=L) - z(\tau=0) =   \bm e_a -  \bm e_{a+1} = \bm \alpha_a, \qquad a=1, \ldots, N
 \label{tunnelings0}
\end{align}
where   $\bm e_{N+1} \equiv   \bm e_{1} $ and   $\bm \alpha_N= -  \sum_{a=1}^{N-1} \bm \alpha_a$ is the affine root of $SU(N)$ Lie algebra. 
These fractional instanton events ${\cal I}_a$ have topological charge $W=1/N$ and action $S=\frac{S_I}{N}=\frac{4\pi}{g^2N}$  and are the semi-classical ingredients of our analysis.  The fugacity associated with these events are
\begin{align}
{\cal I}_a \sim \rme^{-\frac{4\pi}{g^2(\mu) N} + \im \frac{\theta}{N}}=  \Lambda \mu^{-1}
\end{align}
These are minimal tunneling events associated with   $\bm \alpha_a$.  More generally, the set of tunneling events obeying BPS-type equation  are $N^2 -N$ types, and associated with charges 
\begin{align}
 \bm \alpha_{ab} =   \bm e_a- \bm e_b 
 \label{tunnelings1}
 \end{align}
  with topological charge   and action
  \begin{align}
  W=(a-b)/N, \qquad S= |a-b|\frac{S_I}{N}
\label{general}
  \end{align}
The usual instanton in 2d  \cite{Polyakov:1975yp} corresponds to ${\cal I}= \prod_{a=1}^N {\cal I}_a$ with charge  $ \bm \alpha_1 + \ldots +  \bm \alpha_N=0$ configurations. We can also construct infinite towers of the fractional instantons with charge  $ \bm \alpha_a$ by merging it with 2d instantons,  
${\cal I}_a ({\cal I})^q, \; q \in \Z$.  

As we emphasized, there are also important bion configurations, minimal versions of which are 
$[{\cal I}_a \bar {\cal I}_{a+1}]$. These have "Coulomb"  charges   $\bm \alpha_a  - {\bm \alpha}_{a+1} $, topological charge 
$W=0$, and action $S=2S_I/N$. There are also neutral bions  $[ {\cal I}_a \bar {\cal I}_a]_{\pm}$ which have zero "Coulomb" and topological charge, and  $S=2S_I/N$ action.

\subsection{Interpolation between weak and strong coupling holonomy potentials} 
\noindent
{\bf Small  $L$ holonomy potential:}
In the $\mathbb {CP}^{N-1} $ model, the  gauge field $a$  does not have a kinetic term in the  the UV-Lagrangian, and is not an independent degree of freedom.  It is related to the dynamical fields $z(x_1, x_2)$ through equations of motions,  
\begin{align}
a= \im \bar z \diff z 
\label{not-ind}
\end{align}
In the small $L \Lambda \lesssim 1$ regime, we can write down the potential for the holonomy field $\rme^{\im \oint_{S^1} a}$, both with and without the insertion of $\Omega_F$.  
They exhibit a striking difference:  
  \begin{align} 
 V_{\rm thermal}(a)    & =  - \frac{ 2 N}{ \pi  L^2}  \sum_{n=1}^{\infty}  \frac{1}{n^2}  
      \cos( a L n) \sim  -    \frac{ 2 N}{ \pi  L^2}   \cos( a L ) \cr 
 V_{\rm \Omega_F}(a)    
 & =  -  \frac{1}{ \pi  L^2 }  \sum_{n=1}^{\infty}  \frac{1}{n^2}   \left( \tr (\Omega_F^n)     \rme^{ \im aL  n }  + {\rm c.c.} \right)  \cr
 &=  - \frac{2}{ \pi  L^2 } \frac{1}{N} \sum_{n=1}^{\infty}  \frac{1}{n^2}       \cos { (N a  L n)}  \sim 
  - \frac{2}{ \pi  L^2 } \frac{1}{N}    \cos( N aL )
 \label{1-loop}
 \end{align} 
 where $aL$ is periodic by $2\pi$. 
 In thermal case, there is a  unique minimum in the fundamental domain,  and  a potential barrier as $a \rightarrow a+2\pi$ corresponding precisely to tunneling events with $W=1$. 
  
 In the presence of the $\Omega_F$ twist, we obtain a potential $\sim \cos( N L  a )$ just like our $T_N$ quantum mechanics,     which  has $N$-minima at 
 $La= \frac{2\pi}{N} j,  \;j=0, 1, \ldots, N-1$. The interpolation between these configurations correspond to 
 $W=1/N$ fractional instanton events.  
 
  It is important to note that, since  $a$ is not an independent degree of freedom, these minima are some configurations of $z$-field, and tunneling is between $z$ field configurations, which translates to $a$ through 
  \eqref{not-ind}.
  The  profiles of these fields  are shown in Fig.\ref{fig:fraction} borrowed from \cite{Dunne:2012ae}.

 Note that  the $\Omega_F$ twist is an $SU(N)$ background. We can also turn an $SU(N)/\Z_N$  background  
 by coupling to $B^{(2)}  =A^{(1)}_{\rm 1d}  \wedge L^{-1} \diff x_2 $. Then 
  $\ell$ units of    
 $B^{(2)}$ flux  amounts to considering  $ \tr [  \rme^{- L_1 H}  \hat  \Omega_F  \mathsf U ^{\ell} ]  $ which strictly  enforces that a transition between  $j$ and $j+ \ell$ takes place with topological  charge  $W= \frac{\ell}{N}$ and action $S = \frac{\ell}{N} \frac{4 \pi}{g^2}$. 
 
 \vspace{0.5cm}
\noindent
{\bf Large  $L$ holonomy potential:}
 Now, let us consider large-$L$ limit, where the  asymptotically free $\mathbb {CP}^{N-1}$ coupling $g^2(\mu)$ necessarily becomes large at distances  $|x| \geq \Lambda^{-1}$.   A useful handle in this regime is to consider large-$N$ limit. At large-$N$, the theory is solvable \cite{Polyakov:1987ez}.  It can be shown to acquire a mass gap dynamically,  
 \begin{align}
 M=\Lambda = \mu\,e^{-4\pi/Ng^2(\mu)} 
 \label{largeN-gap}
 \end{align}
 which very much looks like a fractional instanton effect. 
    But the  large-$N$ solution is very shy about  giving  us any insights concerning the  microscopic origins of mass.

    The saddle of the large-$N$ solution on $\R^2$ is  $a=0, \lambda= M^2$, where $\lambda$ is Lagrange multiplier that appears in the action as
   $ \int \lambda( \bar z z-1)$ imposing the constraint  $|z|^2 =1$. A remarkable fact  about  $\mathbb {CP}^{N-1}$  model  is large-$N$ volume independence 
in the  $\Omega_F$ twisted background.   The large-$N$ saddle point equation on $\R^2$ may be derived precisely at any size $\R \times S^1$
if and only if one also takes the background $\Omega_F$ \cite{Sulejmanpasic:2016llc, Dunne:2012ae}. 
 
 At large-$N$,  we may say that $z$ particles become massive.   Ref.~\cite{Witten:1978bc} showed  that these particles  generate a kinetic term for the gauge field at one-loop level on $\R^2$.  The effective Lagrangian is 
\begin{equation}
L_{\text{eff}} = 
\frac{1}{e^2} (\diff a)^2 + |Dz|^2+ M^2 |z|^2, 
\label{Lag} 
\end{equation}
where $e^2 =  \frac{48\pi M^2}{N}$, and the effective theory is just   $N$- component scalar QED. Consider this theory on $\R \times S^1_L$. At low temperature $(L \rightarrow \infty)$ for thermal theory, and  
$\Omega_F$-twisted (non-thermal) theory, 
we obtain the holonomy potentials  as 
  \begin{align} 
 V_{\rm thermal}(a, M)    &  =
- \frac{ 2 N}{ \pi  L^2}  \sum_{n=1}^{\infty}  \frac{f_n}{n^2}  \cos(a L  n)  
\approx   -\frac{2 N}{L^2}  \left(  \frac{ML}{2\pi} \right)^{1/2}  \rme^{-ML}   \cos( a L)   
    \cr
 V_{\rm \Omega_F}(a, M)    
 & 
 =   - \frac{1}{ \pi  L^2 }  \sum_{n=1}^{\infty}  \frac{f_n}{n^2}   \left( \tr (\Omega_F^n)     \rme^{ \im a L n }  + {\rm c.c.} \right)  \cr
 &= - \frac{2}{ \pi  L^2 } \frac{1}{N} \sum_{n=1}^{\infty}  \frac{f_{Nn}}{n^2}    \cos { (N  a L n)}  
\approx   - \frac{2}{NL^2}    \left(  \frac{MLN}{2\pi} \right)^{1/2} \rme^{-MLN}   \cos( N a L)  \qquad \qquad   
 \label{1-loop-3}
 \end{align} 
 where $f_n = (nLM)K_1( LMn)$ where  $ K_1$ is the modified Bessel function of the first kind, and 
  in the simplified  form, we used asymptotic behavior  $ K_1( z) \sim \sqrt{ \frac{\pi}{2z}} e^{-z},    z\rightarrow \infty.$  

The untwisted case is derived by Affleck, who showed the existence of a zero temperature  $(L=\infty)$ phase transition, as can be seen by the non-commutativity of $L \rightarrow \infty$, $N \rightarrow \infty$ limits.  
In the $\Omega_F$-twisted case,  Sulejmanpasic showed the  commutativity of the two limits  \cite{Sulejmanpasic:2016llc}. In the  $\Omega_F$ twisted case, we again end up with our $T_N$ model, with $N$ inequivalent minima, and instantons again split up to fractional instantons with winding number 
$ W=1/N$.  

The whole discussion in the $L\Lambda \lesssim 1$ domain and $L MN \gg1$ domains are extremely similar. 
Note that if we interpret the instanton in  $V_{\rm thermal}(a)$ in  \eqref{1-loop-3} as   quantum instantons in the effective theory according to  Affleck with $S \sim N \frac{1}{ML} e^{-ML/2}$, we should view  the instantons in the $V_{\rm \Omega_F}(a)$ as  fractional quantum instantons  with action  $ S \sim \frac{1}{ML} \rme^{-MLN}$. 
 However, neither detail is actually too important for our purpose.  The $a$-field in this description is to a certain extend a place holder, because $e^2 \sim \frac{M^2}{N}$ and electric coupling tend to zero in the large-N limit.  Therefore,  $a$ is  actually a mnemonic  of  what is going on in the world of  $z$-field   through the relation \eqref{not-ind}.  

The main point is,  on a spacetime manifold   ${\bf T^2}$, viewed as    finite-volume regularization of  $\R^2$,  \eqref{tpf-7} tells us that there are configurations with fractional winding number  
$\frac{1}{N} \Z$ and action $S=  \frac{4\pi}{g^2N}$, for which $a$ lives in the $U(1)/\Z_N$ bundle. 
To see the existence of these configurations without the insertion of 
 $(B^{(2)}, B^{(1)})$ flux, by  just using the $SU(N)$ twist is also possible, and already shown  in  \cite{Dunne:2012ae} in semi-classical domain. However, $(B^{(2)}, B^{(1)})$ flux tells us that these configurations are there both at  weak or strong coupling with certainty.   This is just a statment  concerning field configuration topology. As we have shown in $T_N$ quantum mechanics, we can use these configurations to built  fractional action and integer winding  configurations which live in the $U(1)$ bundle. 
Next, we show that a description based  on this idea is actually taking place in 2d  softly broken 
${\cal N}=(2,2)$  $\mathbb {CP}^{N-1}$ model.

\subsection{${\cal N}=(2,2)$  $\mathbb {CP}^{N-1}$ model on $\R^2$ with soft mass   } 
Consider  the $\N=(2,2)$ supersymmetric $\mathbb {CP}^{N-1}$ model. This model has a $\Z_{2N}$ chiral symmetry which is broken dynamically to  $\Z_{2}$ by the formation of  fermion-bilinear condensate. 
 \begin{equation}
  \label{cc1}
 \langle  k|   \psi_{-}   \psi_{+ } |k \rangle  = N \Lambda\, e^{i \frac{2 \pi k}{N}}, \qquad  k=0,1, \ldots, N-1 
 \end{equation}
  leading to $N$ isolated vacua,   consistent with the  index $I_W= \tr((-1)^F)=N$ \cite{Witten:1982df}.

Consider adding a soft mass term to the Lagrangian,   $m \psi_{-}   \psi_{+ } + {\rm h.c.}$. 
 In this case,  the $k^{\rm th}$ vacuum energy density  is modified into 
\begin{align} 
{\cal E}_k = -m  \langle \psi_{-}   \psi_{+ }  \rangle_k + {\rm c.c.} =  -2 m  N  \Lambda   \cos  \frac{ \theta+ 2 \pi k} {N}
\end{align}
 at leading order in $m$.  Here $\Lambda = \mu \rme^{-S_I/N} =  \mu \rme^{- \frac{4 \pi }{g^2(\mu)N}} $ is the renormalization group invariant  strong scale  of the theory (at one-loop order),  
 and $\mu$ is the (Pauli-Villars) renormalization scale. 
  The  graded partition function for the mass deformed theory is
\begin{align}
Z(m, \theta) = \tr [(-1)^F \rme^{ - LH}]\;, 
 \end{align}
 and   on a 2-manifold, it  is modified as 
\begin{align}
Z(m, \theta)= \sum_{k =0}^{N-1}  \rme^{ 2 m  N  \Lambda V_2  \cos  \frac{ \theta+ 2 \pi k} {N} }
\end{align}
where $V_2$ is the volume of the 2-manifold and $m$ is small. (Assume $V_2$ is large compared to strong  length scale.)
 Based on our quantum mechanical example, we can rewrite the partition function as 
\begin{align}
Z(m, \theta)&= N\sum_{ W \in \Z}  \sum_{n=0}^{\infty}   \sum_{\bar n =0}^{\infty}   \frac{1}{n!} \frac{1}{\bar n !}  \left( m N \mu V_2   \rme^{-\frac{S_I}{N} + \im  \frac{ \theta} {N}}  \right)^{n }     \left( m N \mu  V_2   \rme^{- \frac{S_I}{N} -  \im \frac{\theta} {N}}  \right)^{\bar n }     
 \delta_{n-\bar n  - W N, 0} \cr
& =N \sum_{W \in \Z}   \left[ \sum_{\bar n=0}^{\infty}  \frac{  ( m N \mu V_2  \rme^{-\frac{S_I}{N}})^{2\bar n +  WN
}}{{\bar n}!  (\bar n+ WN)!   } \right] \rme^{\im  W \theta }  \cr
 &= N \sum_{W \in \Z}   \left[ I_{NW} (2 m N \mu V_2  \rme^{-\frac{S_I}{N}} ) \right]  \rme^{\im  W \theta }    \qquad ({\rm strong\; coupling, \;no\;}   B^{(2)} {\; \rm  flux})
\label{frac.gas}
\end{align}

The interpretation of this formula in the light of our discussion of $\mathbb {CP}^{N-1}$  model is following. 
\begin{itemize}
\item  Terms in the sum are sourced by the solution of self-duality equation  in the  
$\mathbb {CP}^{N-1}$ theory with $PSU(N)$ background. However, these  cannot  directly contribute to the partition function as they live in a part of $PSU(N)$ bundle that cannot be lifted to $SU(N)$ bundle. Happily,  the  constraint tells us that they do not.

\item The constraint  $\delta_{n-\bar n  - W N , 0}$ guarantees that  non-BPS configurations in the $PSU(N)$ bundle that can be  uplifted to $SU(N)$ bundle contribute to the sum.   The  sum  consistently reduces to a sum over integer topological charge configurations  $W \in \Z$, but there are fractional action configurations  $S= \frac{2}{N}S_I$ contributing to it.

\item In the  massless theory,   the chiral symmetry is $\Z_{2N}$. In the  $PSU(N)$ background,
 each of these objects carries  $2$ zero modes.  This is not inconsistent with ABJ anomaly thanks to the constraint $\delta_{n-\bar n  - W N , 0}$.  
  The $2$ zero modes  is also a natural reflection of mixed anomaly between  $PSU(N)$ and $\Z_{2N}$.  
    If we were to gauge  $PSU(N)$, we would indeed end up with $2$ zero modes with each of these configurations.  These zero modes are lifted by the insertion of mass term $m$.

\item The solutions in the $PSU(N)$ background   have 2  bosonic zero modes, and altogether, they should have $2N$ bosonic zero modes as a $2d$ BPST instanton.   $V_2$ may be viewed as the volume of the bosonic moduli space.   $\mu$ is Pauli-Villars regularization scale. It appears with the power 
$\mu^{n_b -  \frac{n_f}{2} } = \mu^1$ where $n_b=2, n_f=2$ are the numbers of bosonic and fermionic moduli. 

\item The factor $N$ in the argument and order of Bessel function arises because there are $N$ different types of fractional instantons in the $PSU(N)$ background.  
Note that at $m=0$, supersymmetric point,  modified Bessel obeys  $ I_{j}(0)= \delta_{j, 0}$ and  $Z(m=0, \theta) =N$ which is just supersymmetric  index \cite{Witten:1982df}.  At $m=0$, the only sector contributing to graded partition function is the zero topological charge sector.  
\end{itemize}

Again, we reach to the same conclusion in $\N=(2,2)$ $\mathbb {CP}^{N-1} $, as in  the bosonic model. 
Even when we consider $\mathbb {CP}^{N-1} $  model without $  B^{(2)} $  flux, we should still consider the field space in the $U(1)/\Z_N$ bundle, where both 
 topological charge and action are fractional, and sum up over configurations that can be lifted to $U(1)$ bundle. 
 This tells us that the non-perturbative expansion parameter in the theory is $  \rme^{-  S_I/N} $ which is  exponentially more important than 2d instantons \cite{Polyakov:1975yp}. 

\vspace{0.5cm}
\noindent
{\bf Topological susceptibility and how to rescue naive instanton analysis:} 
\label{naive}
Despite the fact that \eqref{frac.gas} is a sum over $W \in \Z$  configurations, it leads to 
to topological susceptibility that is in qualitative agreement with the purely bosonic theory 
\begin{align} 
\chi_{\rm top.}^{\rm soft} = \frac{ \partial^2 {\cal E}} {\partial \theta^2} \Big|_{\theta=0}  =  \frac{2 m \Lambda}{  N} \qquad  {\rm vs.} \qquad  \chi_{\rm top.}^{\rm bos.}  =   \frac{3 \Lambda^2}{\pi  N}
\label{soft}
\end{align}
The result in bosonic  theory is based on large-$N$ solution (see  \cite{DAdda:1978vbw}, and   \cite{Marino:2015yie} for a pedagogical description)  
and lattice simulations 
\cite{Vicari:2008jw}. 
   We expect, once,  $m \geq  \Lambda$,  $\chi_{\rm top.}^{\rm soft}$ to   saturate 
to  $\chi_{\rm top.}^{\rm bos.}$. The important things is both are of order $1/N$ as they should, as opposed to  $\rme^{-O(N)}$ as predicted by naive instanton analysis.

In the standard instanton analysis in 2d that we learn from textbooks, one is instructed to  sum 
  over all integer topological charge  $W \in \Z$ configurations  with action $S= |W|S_I$ 
  \cite{ Coleman198802,  Schafer:1996wv}:
    \begin{align} 
 Z_{\rm ins. \; gas} (\theta) 
&=  \sum_{n=0}^{\infty}   \sum_{\bar n =0}^{\infty}   \frac{1}{n!} \frac{1}{\bar n !}  \left( V_2 K  \rme^{-S_I + \im \theta}  \right)^{n}  \left( V_2 K  \rme^{-S_I - \im \theta}  \right)^{\bar n}    \cr 
&=  \sum_{W \in Z}  \left[  \sum_{\bar n =0}^{\infty}   \frac{1}{(\bar n +W)!} \frac{1}{\bar n !}  \left( V_2 K  \rme^{-S_I}  \right)^{2 \bar n +W}  \right] 
\rme^{\im W \theta}   \cr
&=  \sum_{W \in Z}  I_W \left( V_2 K  \rme^{-S_I}  \right) \rme^{\im W \theta}   \cr
&= \rme^{V_2 K  \rme^{-S_I} \cos \theta }
\label{ins.gas}
\end{align}
leading to  vacuum energy density  and topological susceptibility 
    \begin{align} 
{\cal E} (\theta) \sim - K  \rme^{-S_I} \cos \theta \sim  - K  \rme^{-O(N)} \cos \theta, \qquad  \;\;   \chi_{\rm top.}^{\rm naive} \sim  \rme^{-O(N)} 
\label{ins.gas-2}
\end{align}
which are  in clear contradiction with  the correct result \eqref{soft}. 

In principle, in both \eqref{ins.gas} and \eqref{frac.gas}, we are summing over only $W \in \Z$ configurations. What went so wrong in the naive instanton analysis? 
Of course, the answer is  clear.    In \eqref{frac.gas}, we used the configurations with fractional action and topological charge, but imposed a global constraint on topological charge which restricts the sum to $W \in \Z$. This is equivalent to the statement that we used saddles in the 
 $U(1)/\Z_N$ bundle thanks to  TQFT coupling and then,  lifted the configurations that obey  \eqref{frac.gas} constraint to the $U(1)$ theory. 
In this way, we obtain the  $\theta$ angle dependence  and topological susceptibility correctly.  This settles  the issue raised in 
\cite{DAdda:1978vbw} concerning the role of fractional instantons in determination of topological susceptibility.

\section{Continuation of $\mathbb{CP}^{N-1}$ semiclassics to strong coupling}
The discussion of this section is the heart of our idea of connecting weakly coupled calculable domain to strongly coupled domain. Hence, it is important to  summarize what we did so far for $\mathbb {CP}^{N-1}$, and state what we will do next. 

{\bf 1)} We  showed  the existence of smooth field configurations with fractional $W=1/N$  topological charge in the presence of background gauge fields  for global symmetries.   Their existence follows from the usual Bogomolny argument showing they have  minimum action proportional to their (fractional) topological charge.

{\bf 2)} The backgrounds in which you see this fractionalization are topological  in the sense that they are related to background gauge fields for discrete $ \mathbb Z_N$ symmetries. One can also   gauge the discrete symmetry in question by summing over their background fields weighted by a topological phase called discrete theta angle  which effectively projects out all but one fractional topological charge sector, and so gives the same information.

{\bf 3)} There are then configurations which are approximately the sum of many 
widely-separated fractional instantons (ie, governed by ``saddle points at infinity") in the semi-classical domain \cite{Behtash:2018voa}. Below, we will present the  strong coupling realization of this concept. 

{\bf 4)} These configurations can then contribute to the path integral with fractional-instanton action even if background and boundary conditions only allow integer topological charge.

{\bf 5)} It is a matter of dynamics whether these fractional-instanton-action configurations do or do not contribute in an important way to the path integral.  But analytic continuation from weak coupling semiclassical expansion (familiar from resurgent transseries expansion \cite{Dunne:2012ae}) suggests that at strong coupling there might be an effective description in terms of fractional-instanton local fields (ie, fields which create local excitations with the quantum numbers of the fractional-instanton bosonic and fermionic  zero modes).  This is  familiar from Polyakov model \cite{Polyakov:1975rs,  Polyakov:1987ez} and deformed Yang-Mills theory on $\R^3 \times S^1$ \cite{Unsal:2008ch,  Unsal:2007jx}   where dual theory is constructed based on monopole-instantons. However, although  
this information is  extremely crucial for us,  this is {\bf not}  the approach we will pursue. \footnote{In particular, weak coupling semi-classical expansions  is always hierarchical. On the other hand, strong coupling descriptions are egalitarian. We will comment on this, in retrospect,  after our construction is over.}
    Hopefully we will do something more powerful. 


{\bf 6)} In Refs.\cite{Berg:1979uq, Fateev:1979dc}, Fateev, Frolov, Schwarz, and   Berg,   L\"uscher,  considered   multi-instantons in 
$\mathbb {CP}^{1}$ on   $\R^2$, and parametrized the moduli space of  $n$-instanton, which is $4n$ dimensional, as $2n$ complex coordinates.  Gross    uses this parametrization  earlier  in  \cite{Gross:1977wu}  to show that the interaction between the instanton  and anti-instanton  is a dipole-dipole interaction in $2d$, where each dipole is  as if it is composed of vortex-anti-vortex pair. 
There is no explicit physical splitting in this parametrization, (let us call it X-parametrization for brevity where X is  short-hand for position moduli in terms of fractional vortex instantons)
 for example, a single instanton is still a single lump and {\bf not} two lumps!       The magical thing of this parametrization    is that the determinant of the fluctuation operator can be calculated  {\it exactly}, and muti-instanton gas behaves as if it is a Coulomb gas of vortices  interacting each other depending on their charges and position  moduli. 
  We will give a review of these works,  and close its shortcomings.   Of course, one of our main motivation is the consistency between  fractional quantization of action and topological charge  induced by TQFT coupling, fractional quantization that emerges in weak coupling resurgent expansion and the analysis of determinant of the fluctuation operator  in X-parametrization 
  admitting an interpretation as if the theory on $\R^2$ can be described in terms of these fractional defects. Our goal is to make this correspondence precise.

{\bf 7)} Inspired from  the analysis of  Refs.\cite{Berg:1979uq, Fateev:1979dc},  consistency in  weak coupling semi-classical limits  and resurgent 
expansions, matching of the global symmetries and mixed  't Hooft anomalies,    and  by the 
 consistency (stability) under RG flow of these descriptions,  we propose 
\begin{itemize}
\item massive $N$-flavor Schwinger = Abelian bosonization  =  mass-deformed $SU(N)_1$ WZW 
\end{itemize}
models as low-energy description of asymptotically free $\mathbb  {CP}^{N-1}$ model on $\R^2$. 

\vspace{0.5cm} 
\noindent
{\bf Remarks:}
The fractional vortex  instantons in all  cases are classified with the roots of $SU(N)$ algebra. In semi-classical limits in each case, the elementary tunneling events are given by:
          \begin{align}
        \Delta_{\rm affine}^{(1)} =  \left\{ \bm \alpha_1, \ldots,    \bm \alpha_{N}  \right\}  & \cr
{\bm \alpha}_a \in \Gamma_{r}  \; {\rm of} \; SU(N), & \qquad   {\cal I}  =   \prod_{a=1}^N   \widetilde {V}_a 
\end{align}
where $\bm \alpha_N = - \sum_{a=1}^{N-1} \bm \alpha_a$ is the affine root. 
 We can in principle write a dilute gas based on these charges, and also include  their correlated events such as  $[\widetilde {V}_a  {\widetilde {V}}_b]$  or 
 $[\widetilde {V}_a  \bar {\widetilde {V}}_b]$ etc,  but instead we do something more powerful.  We  write theories for which all orders non-perturbative expansions are the same, according to our understanding of resurgence.  We can call the operators 
 $\widetilde {V}_a$  associated with tunneling  the {\it dual vertex operators}.  
 
 The action of  the abelian bosonization of Schwinger model   will be written in terms of vertex operators (not dual vertex operators),  $ {V}_a$, associated with the  ${\bm \nu}_a \in \Gamma_{w} $ weights of  $SU(N)$ Lie algebra. As asserted, tunnelings in the 
abelian bosonized description are associated with the ${\bm \alpha}_a \in \Gamma_{r} $. In particular, abelian bosonization, its non-abelian version, and massive Schwinger model  are not effective theories describing proliferations of fractional vortex instantons. The fractioal vortex instantons in these theories are same as the one in $\mathbb  {CP}^{N-1}$ model in weak coupling limit. 


 
 There is a necessary matching of scales between different theories.  For example, the non-perturbative  expansion parameter in $\mathbb {CP}^{N-1}$ is  $\rme^{-S_I/N}$, which controls the fugacity of vortex instantons with charges $\bm \alpha_a \in \Gamma_{r}$ in semi-classical domain.  This maps to mass of the fermion in the Schwinger model,  $m_\psi \sim \rme^{-S_I/N}$, which in turn becomes mass of the  rebosonized field in bosonization construction, or mass deformation of the WZW model. 
 

The above description is self-consistent,  because  the mass gap cuts off the IR divergence in the fractional-instanton size modulus, so justifies being analytically connected to a dilute fractional-instanton gas (ie, semiclassical) picture. The renormalization group analysis is independent of $\theta$-angle. In particular, it gives a prediction of gaplessness  for    $\mathbb {CP}^{1}$ at  
$\theta=\pi$ as we show in detail. 
As a result, gap,  gaplessness and the  $\theta$-angle dependence of mass gap  that we find in our formalism  are reliable qualitative predictions.

\subsection{Fateev, Frolov, Schwarz, and    Berg,   L\"uscher, in retrospect}
In 1979, Refs.\cite{Berg:1979uq, Fateev:1979dc}  came up with a very    clever construction  in   the  $\mathbb {CP}^{1}$ model, though as they point out themselves, their method has  some issues. (See Polyakov's textbook for an assessment \cite{Polyakov:1987ez}.)  
 Our construction  partially justifies  their work on rigorous grounds, and improves it. 
    We feel it is extremely important to understand this  discussion in order to perform a similar analysis in gauge theories on $\R^4$.  So, below, we will  explain and improve  the ideas of Refs.\cite{Berg:1979uq, Fateev:1979dc,  Polyakov:1987ez} and bring our own perspective. 

Let us outline Refs.\cite{Berg:1979uq, Fateev:1979dc}  from our perspective. 
 They consider a $q$-instanton configurations in   $\mathbb {CP}^{1}$ model. 
  For $q=1$, 
  an instanton on $\R^2$ has 4 bosonic zero modes, which are usually parametrized  as 
  as 2 position moduli  for the center of instanton $a^{\rm c}$, one size moduli $\rho$, and 
 a $U(1)$  angular moduli $\phi$. The $q$  multi-instanton solution  in $\mathbb {CP}^{1}$ model  can be written in terms of  a holomorphic function  
\begin{align}
w(\z) = \prod_{i=1}^{q} \frac{\z-a^1_i}{\z-a^2_i}
\label{multi-inst}
\end{align}
  where   
  \begin{align}
  \{a^1_1, \ldots, a^1_q\},  \qquad   \{a^2_1, \ldots, a^2_q\} 
  \label{moduli}
  \end{align}
   are just $4q$ complex moduli.  We will refer to this as X-parametrization, for short. 
    It turns out that this 
  is 
  a very useful   parametrization for  the combination of the following two reasons. reasons that will become clear.

{\bf 1)}  Despite the fact that, what one may think at first sight,  the exact solution  for instanton ($q=1$)  does {\bf not} correspond to   physical  splitting 
   of  an instanton  into two distinct lumps.   
 If one inspects the topological charge density (integrand in \eqref{top-den})  for the $q=1$ solution,  
 it is clear that this is a one-lump configuration:
 \begin{align}
Q= \frac{1}{\pi} \int {d^2x}  \frac{ |a^1- a^2|^2 }{ (|\z- \half ((a^1+ a^2)|^2 + |a^1- a^2|^2)^2}  = 
\frac{1}{\pi} \int {d^2x}  \frac{ \rho^2 }{ (|\z- a^{\rm c} |^2 + \rho^2)^2}  =1
\label{top-den}
\end{align} 
centered at $a^{\rm c}$, with size moduli   $\rho$,  and {\bf not} two lumps   located at $a^1$ and $a^2$:
\begin{align} 
a^{\rm c}= \half (a^1+ a^2), \qquad  a^{\rm r}=   \rho \rme^{\im \phi} =  (a^1- a^2)
\label{coord}
\end{align}
Compare this with the  configurations found on $\R \times S^1$ in the $\Omega_F$ background. There,  
when  the size moduli $\rho$   is smaller than the length scale set by  $\Omega_F$, there is a single smooth lump. 
When  the size moduli $\rho$   is larger than  the scale set by  $\Omega_F$,  the instanton fractionates to 
$N$ distinct smooth lumps. 
See Fig. \ref{fig:fraction} and Ref.\cite{Dunne:2012ae}. 

%

{\bf 2)}   Despite the fact that $q=1$ configuration  in \eqref{multi-inst} is a single lump, 
  when the  quantum fluctuation  determinant around  multi-instanton configuration is computed exactly at one-loop order, it  admits an elegant interpretation in terms of a Coulomb gas in 2d,  as if each  instanton is a dipole of two fractional vortex instantons. 
The partition function (counting only  instantons)  can be written as \cite{Berg:1979uq, Fateev:1979dc, Polyakov:1987ez}: 
\begin{align}
Z_{\rm inst} &= \sum_{q=0}^{\infty} \frac{(K \rme^{-S_I/2})^{2q}}{(q!)^2} \int \prod_{j=1}^{q} d^2a^1_j  \;  d^2a^2_j   \cr 
&\times  \exp \left[ \sum_{i < j} \log |a^1_i- a^1_j|^2 +   \sum_{i < j} \log |a^2_i- a^2_j|^2 -   \sum_{i, j} \log |a^1_i- a^2_j|^2 \right] 
\label{sum-vortex}
\end{align}
This suggests the following  picture:   It is as if we are supposed to think of 
 instantons  ${\cal I}$   as the composites of  fractional vortex instantons on $\R^2$, 
$ {V}^1$, and  ${V}^2$.   In some sense, the formalism presented 
in  Refs.\cite{Berg:1979uq, Fateev:1979dc} can be viewed as ``\emph {fractionalization 
  without fractionalization}". 
One can formally write ${\cal I}  \sim {V}^1  {V}^2$, 
where  we assign
$  {V}^1 $    Coulomb charge  +1 ,  and  $  {V}^2 $  charge  $-1$.  
We will learn that    $  {V}^2 $  cannot be interpreted as  anti-vortex, because  if we incorporate theta angle, both  ${V}^1$  and  ${V}^2$ acquires phases $\rme^{\im \frac{\theta}{2}}$.\footnote{The sum which takes only instantons into account  (without anti-instantons) is actually meaningful here, but not always meaningful. 
  For example, in 3d Polyakov model, it would not be meaningful because one cannot impose global charge neutrality with only charge +1 monopoles.  
  (Coulomb system with only + charges is not stable.)
   Here, one   can guarantee global charge neutrality thanks to the fact that there are two types   of fractional vortex instantons  with opposite electric charges, and one can have charge neutrality without taking anti-instantons into account.  
 On $\R^3 \times S^1$ deformed Yang-Mills theory (unlike Polyakov model on $\R^3$), taking only monopole-instantons into account (without anti-monopoles)  is also meaningful for the same reason. There are two kinds of monopoles with charges $+1$ and $-1$, and one does not need anti-monopoles to have global charge neutrality. In fact, the $SU(N)$  theory  with only monopoles is integrable. 
 It maps to  complex affine Toda theory, which happens to have a real spectrum (at $\theta=0$) and can be viewed as a Hirota bilinear form.  See   \cite{Unsal:2008ch}  on these subtle issues.   Once, one includes both monopoles and anti-monopoles, one obtains  \eqref{master} that we discuss in the next section which is no longer integrable.}
 There are actually 
$ \bar {V}^1$  and  $ \bar  {V}^2 $ in the game, that are not included at this stage, but easy to incorporate (see below.)
So, the system behaves as a Coulomb gas with $\pm$ charges, with fugacities  $e^{-S_I/2}$ for fractional vortex-instantons. 
    The Coulomb  systems in 2d  have two phases, 
 a gapless molecular phase and gapped  plasma phase.   $\mathbb {CP}^{1}$  (at $\theta=0$)  happens to be in the gapped plasma phase. 
 The fugacity of the fractional instantons  emerges as the inverse correlation length of the system, $\rme^{-S_I/2} \sim \Lambda $, the strong scale of the theory.   Now, interestingly, despite the fact that these arise from smooth 
 instanton configurations with  X-parametrization, the final form of the action coincide with the one obtained by Gross in 
 \cite{Gross:1977wu} obtained by using merons and intuition.

  The instanton partition function $Z_{\rm inst}$ can be reproduced by using
 a free massless Dirac field 
$ \psi=   {\psi_L \choose  \psi_R} $, and     bosonization formulas  in $2d$  (see Sec. 32 of \cite{ZinnJustin:2002ru}) 
\begin{align}
\sigma_{+} (x)= \bar \psi_L \psi_R (x), \qquad 
\sigma_{-} (x)= \bar \psi_R \psi_L (x), \qquad \qquad   \bar \psi   \psi = \sigma_{+} + \sigma_{-}  
\end{align}
and observing that  
\begin{align}
&\Big\langle \sigma_{+} (a^1_1) \ldots \sigma_{+} (a^1_q)  \sigma_{-} (a^2_1) \ldots \sigma_{-} (a^2_p) \Big \rangle = 
\prod_{i < j}  |a^1_i- a^1_j|^2  |a^2_i- a^2_j|^2   \prod_{i, j}  |a^1_i- a^2_j|^{-2}  \delta_{pq} \;  \equiv  \rme^{- V_{\rm Coulomb}} 
 \delta_{pq}
 \cr
&=  \delta_{pq} \exp{  \left[ \sum_{i < j}  \log  |a^1_i- a^1_j|^2   +\sum_{i < j}   \log |a^2_i- a^2_j|^2   - \sum_{i, j}   \log  |a^1_i- a^2_j|^{2}  \right] }
\label{rule}
\end{align}  
where $\langle  \cdot \rangle$ is evaluated by using massless Dirac action.  
The  $\delta_{pq}$ is due to conservation of chirality. As a result,  the instanton  partition function \eqref{sum-vortex}  can be written as 
 \begin{align}
 Z_{\rm inst} = \int \Diff  \bar \psi \Diff  \psi  \; \; \exp \left[- \int    \left( \bar \psi  \im \gamma_\mu  \partial_{\mu}  \psi + 
K \rme^{ - {S_I \over 2}}   \bar \psi   \psi  \right) \right]
\label{FFSBL}
\end{align}  
 Despite the fact that the inclusion of anti-instantons 
 is presented in   Refs.\cite{Berg:1979uq, Fateev:1979dc,  Polyakov:1987ez} as a conceptual  difficulty, it is actually a relatively simple matter now, especially after understanding the role of fractional instanton-antiinstantons in resurgence \cite{Dunne:2012ae},   and we describe their inclusion below.  There are  few issues  with the  above formula.  
\begin{itemize} 
\item  The reliability (or at least self-consistency)  of semi-classical expansion.
\item The fermionic theory  (at least naively) has a global   $U(1) $  symmetry  that does not match with  the $SU(2)/\Z_2$ symmetry of the original 
$\mathbb {CP}^{1}$ theory.\footnote{Thanks to Aleksey Cherman for discussions about this point.} Effective field theories must respect the global symmetries of microscopic theories. 
On the other hand, there is actually hope that this point may be fixable, because abelian bosonization (as opposed to non-abelian bosonization) makes some symmetries extremely non-obvious \cite{Coleman:1976uz}. 
There is clearly something to be figured out. 
\item If we incorporate $\theta$, it will not produce $\theta$ angle physics correctly. But this is obvious, since we did not include anti-instantons.   
 \end{itemize}
 
 \noindent
{\bf What FFS-BL did not know and resurgence at help.}  The coupling  a $\Omega_F$-twist and  $\Z_N$  TQFT makes it manifest that the topological defects with fractional charge $W=1/N$ and action  $S_I/N$ exist both in the weak coupling semi-classical domain on $\R \times S^1$ and on arbitrary $T^2$ at strong coupling.   Furthermore,  unlike the past times \cite{Berg:1979uq, Fateev:1979dc}, we  no longer  feel uncomfortable   about (fractional or not)   instanton anti-instanton pairs 
because 
  resurgence theory tells us what their effects are, and how they  are incorporated correctly and cancel ambiguities in perturbation theory  \cite{Dunne:2012ae}.  In particular, when we do semi-classics, we are forced (this is not a choice) to consider the complexification of field space. In the old days, it was thought  that the   instanton-antiinstanton  configurations could not be distinguished from perturbative vacuum, and the 
   quasi-zero mode integration  between a fractional     instanton-antiinstanton  would be ill-defined. Now, we know  two facts.  Fractional instanton antiinstanton  configurations are critical points at infinity (see around 
   \eqref{cluster} for a reminder),    
    and in order to calculate their contribution to path integral, we need to do path intergal over the Lefschetz thimble, which is meaningful \cite{Dunne:2012ae, Behtash:2018voa}.  Lefschetz thimble integration at second order in semi-classic  gives two types of contributions, uncorrelated  and correlated pairs     
$    \frac{\beta^2}{2!}   [ {\cal I}_a]  [\bar {\cal I}_a]+  \beta [ {\cal I}_a \bar {\cal I}_a]_{\pm}$.
 In other words, thimble integration is a sophisticated  version of cluster expansion, which gives the effect of both uncorrelated pairs as well as correlated pairs.  
  The effect of correlated pair is 
   two-fold ambiguous  $\Im  [ {\cal I}_a \bar {\cal I}_a]_{\pm} \sim  \pm  \rme^{-2S_I/N}$. 
 But there is another ambiguity in the theory that comes upon 
   lateral (left/right) Borel resummation of perturbation theory,  which we denote as $\mathbb B_{\pm}$. Remarkable fact that comes from resurgent analysis of $\mathbb {CP}^{N-1}$ is that 
 \begin{align}  
 \Im \mathbb B_{\pm} + \Im [ {\cal I}_a \bar {\cal I}_a]_{\pm} =0
\end{align}   
i.e., the imaginary ambiguities  cancel  \cite{Dunne:2012ae}.   This ambiguity cancellation is  at order $\rme^{-2S_I/N} \sim \Lambda^2 $ and is 
the semi-classical realization of   't Hooft's   famous  IR renormalon  puzzle \cite{tHooft:1977xjm}  (see also  \cite{Munster:1981zn}) in  $\mathbb {CP}^{N-1}$ model. 
 It should be noted in the analysis below, we do not use resurgence, in fact,  
  we do not even use second order correlated pairs in semi-classics 
 except one occasion where it is strictly necessary to determine the IR-dynamics  in $\mathbb {CP}^{1}$ at $\theta=\pi$ around \eqref{ambiguity-2}.    But the advantage that resurgence provides is clarity it provides in knowing what various non-BPS configurations are,  and what the corresponding operators  and their 
 physical effects are, and we do not need to need all thanks to renormalization group. 
  Therefore, in the light of these progress progress in resurgent semi-classical analysis,  we feel encouraged to   reconsider the  FFS-BL analysis \cite{Berg:1979uq, Fateev:1979dc} on $\R^2$, and improve it.

%

\subsection{Incorporating $\theta$ and  fractional anti-instantons into FFS-BL}

Based on \eqref{sum-vortex},  we view the instanton on $\R^2$ as a composite of the two fractional (vortex) instantons. But recall that the topological charge density or action density 
\eqref{top-den} are  not a two lump configuration. They  can be smoothly be split to two-lumps either using   $\Omega_F$ twist background, or in the background of $B^{(2)}$ 
$\Z_2$-flux  TQFT coupling. 
 Both of these backgrounds introduce a scale and if the moduli parameter becomes larger than a certain scale set by the background, we  see fractionalization explicitly.    The remarkable fact about the \cite{Berg:1979uq, Fateev:1979dc}  analysis is that it reveals something deep about the inner structure 
of 2d instanton via a parametrization and exact computation of the fluctuation determinant. 

 Due to its extreme importance in the story, we will also incorporate $\theta$ angle dependence from now on. 
First, note that incorporating  $\theta$ angle in \cite{Berg:1979uq, Fateev:1979dc} 
 modifies  \eqref{FFSBL} 
 as $ \rme^{-\frac{S_I}{2} }  \rightarrow  \rme^{-\frac{S_I}{2} + \im \frac{\theta}{2}} $. Expanding the exponentials, we can perform the  path integral 
 by using  \eqref{rule} which tells us that when the  
 number of  $\sigma_{+} $ and $\sigma_{-}$  insertions are not  equal, the integration vanishes due to excess chirality. 
 As a result, the   fractional vortex instanton sum  \eqref{sum-vortex}  is modified 
 into $\sum_{q=0}^{\infty}  (\ldots) \rme^{ \im  q \theta} $, where  $q =0,1, 2, \ldots $ which indicates that the sum 
is  only  over non-negative  integer winding  instantons (without anti-instantons).  The inclusion of $\theta$ 
modifies  the action  of Refs~\cite{Berg:1979uq, Fateev:1979dc}  into 
\begin{align}
S= 
 \bar \psi  \im \gamma_\mu  \partial_{\mu}  \psi + 
K \rme^{ - {S_I \over 2}}   \rme^{ \im  \frac{\theta}{2}}   \bar \psi   \psi 
\label{FFSBL2}
\end{align}
which looks a bit awkward, because $\rme^{ \im  \frac{\theta}{2}} $ appears in front  of the Dirac mass term.

If we were to incorporate our sum based on proliferation of $  {V}_1 $,  $ \bar {V}_1$, (instead of   $  {V}_1 $,  $  {V}_2$ pair),    the effective action would be  replaced by  
$\bar \psi  \im \gamma_\mu  \partial_{\mu}  \psi  +  \rme^{ \im  \frac{\theta}{2}}   \bar \psi_{\rmL} \psi_{\rmR}   +  \rme^{ - \im  \frac{\theta}{2}}   \bar \psi_{\rmR} \psi_{\rmL} $.  This can only  describe  topological charge $Q=0$ sector of the theory.    


 Let us know take into account both instantons and anti-instantons.    
 We view ${\cal I} \sim {V}_1  {V}_2$ where ${V}_a$ stands for {\it fractional vortex instanton}:  
  \begin{align}
&{\cal I} \sim {V}_1  {V}_2,  \qquad  {V}_1 \sim  \rme^{-\frac{S_I}{2} + \im \frac{\theta}{2}} \sigma_{+}, 
\qquad {V}_2 \sim  \rme^{-\frac{S_I}{2} + \im \frac{\theta}{2}} \sigma_{-} \cr
& \bar {\cal I} \sim \bar {V}_1  \bar {V}_2,  \qquad  \bar {V}_1 \sim  \rme^{-\frac{S_I}{2} - \im \frac{\theta}{2}} \sigma_{-},  \qquad  \bar {V}_2 \sim  \rme^{-\frac{S_I}{2} - \im \frac{\theta}{2}} \sigma_{+} 
\label{ops}
  \end{align}  
 The topological and Coulomb charges of these configurations are: 
   \begin{align}
&{V}_1: \left(+\half, +1 \right), \qquad     {V}_2:  \left(+\half, -1 \right)   \cr 
&\bar {V}_1: \left(-\half, -1 \right), \qquad    \bar  {V}_2:  \left(-\half, +1 \right) 
    \end{align} 
    Note that Coulomb charges of ${V}_1$ and  $\bar {V}_2$  are same, but their topological charge differ in sign. 
They   interact with each other according to  Coulomb's law in $2d$.  The proliferations of these events on $\R^2$ produce 
a modified mass term: ${V}_1 +   {V}_2 + \bar {V}_1 +   \bar {V}_2$.
As a result, we obtain the  generalization of  \eqref{FFSBL2},  
 \begin{align}
 Z_{\rm inst/anti-inst} &= \int \Diff  \bar \psi \Diff  \psi  \; \; \exp\left[- \int    \left( \bar \psi  \im \gamma_\mu  \partial_{\mu}  \psi + 
2 K \rme^{ - {S_I \over 2}}   \cos\frac{\theta}{2}   ( \bar \psi   \psi )  \right) \right]  
   \label{full}
\end{align}  
which is now   invariant under $\theta \rightarrow - \theta$.  This path integration does reproduce the proliferation of both vortices and anti-vortices with the correct $\theta$ angle dependence, and incorporate all $W \in \Z$ sectors.    However, it is not only a sum over configurations of the form ${\cal I}^n, \bar {\cal I}^n$, it also include contribution of configurations such as $ {V}_1  \bar {V}_1 $ and  $ {V}_2  \bar {V}_2 $.

Let us now investigate the  configurations contributing to the full partition function carefully. 
 \begin{align}
 Z & =  \int \Diff  \bar \psi \Diff  \psi  \; \;   \rme^{- \int    \left( \bar \psi  \im \gamma_\mu  \partial_{\mu}  \psi +  \lambda ( \rme^{\im \frac{\theta}{2} }   \bar \psi_L \psi_R +   \rme^{\im \frac{\theta}{2} }   \bar \psi_R \psi_L +   \rme^{-\im \frac{\theta}{2} }   \bar \psi_R \psi_L +  \rme^{-\im \frac{\theta}{2} }   \bar \psi_L \psi_R )   \right)}   \cr
& = \int \Diff  \bar \psi \Diff  \psi  \; \; \rme^{- \int    \left( \bar \psi  \im \gamma_\mu  \partial_{\mu}  \psi \right)  }   \cr
 & \times  \sum_{n_1=0}^{\infty}  \frac{ \left( \rme^{\im \frac{\theta}{2} }   \lambda \right)^{n_1} }{n_1!} 
 \int  \left( \prod_{j=1}^{n_1}  d^2a^{1}_j   \right) \sigma_{+} (a^{1}_1)   \ldots \sigma_{+} (a^{1}_{n_1})      \cr
&\times   \sum_{n_2=0}^{\infty}  \frac{ \left( \rme^{\im \frac{\theta}{2} }   \lambda \right)^{n_2} }{n_2!}  
  \int    \left( \prod_{j=1}^{n_2}  d^2a^{2}_j   \right) 
   \sigma_{-} (a^{2}_1)   \ldots \sigma_{-} (a^{2}_{n_2})   \cr
& \times   \sum_{ \bar n_1=0}^{\infty}  \frac{ \left( \rme^{ - \im \frac{\theta}{2} }   \lambda \right)^{ \bar n_1} }{\bar n_1!} 
 \left( \prod_{j=1}^{\bar n_1}  d^2b^{1}_j   \right)   \sigma_{-} (b^{1}_1)   \ldots \sigma_{-} (b^{1}_{\bar n_1})      \cr
 &\times    \sum_{ \bar n_2=0}^{\infty}  \frac{ \left( \rme^{ - \im \frac{\theta}{2} }   \lambda \right)^{ \bar n_2} }{\bar n_2!} 
 \left( \prod_{j=1}^{\bar n_2}  d^2b^{2}_j   \right)     \sigma_{+} (b^{2}_1)   \ldots \sigma_{+} (b^{2}_{\bar n_2})      
   \label{full-2}
\end{align}  
A  typical term in the sum has $(n_1, n_2, \bar n_1, \bar n_2) $  many   $({V}_1, {V}_2, \bar {V}_1, \bar {V}_2)$ insertions, respectively.  
In \eqref{full-2},  the number of $\sigma_{+}$ insertions  is  $n_1 +\bar n_2 $ and the number of $\sigma_{-}$ insertions  is  $n_2 +\bar n_1 $. 
Therefore, using   \eqref{rule}, we observe that the sum is non-zero if and only if  
\begin{align}
n_1 -\bar n_1 = n_2 -\bar n_2 = W \in \Z 
\label{const-5}
\end{align}
 constraint is obeyed.  
We can solve this one constraint to turn four sums into triple sum one of which can be identified as a topological charge $W\in \Z$: 
 \begin{align}
Z &     =  \sum_{W \in \mathbb Z}  \left[\sum_{\bar n_1=0}^{\infty} \sum_{\bar n_2=0}^{\infty}  \frac{\lambda^{2 \bar n_1 + 2 \bar n_2 + 2 W}}{(\bar n_1+ W)! \bar n_1!  (\bar n_2+ W)!  \bar n_2! }      \int  \prod_{j=1}^{ n_1}  d^2a^{1}_j   
 \prod_{j=1}^{ n_2 }  d^2a^{2}_j  \prod_{j=1}^{\bar n_1 }  d^2b^{1}_j    \prod_{j=1}^{\bar n_2 }  d^2b^{2}_j  \;  \rme^{-S_{\rm int}} \right]_{n_a= \bar n_a+ W}
\rme^{ i W \theta} 
 \end {align}
 where the interaction action is: 
  \begin{align}
S_{\rm int}=  4  \pi \sum_{i=1}^{2}  q_i \cdot q_j \left[ \sum_{k=1}^{n_i} \sum_{l=1}^{n_j} G(a^i_k - a^j_l)  +  \sum_{k=1}^{\bar n_i} \sum_{l=1}^{ \bar n_j} G(b^i_k - b^j_l)   -  2  \sum_{k=1}^{ n_i}  \sum_{l=1}^{\bar n_j} G(a^i_k - b^j_l)    \right]_{n_i= \bar n_i +W}
 \end {align}
 where $G(a) = -  \frac{1}{4 \pi} \log |a|^2$ is 2d Green's function.

 Now, if we take \eqref{full} seriously, assuming the semi-classics that leads to it is reliable (we will question this later),  it  tells us that the $\mathbb {CP}^1$ theory on $\R^2$ develops a finite correlation length for $\theta \neq 0$,  a mass gap, of the form: 
\begin{align}
m_{\rm gap} (\theta)  \;  \underbrace{=}_? \;  \Lambda \left| \cos\frac{\theta}{2}  \right|
\label{gap}
\end{align}
This is, of course, a beautiful  result at first sight, compliant with all of our  educated guesses, a  non-perturbative mass gap at $\theta\neq \pi$ and  gaplessness at   $\theta=\pi$, and a wonderful confirmation of the formalism  described in 
Refs~\cite{Berg:1979uq, Fateev:1979dc}, and \cite{Polyakov:1987ez}. But  this  formula (not its consequences)  \eqref{gap} is actually not quite correct, because the formalism we set-up to this point, has some  flaws.

First, let us describe why we  got an ``correct looking" result. 
 The partition function has the form  $Z      =  \sum_{W \in \mathbb Z}  Z_W  \rme^{ i W \theta} $ where  $Z_W$   is the partition function of a fixed $W \in \Z$ sector.  But contributing configurations is  richer than just naive  instanton sums.  
Clearly, there are configurations of the form 
 ${V}_1   \bar {V}_1 $ with action $\rme^{-2 \times \frac{S_I}{N} } $ which happens to be just instanton action for $N=2$, but for general $N$, fractional action configurations.   At any rate, even for $N=2$, this configuration is distinct from 2d  instanton \cite{Polyakov:1975yp}, which is ${V}_1   {V}_2 $.   The fact that configurations such as  ${V}_1 $ here, and $  \bar {V}_1 $ there   contribute  to path integral is responsible for the appearance of $\cos\frac{\theta}{2}$  in the effective action. With the usual instantons, we could never get such a factor. This is a good part of the truth.

Polyakov, in his book \cite{Polyakov:1987ez},  following a  proposal of   Ref.\cite{Bukhvostov:1980sn}, 
argues that to incorporate anti-instantons,  one should include a  second Dirac fermion and couple the two in a special way. 
This is by no means necessary. The Coulomb charge of ${V}_1$ is same as the Coulomb charge of 
$\bar {V}_2$. At $\theta=0$, there is no distinction between the corresponding defects and the corresponding  operators. 
Only when we turn on    $\theta$, we see their difference due to topology 
$ {V}_1 \sim  \rme^{-\frac{S_I}{2} + \im \frac{\theta}{2}} \sigma_{+},   \; 
\bar {V}_2 \sim  \rme^{-\frac{S_I}{2} - \im \frac{\theta}{2}} \sigma_{+}$, but this still does not demand introduction of a second fermion, since the Coulomb charges are same.  
Hence, to produce the correct sum over vortex instantons and anti-instantons,  \eqref{full} is  perfectly sufficient as we showed.  

A possible   issue with  \eqref{FFSBL} (or even \eqref{full}) is that, at least naively,  it  does not  look  consistent  
with  the global symmetry of the original $\mathbb {CP}^{1}$ model, which is $SU(2)/\Z_2$.  Rather, it has  an obvious  $U(1)$ symmetry.    But effective field theories {\bf must}
respect the global symmetries of microscopic theories, and there is no negotiation room there. However, similar to abelian bosonization, sometimes symmetries may be obscured. Still, we want to make global symmetry manifest.

Furthermore, the summation over fractional vortex instantons $V_a$, $\bar{V}_a$ is a good approximation if there exists a weak coupling semi-classical domain of the theory, similar to small $S^1 \times \R$ as in \cite{Dunne:2012ae}. 
But here, we are in the strongly coupled domain of the $\mathbb CP^{1}$. Is it possible in some way to take into account correlated events such as  $[{V}_a \bar {V}_1]_{\pm},   [{V}_a \bar{V}_b],  [{V}_1 {V}_1 {V}_1 ],   \ldots $ etc. Note that this is not a quantitative, but qualitative  issue. For example, does the gaplessness \eqref{gap} persist once higher  order effects are taken into account?  Or can the theta dependence include terms of the form 
$\cos \frac{3 \theta}{2}$, or include fractional powers  such as $\left(\cos \frac{\theta}{2} \right)^{1/3}$?  
 If one can achieve  this, there is a hope to go beyond semi-classical ideas via such construction.

\subsection{Fermionization, Abelian rebosonization and mass gap  at arbitrary $N$ }
We are largely inspired from the fermionic description of the fractional vortex instanton gas \cite{Berg:1979uq, Fateev:1979dc} in $ \mathbb {CP}^{1}$ model. However, this construction comes with a number of deficits, a number of which we  already fixed. However, probably the most important issue is that original $ \mathbb {CP}^{1}$  has a 
$SU(2)/\Z_2$ symmetry, which is not manifest in   \eqref{FFSBL} (or \eqref{full}). In more general  $ \mathbb {CP}^{N-1}$ case, this symmetry is  $SU(N)/\Z_N$, and it is important to capture the mixed anomaly structure of these theories correctly. Therefore, we introduce $N$ massive fermions and gauge over-all $U(1)$ in the general case. 
Therefore, we land on  the $N$-flavor  Schwinger  model with an $SU(N)$  invariant  mass term. 
\be
S={1\over 2e^2}\int_{M_2} |\diff a|^2 +{\im \theta\over 2\pi}\int_{M_2} \diff a
+\int_{M_2} \diff^2 x\,  \overline{\bm \psi} \gamma^{\mu}(\partial_{\mu}+  \im a_{\mu}) {\bm \psi} + 
m   \overline{\bm \psi}  {\bm \psi} 
\label{eq:Schwinger-2}
\ee
%
The global symmetry, mixed anomalies and global inconsistencies, weak coupling semi-classical description 
 of the $N$-flavor Schwinger model match exactly with the 
 $ {\mathbb CP}^{N-1}$ model.  Furthermore,  $2$-flavor Schwinger model  provides the correct fix  for the 
 result of Refs~\cite{Berg:1979uq, Fateev:1979dc} on $\R^2$ based on the X-parametrization of moduli space. 
 Because of  these matching,  it is natural that  the low energy theory for the  $ \mathbb {CP}^{N-1}$ model is described by massive Schwinger model, or equivalently, its abelian bosonization or its non-abelian bosonization given by mass deformed $SU(N)_1$ WZW model with an extra scalar. 
The global symmetry in all cases is 
$G = SU(N)/\Z_N$
These theories also have charge conjugation symmetry  ${\mathsf C}$  at $\theta=0, \pi$.  The mixed anomaly and global inconsistency are exactly the same in these theories \cite{Dunne:2018hog, Misumi:2019dwq}.

 We will now show that, upon bosonization, the  tunneling events in this theory are described by the same set as in 
  $ \mathbb {CP}^{N-1}$ model, by   \eqref{tunnelings0}  and   \eqref{tunnelings1}.   Later, when we describe semi-classics, we will observe that the mass of the fermion  $m$ and the coupling $e^2$  in Schwinger model are  controlled by strong scale 
 $ \Lambda =  \mu \rme^{-\frac{4\pi}{g^2(\mu) N} } $   of the    $ \mathbb {CP}^{N-1}$  model.  In particular, $m \sim \Lambda$ and $e^2 =  \frac{48\pi \Lambda^2}{N}$.

First, let us review  Coleman's celebrated result, by following his footsteps in \cite{ Coleman:1976uz}, with some minor remarks sprinkled here and there.  For $N$-Fermi fields, $N$-Bose fields are introduced. 
The  bosonization dictionary is: 
\begin{align} 
\overline{\psi}_a  \gamma^{\mu} \psi_a \leftrightarrow \pi^{-1/2} \epsilon_{\mu \nu} \partial_\nu \phi_a \cr 
\overline{\psi}_a  \psi_a  \leftrightarrow  -c m N_m \cos \left( \sqrt {4 \pi}  \phi_a \right)
\end{align} 
where $N_m$ denotes normal ordering with respect to mass $m$, and $c$ is numerical constant. There are two useful identities proven  in \cite{Coleman:1974bu} and used in  \cite{ Coleman:1976uz}.  $N_m[ \Pi^2 + (\partial_1 \phi)^2] = N_\mu [ \Pi^2 + (\partial_1 \phi)^2  + \rm constant]$  (just an additive shift to ground state energy, which we can be ignorant about in this work),  and more importantly  
\begin{align}
N_m \cos (\beta \phi)  =  \left(\frac{\mu}{m}\right)^{\beta^2/4\pi}  N_\mu  \cos (\beta \phi) 
\label{renormal}
\end{align}
This relation encodes multiplicative remormalization of the  $\cos (\beta \phi)$ operator, 
 to account for the change of the normal ordering scale.  

To obtain the bosonized action, we use the dictionary and integrate out  the gauge field.  
\be
L=\half \sum_{a=1}^N (\diff \phi_a)^2  -  c m^2   \sum_{a=1}^{N} \cos \left( \sqrt {4 \pi}  \phi_a \right) +  
 \frac{e^2}{2 \pi}  \min_k  \Big( \sum_{a=1}^N  \phi_a - \frac{1}{\sqrt{4 \pi} } ( \theta +  2 \pi k)  \Big)^2
\label{coleman0}
\ee
This is same as Coleman's original \cite{Coleman:1976uz} for $N=2$, except that we took into account the global aspects of gauge field in integrating it out, hence, $\min_k$ structure in the potential.  
As Coleman instructs us, the  $SU(N)$ invariance  of the    theory is  obscured in this presentation, but not lost.  If we wanted to keep $SU(N)$  invariance manifest throughout the discussion,  we have to  use non-abelian bosonization and mass perturbation  of the WZW action.  Because of its utility, we will return to this perspective as well.

Define the following  transformation:
\begin{align}
\widetilde \phi &= \frac{1}{\sqrt N}  \sum_{a=1}^N  \phi_a  \equiv  \frac{1}{\sqrt N} {\bm e}_0 \cdot  \bm \phi 
 \cr  
\widetilde \phi_a &=  \phi_a -  \frac{1}{ N} \sum_{a=1}^N  \phi_a  =  {\bm \nu}_a \cdot \bm \phi   \qquad a=1, \ldots, N  \qquad   \sum_{a=1}^{N} \widetilde \phi_a  =0
\end{align}
where now  $N$-fields $\widetilde \phi_a$ obeys a  constraint,   hence, representing $N-1$ fields, we can write the kinetic term as 
$  \sum_{a=1}^N  (\diff  \widetilde \phi_a)^2  +  \frac{1}{2}   (\diff  \widetilde \phi)^2 =  \frac{1}{2}  \sum_{a=1}^N  (\diff \phi_a)^2$. As a result, the lagrangian
 can be written as 
\begin{align}
L&=\frac{1}{2}  (\diff  \widetilde \phi)^2  +  \frac{1}{2}   \frac{Ne^2}{\pi}  \min_k \left(\widetilde \phi -   \sqrt {\frac {1} {4 \pi N} } (\theta + 2 \pi k)  \right)^2+  \cr
&+ \frac{1}{2}  \sum_{a=1}^{N}  (\diff  \widetilde \phi_a)^2   -  c m^2   \sum_{a=1}^N      \cos \left( \sqrt {\frac{4 \pi}{N} }  \widetilde \phi   + \sqrt {4 \pi  }  \widetilde \phi_a
 \right) 
\end{align}
$\widetilde \phi$  particle has a mass  square $M^2 = \frac{Ne^2}{\pi} =  48 \Lambda^2$. In Schwinger model, this mode can be made parametrically heavy compared to other modes.  However, since we are performing a matching between 
$\mathbb {CP}^{N-1} $ and Schwinger model to match their semi-classical description in terms of tunneling events,  we know that this mode cannot be made parametrically heavier than the lower lying modes. This mode is  likely the $SU(N)$ singlet mode in the spectrum of $\mathbb {CP}^{N-1} $,  
  $\lim_{x \rightarrow y} \bar z_i(x) \rme^{\im \int_x^y a }  z^i (y) $,    which differs from  the adjoint by an order  
  $O(1/N)$   splitting \cite{Witten:1978bc}.  So, this splitting is more pronounced for $N=2,3$, and not so important at large-$N$ where indeed, the adjoint and singlet become degenerate. With this in mind, let us integrate out the singlet field. 
  Using  the renormal ordering (or matching) prescription   \eqref{renormal} of Ref.\cite{Coleman:1976uz}, 
the  low-energy theory takes the form:
\begin{align}
L &= \frac{1}{2}  \sum_{a=1}^N  (\diff  \widetilde \phi_a)^2   -c m^{\frac{N+1}{N}} M^{\frac{N-1}{N}}   \sum_{a=1}^N    \cos \left( \frac{\theta + 2 \pi k}{N}    + \sqrt {4 \pi} \;   \widetilde \phi_a  \right),  \qquad   \widetilde  \phi_N= -  \sum_{a=1}^{N-1} \widetilde \phi_a
 \label{Coleman1}
\end{align}
%
Two mass parameters in this action can be further merged to one, by re-using a renormal-ordering  with respect to mass $m'$. 
\begin{align}
L&=\half  (\diff \bm \phi)^2  - (m')^2   \sum_{a=1}^{N} \
\cos \left(   \sqrt {4 \pi}  {\bm \nu}_a \cdot \bm \phi   \right)  + \ldots \cr
 m' (\theta) &= \left( c m^{\frac{2}{N}} M^{\frac{N-1}{N}} \cos  \frac{\theta + 2 \pi k}{N}    \right)^{\frac{N}{N+1}} 
 \label{Coleman2}
\end{align}
This result agrees with  Refs.\cite{Coleman:1976uz} for $N=2$ and  \cite{ Smilga:1992hx, Hetrick:1995wq, Hosotani:1998za} for general $N$.   Here, we can regard  $\bm \phi$ as an $N$-component vector. The mode associated with  
${1 \over {\sqrt N}}  \bm e_0 \cdot \bm \phi$ decouples from the rest, making \eqref{Coleman2} particularly easy formulation to work with. \footnote{ In the process of  completing  this paper,  we became aware of Ref.\cite{Wamer:2020hmv},  which  
gives a description of  flag manifold sigma models, where fractional vortex instantons associated with   ${\bm \nu}_a \in \Gamma_w$  gaps out the theory.  It is easy to make a connection between that work and ours by using $SU(N)_1$  WZW formulation \eqref{eq:general_bosonization},   and using  double-trace deformations \cite{Unsal:2008ch} to connect WZW to flag manifold, generalizing \cite{Tanizaki:2018xto}. }

\vspace{0.5cm}
\noindent
{\bf Tunneling events in abelian bosonization description:}
The elementary  tunneling events in \eqref{Coleman2} can be described similar to \eqref{tunnelings0}, and they correspond to  
\begin{align}
\Delta \bm \phi = \bm \phi(\tau=L) - \bm \phi(\tau=0)  = \sqrt \pi \bm \alpha_a,    \qquad a=1, \ldots, N
 \label{tunnelings3}
\end{align}
the simple roots  and affine root of $SU(N)$ algebra living in the affine root system.  
     \begin{align}
        \Delta_{\rm affine}^{(1)}=  \left\{ \bm \alpha_1, \ldots,    \bm \alpha_{N}  \right\} 
       \end{align} 
The crucial point is that the action of these tunneling events is controlled by $m'$. Using the matching conditions with 
 $\mathbb {CP}^{N-1}$ for $m$ and $M$,  we observe that the tunneling amplitude is  controlled by  $\rme^{- S_I/N}$ in the bosonized version of the Schwinger model. 
 
We can also   define tunneling events associated with higher roots, and define the orbits as:
  \begin{align}
            \Delta_{\rm affine}^{(k)}&=  \left\{ \bm \alpha_a + \bm \alpha_{a+1} + \ldots   \bm \alpha_{a+k-1}    | a=1, \ldots N \right\}, \qquad  k=1, \ldots, N-1
       \end{align} 
 corresponding to deomposition of $N^2 -N$ roots in total to $N-1$ orbits.
  Note that  $  \Delta_{\rm affine}^{(N)}=  \bm \alpha_a + \bm \alpha_{a+1} + \ldots   \bm \alpha_{a+N-1}  =0. $
The events in the $ k^{\rm th}$  orbit have fugacity $\rme^{- k S_I/N}$

 The proliferation of the vortices describe a 
  classical  two-dimensional Coulomb gas of $N^2-N$ types of  charges  associated with charges 
$   \bm \alpha_{ab} =  \bm \nu_a -  \bm \nu_b=  \bm e_a- \bm e_b$ interacting with each other via 
 via  the 2d Coulomb's law,  
   \begin{align}
   \pm  \bm \alpha  \cdot  \bm   \beta   \frac{1}{4 \pi} \log |  \bm a_{\bm \alpha} -\bm a_{\bm \beta} |^2
   \end{align}
In order to define the operators associated with tunneling events, we need to use abelian duality in 2d. Using 
$\diff \bm \sigma = \star \diff \bm \phi$, we can write the  tunneling events as: 
\begin{align}
\widetilde {V}_{\bm \alpha_{ab}} = \rme^{- \frac{  |a-b| S_I}{N}} \rme^{\im   \sqrt \pi  \bm \alpha_{ab}  \cdot  \bm \sigma }
\label{dualdual}
\end{align}
Note that these events are hierarchical in a semi-classical domain, for example, once we introduce the  $\Omega_F$  
  background.  

Since both semi-classical expansion and  the mixed anomalies of the massive Schwinger model  are same with the  $\mathbb {CP}^{N-1}$,  the massive Schwinger model will provide an qualitatively accurate description of the infrared physics of  $\mathbb {CP}^{N-1}$.

\subsection{Fractional instanton renormalization group (FIRG)}
 Since $ \widetilde \phi_a$ is a constrained field, we need to be careful in reading off  the dimension of the $\cos(\ldots)$ operator. For example, setting $N=2$ on \eqref{Coleman1}, case examined in \cite{Coleman:1976uz},  $ \widetilde \phi_2= -\widetilde \phi_1$, where $ \widetilde \phi_1 $ is unconstrained. Then, properly normalizing the kinetic term  by a field redefinition, 
\begin{align}
L &= \frac{1}{2}   (\diff  \widetilde \phi_1)^2   -2 (m')^2 
\cos \left( \sqrt {2 \pi} \;   \widetilde \phi_1  \right), 
 \label{Coleman1N2}
\end{align}
and dimension of the $\cos(\ldots)$ operator   is $\Delta[\cos ( \sqrt {2 \pi} \;   \widetilde \phi_1  ) ]= \frac{\beta^2}{4\pi} = \half$, in agreement with  \cite{Coleman:1976uz}.   For  general $N$, it is slightly tricky to read off the dimension of $\cos(\ldots)$ operator 
and for this, the way we parametrized things in  \eqref{Coleman2} is useful. In this formula, $\bm \phi$ field is unconstrained, 
but there is an extra redundant mode in it that decouples from the dynamics. As a result, the 
 dimension of   vertex operators are easy to read: 
\begin{align} 
{\rm dim}({V}_a ) = \Delta \left[ \rme^{\im  \sqrt {4 \pi} \;  {\bm \nu_a \cdot  \bm \phi}   } \right]= \frac{\beta^2 {\bm \nu_a}^2 }{4\pi}= 
 \frac{N-1}{N}
  \end{align}
This is consistent with non-abelian bosonization in which the identification is: 
\begin{align} 
\overline{\psi}_{\rmL, a}  \psi_{\rmR,b}   \sim     U_{ab} = 
  \rme^{\im  \sqrt {\frac{4 \pi}{N} }    \widetilde \phi }   g_{ab}
  \end{align}
and  the dimensions of the corresponding  operators are 
$  \Delta \Big[    \rme^{\im  \sqrt {\frac{4 \pi}{N} }    \widetilde \phi }  \Big]  = \frac{1}{N}$ and    $\Delta \Big[ g_{ab}  \Big]=  \frac{N-1}{N}$
where $g$ is $SU(N)$ valued.   

Note that IR scaling in the vertex operator in the large-$N$ limit goes to 1,  same as free field or UV of a fermion bilinear in 2d. 
The most dramatic change in the scaling dimension happens for $N=2$,  in that case, the scaling of fermion bilinear drops from 
 $1$ in the UV to $\frac{1}{2}$ in the IR.     This has an interesting effect that we discuss separately for $N=2$ case with $\theta$  angle.

%
%


The mass matrix for the $ \bm  \phi$  field can be found by diagonalizing  $\sum_a ({\bm \nu}_a \cdot \bm  \phi)^2$ matrix.  The mass square matrix components are
$M_{ij}^2 = {\bm \nu}_a^{i}  {\bm \nu}_a^{j}   = \delta_{ij} - \frac{1}{N}$
\begin{align}
M^2 =\bm{1}_N - \frac{1}{N} {\bm J}_N
\end{align}
where  $\bm J_N$ is all-one matrix and  $\bm{1}_N$ is identity matrix.  
 $M^2$ has $N-1$ eigenvalues which are equal to $1$, for which the eigenvectors are $\bm \alpha_a, a=1, \ldots   N-1$ and $1$ eigenvalue equal to $0$ with eigenvector $\bm e_0$.   
%


\vspace{0.5cm}
\noindent 
{\bf   Relevants and irrelevants in infrared dynamics (IR scaling dimensions):}
Let us list  the scaling dimension of various vertex operators.\footnote{The  vertex operators in the effective action \eqref{Coleman3}  can be viewed 
as tunneling events of the effective action based on proliferation of the vortices with charges  \eqref{tunnelings3} in the 
$\bm \alpha \in \Gamma_{r}$, based on the dual field  $\bm \sigma$ in   \eqref{dualdual}. }
  This will help us to identify which of these operators are relevant in the infrared. It will also help us "tower" of interesting operators which cary same Coulomb charge (scaling dimension), but higher topological charge.  
\begin{align}
{\rm dim}(1)&= 0 \cr 
{\rm dim}({V}_a )&=   \Delta \left[ \rme^{\im  \sqrt {4 \pi} \;  { \bm \nu_a   \cdot  \bm \phi}   } \right]=  (\bm \nu_a)^2 =  \Big(1-\frac{1}{N} \Big)   \cr
{\rm dim}({V}_a {V}_{b} )&= 
\Delta \left[ \rme^{\im  \sqrt {4 \pi} \;  { (\bm \nu_a + \bm \nu_b)  \cdot  \bm \phi}   } \right]=  (\bm \nu_a + \bm \nu_b)^2  = 
 \left\{ \begin{array}{ll}
    2 \Big(1-\frac{2}{N} \Big) &  \qquad a \neq b \cr
   4 \Big(1-\frac{1}{N} \Big)  &  \qquad   a=b \qquad 
\end{array} \right.  \cr
{\rm dim}({V}_a  \overline {V}_{b} )&= 
\Delta \left[ \rme^{\im  \sqrt {4 \pi} \;  { (\bm \nu_a -  \bm \nu_b)  \cdot  \bm \phi}   } \right]=  (\bm \nu_a - \bm \nu_b)^2  =  2  \cr
{\rm dim}({V}_a {V}_{b} {V}_{c}  )&= 
\Delta \left[ \rme^{\im  \sqrt {4 \pi} \;   { (\bm \nu_a + \bm \nu_b +\bm \nu_c  )  \cdot  \bm \phi}   } \right] =  (\bm \nu_a + \bm \nu_b + \bm \nu_c)^2   \geq 
  3 \Big(1-\frac{3}{N} \Big)   \cr 
  {\rm dim}({V}_{a_1}  \ldots  {V}_{a_k}  )&= 
\Delta \left[ \rme^{\im  \sqrt {4 \pi} \;   { (\bm \nu_{a_1}  + \ldots  +\bm \nu_{a_k}  )  \cdot  \bm \phi}   } \right] =  (\bm \nu_{a_1} + \ldots  + \bm \nu_{a_k})^2   \geq 
  k \Big(1-\frac{k}{N} \Big)   \cr 
    {\rm dim}  \left( \prod_{a=1}^N {V}_a \right) & = \Delta \left[ \rme^{\im  \sqrt {4 \pi} \;   { (\bm \nu_{a_1}  + \ldots  +\bm \nu_{a_N}  )  \cdot  \bm \phi}   } \right]  = 0  \cr
  {\rm dim}\left(  \left( \prod_{a=1}^N {V}_a \right)^q   {V}_{b}  \right)&= 
\Delta \left[  \rme^{\im  \sqrt {4 \pi} \;  {   \bm \nu_b  \cdot  \bm \phi}   } \right]= \Big(1-\frac{1}{N} \Big)    \cr 
  {\rm dim}({V}_{a_1}  \ldots  {V}_{a_k}  )&=    {\rm dim}({V}_{a_{k+1}}  \ldots  {V}_{a_{N}}  )
\label{relevance-stuff}
  \end{align}
 
To be relevant in the renormalization group sense  \cite{Kosterlitz:1973xp},  an operator must have 
$\Delta <2$. These are vertex operators   ${V}_a $ with   $(\Delta \sim 1)$  and two-vertex  operators  
$[{V}_a   {V}_{b}]$ and $  [{V}_a  \overline {V}_{b}]$ for which   $(\Delta \sim 2)$.
We can ignore others $(\Delta \gtrsim  3)$ for IR physics, but still need to be careful  because of some subtleties, 
such as ${\rm dim}({V}_{a_1}   )=    {\rm dim}({V}_{a_1}  \ldots  {V}_{a_{N-1}} ) \equiv {\rm dim}({V}_{a_N}   )$.

  Finally, in the large-$N$ limit, only  level $k=1$ (and $k=N-1$)  operators are relevant.  $k=2$  and  $k=N-2$  are  marginal, 
 and  the rest is irrelevant.      This will allow us to determine the IR physics of the theory sufficiently robustly.

\subsection{Non-abelian bosonization:  Mass deformed $SU(N)_1$ WZW  }
The abelian bosonization reveals crucial amount of information about dynamics,  but makes the non-abelian global symmetry of the theory non-obvious. 
On the other hand, if we use  non-Abelian bosonization, the matching of the global symmetries become manifest.  Let us describe this briefly. 

Non-abelian bosonization  maps $N$-flavors of free massless  Dirac fermions to $U(N)_1$ WZW model~\cite{Witten:1983ar, Polyakov:1983tt, Polyakov:1984et}. 
The correspondence between operators is 
\be
\psi_\rmL \overline{\psi}_\rmR\sim U,  
\ee
where $U$ is the $U(N)$-group valued scalar field.  
The  $0$-form symmetry of the massless Schwinger model and  $U(N)_1$ WZW models are
$
{SU(N)_\rmL\times SU(N)_\rmR} \over (\mathbb{Z}_N)_\rmV, 
$
acting on the corresponding operators by conjugation $( \cdot) \mapsto V_\rmL  (\cdot) V_\rmR^\dagger $. 
Turning on a mass term for fermion reduce the symmetry to the vectorlike subgroup,  
\begin{align}
 PSU(N) = {SU(N)_\rmV} / (\mathbb{Z}_N)_\rmV 
 \end{align}
 which is the global symmetry of three theories:  $\mathbb {CP}^{N-1}$ model, massive $N$-flavor  Schwinger model, and mass deformation of $SU(N)_1$ WZW model.

The bosonized action of the massive  Schwinger model is given by
\begin{align}
S&={1\over 2e^2}\int_{M_2}|\diff a|^2+{1\over 8\pi}\int_{M_2}\tr\left(|\diff U|^2\right)   -\frac{c}{2} m^2 (\tr U +  \tr U^{\dagger} )
\nonumber\\
&+{\im \over 12\pi}\int_{M_3}\tr\left((U^{\dagger}\diff U)^3\right)+{ 1\over 2\pi}\int_{M_2}\diff a \wedge  (\ln \det(U)  + \im \theta ),
\label{eq:general_bosonization}
\end{align}
The first two terms are kinetic terms, and the third term is mass deformation, the fourth term is 
 the level-$1$ Wess-Zumino term.  
 The last term can enforces the  ABJ anomaly $U(1)_\rmR \to (\mathbb{Z}_{N})_\rmR$ of the massless limit, and incorporates theta
 dependence away from massless point.

Integrating out the gauge field  as in the abelian bosonization discussion, 
we obtain 
\begin{align}
S&= {1\over 8\pi}\int_{M_2}\tr\left(|\diff U|^2\right)   -\frac{c}{2} m^2 (\tr U +  \tr U^{\dagger} )
\nonumber\\
&+{\im \over 12\pi}\int_{M_3}\tr\left((U^{\dagger}\diff U)^3\right)+{ e^2 \over 2\pi}\int_{M_2}  (\ln \det(U)  + \im \theta )^2,
\label{eq:general_bosonization2}
\end{align}
This is the non-abelian version of \eqref{coleman0} and now, the full $PSU(N)$ symmetry is manifest. We can further (locally)  integrate out  mode associated with $\det(U) $ to obtain counterpart of \eqref{Coleman1}. 

According to \eqref{relevance-stuff}, there are very few  relevant or  marginally relevant deformations of level-1 $SU(N)$ WZW.  
 These are  $ \tr U $ with $\Delta \sim 1$ and $ \tr (U^2) $ with $\Delta \sim 2$  in large-$N$ limit, and,
$| \tr U |^2,  (\tr U)^2$  for which $\Delta \sim 2$. In particular, double-trace deformations of the form $| \tr U^n |^2,  (\tr U^n)^2$ for which $\Delta \sim 2n^2$,  are irrelevant in general for $n \geq 2$.

\subsection{Secrets of Coleman's 
 formula: Relation of RG with all orders semi-classics} 
 
In our construction, \eqref{Coleman2} arises as low energy description of the $\mathbb {CP}^{N-1}$ model. For that purpose, we can identify parameters of fermionized low energy limit of $\mathbb {CP}^{N-1}$  as 
$m \sim \Lambda  \sim \rme^{-S_I/N} $ and  $\mu^2 \sim  N e^2  \sim \Lambda^2$ and  hence, mass gap is given by
\begin{align}
 m' (\theta)  = \Lambda \max_k  \left(\cos \frac{\theta + 2 \pi k }{N}  \right)^{\frac{N}{N+1}} 
 \label{Coleman3}
\end{align}
We would like to discuss  rich physics associated with  this result, which will teach us new things concerning extension of semi-classical ideas to $\R^2$. We believe all the major lessons extend  to gauge theory on $\R^4$ and $\R^3 \times S^1$ and has non-trivial implications.

\begin{itemize} 
 
\item Conceptually, the most interesting aspect of  Coleman's result \cite{Coleman:1976uz} is following: 
Naively speaking, if we obtain  \eqref{Coleman1} from semi-classical gas of vortices in 
$S[\bm \sigma]$, we are performing a {\bf  first order} in semi-classic analysis. The Lagrangian has a  
$\cos  \left( \frac{\theta  }{N} \right)   \cos\left(   \sqrt {4 \pi} \;  { \bm \nu_a   \cdot  \bm \phi}     \right)$ factor in it which happily arise from  the leading order fractional  instantons, ${V}_a$. 
However, in \eqref{Coleman2}, we obtain mass gap    $\left(\cos \frac{\theta + 2 \pi k }{N}  \right)^{\frac{N}{N+1}}$  which is  {\bf impossible to obtain at any  finite order} in semi-classical expansion.  It requires infinite order in semi-classics. What happened in between?

\item What Coleman did \cite{Coleman:1976uz}, as he also states, is  renormalization group for topological defects \cite{Kosterlitz:1974sm, Kosterlitz:1973xp}. But renormalization group is smart, even if  
we just write down  few operators that are permitted by symmetries,  it will induce all the other operators that are permitted by symmetries. In \eqref{Coleman3}, it is actually inducing {\bf infinitely many} relevant operators and resumming over them!

\item What are these infinitely many relevant  operators and why did  we forget  about them at our starting point?  
The operator $\left(\cos \frac{\theta + 2 \pi k }{N}  \right)^{\frac{N}{N+1}} \cos \left( \sqrt {4 \pi}   {\bm \nu_a \cdot  \bm \phi}  \right)  $ 
is actually telling us that, 
apart from the  $\rme^{\im    \sqrt {4 \pi}   {\bm \nu_a  \cdot  \bm \phi}   +    i \frac{\theta}{N}  }$, there are infinitely many other operators of the form   $  \rme^{\im   \sqrt {4 \pi}   {\bm \nu_a  \cdot  \bm \phi}   + i \theta ( q + \frac{1}{N}) }$, the scaling dimension (or Coulomb charges)  of which  are  the same,   but topological charges are distinct, associated with 
 topological defects with winding number  $W= \frac{1}{N} (1+ N q),  \; q= 0,1,2, \ldots $   in the semi-classical language.\footnote{For gauge theory on $\R^3 \times S^1$, each monopole instanton has an infinite KK tower associated with it. If we wish to go beyond semi-classics, the above discussion instructs us that we should some over the whole tower.}

\end{itemize}

Let us consider the  $N=2$ case, where the effects is most pronounced.  The mass gap as a function of $\theta$ angle in this case is given by: \footnote{In the vicinity of $\theta=\pi$, our construction implies that the mass gap in the $\mathbb {CP}^1$ model  vanishes in a  particular way: 
\begin{align}
 m_{\rm gap} (\theta) =  \Lambda \left|  \frac{\theta - \pi }{2} \right|^{\frac{2}{3}}, \qquad  \theta \rightarrow \pi
\end{align}
If one can overcome the sign problem due to $\theta$ angle, it would be nice to check this prediction of our formalism.}
\begin{align}
 m_{\rm gap} (\theta)  =  \Lambda \left| \cos \frac{\theta}{2} \right|^{\frac{2}{3}}
 \label{mgap-CP}
\end{align}
 Let us  express    $m_{\rm gap}(\theta)  $ as a  Fourier series:
\begin{align}
 m_{\rm gap} (\theta) \Lambda^{-1}  =  \sum_{q=0}^{\infty} a_q \cos \left( (q + \half) \theta \right)
 \label{gap-corrected-CP1}
\end{align}
where the Fourier coefficients are given by:
\begin{align}
 a_q =   \frac{1}{\pi} \int_{-\pi}^{\pi} \diff \theta   \;  m_{\rm gap} (\theta) \Lambda^{-1} \;   \cos \left( (q + \half) \theta \right) 
= -\frac{2^{\frac{4}{3}}    \pi }{\sqrt{3} \Gamma \left(-\frac{2}{3}\right) \Gamma
   \left(\frac{5}{6}-q\right) \Gamma \left(q+\frac{11}{6}\right)}  \qquad \qquad
\label{gap-corrected-3}
\end{align}
Numerically, we find
\begin{align}
\{a_1, a_2, a_3, a_4, a_5, \ldots  \} =\{ 1.07119,   -0.09738,   0.04009,   -0.02266,   0.01484, \ldots  \}
\end{align} 
which suggest that even in the case where the leading order semi-classics is supposed to be  at its worst $(N=2)$, the numerical error it makes is quite small. In large-$N$,  the leading order semi-classics  capture the whole result \eqref{Coleman3}. 
 
Our intermediate level Lagrangian  \eqref{Coleman3}   is   based on leading vertex operators: 
\begin{align}
{V}_1,  \; {V}_2,  \;  \overline {V}_1, \;  \overline {V}_2,  
\end{align}
But renormalization group  generates all other (relevant and irrelevant) operators permitted by symmetries. Therefore, we ended up 
having a theory which describes proliferation of 
\begin{align}
{V}_1,  \; {V}_2,  \;  \overline {V}_1, \;  \overline {V}_2,   \;   [({V}_1 {V}_2) {V}_1],    \; 
 [({V}_1 {V}_2) {V}_2],   \;   [({V}_1 {V}_2)  \overline{V}_1],     \;   [({V}_1 {V}_2)  \overline{V}_2], \ldots    
\end{align}
topological defects with Coulomb charge $\pm1$ (which determines the relevance), but arbitrary half integer topological charge, 
$W \in \half  + \Z$, which tells us that these  are sourced by topologically distinct defects:
\begin{align}
 [({V}_1 {V}_2)^q {V}_a] \mapsto  \rme^{\im \sqrt {4 \pi}  \widetilde \phi_a}  \rme^{\im \frac{1}{2} ( q + \half) \theta }
\end{align}

  There are infinitely many correlated defects,  which  are relevant in the RG sense.  It is natural  that these defects 
generate the non-analytic structure that we see in  Coleman's theta angle dependence \eqref{Coleman2}.  These contribution arise 
from the different columns  in the resurgence triangle, see  Fig.\ref{fig:saddles} or Ref.\cite{Dunne:2012ae} associated with 
$W \in \half  + \Z$. 

\subsection{$\mathbb {CP}^{1}$ at $\theta=\pi$}
The  semi-classical approximation must be reliable provided 
$\rme^{-\frac{2 \pi}{g^2 (\tilde \ell)}}= \Lambda \tilde \ell \lesssim 1$
where $\tilde \ell$ is the characteristic distance between two  fractional vortex instanton position $(\tilde \ell \sim   |a^1_i- a^1_j|) $.
   So, close fractional vortex  instanton positions  in Euclidean set-up benefit from asymptotic freedom, just like close quarks do.    In principle, we should not use this formula for $\tilde \ell  \gtrsim \Lambda^{-1}$.

   The good news is, the correlation length of the system  (at $\theta=0$) is given by  
$ \Lambda^{-1}$.  After summation over fractional vortex instantons and anti-instantons, the would be long distance divergences are cut-off by the Debye length, and the mass gap can be obtained self-consistently.


Now, this may receive an objection as we change $\theta$. Is it possible that \eqref{gap-corrected-CP1} is  quantitatively  and qualitatively correct for any $\theta$ even when correlation length diverges at $\theta=\pi$?  How  seriously should we take the vanishing of the gap at $\theta=\pi$? 

We encounter  exactly the same situation  in the leading order semi-classics of the deformed Yang-Mills on small $\R^3 \times S^1$, but there, the  gaplessness is lifted at second order (see  \cite{Unsal:2012zj} for details).   Here, we will reach  to a different  conclusion due to renormalization group construction applied to defect theory \eqref{Coleman2}.   
The crucial difference of  Coulomb gas in 3d and 2d is that in the former Coulomb gas is always in the plasma phase (with only one known exception  \cite{Cherman:2017dwt}) 
while in 2d Coulomb gas, the  system has a  Kosterlitz-Thouless transition, so both gaplessness and gap are possibilities. 
See \cite{fradkin_2013}, page 320.
  
This problem is particularly important for  anti-ferromagnetic  spin chains in 1-dimensional spatial lattice. 
These systems at low energies are described by $2d$ $\mathbb {CP}^{1}$ non-linear sigma model. Integer spin is described 
by a sigma model without theta term and half integer spin is described by  $\mathbb {CP}^{1}$ with $\theta=\pi$,  $\theta= 2 \pi S $.
Haldane's conjecture \cite{Haldane:1983ru} states  that integer and half-integer spin chains falls into distinct universality classes, and 
 half-integer spin systems are gapless ($\theta=\pi$), while the integer spin systems are gapped  
($\theta=0$). See also \cite{Affleck:1987ch}.  

Since leading order vertex operators ${V}_a$ produce zero mass gap at $\theta=\pi $, we would like to understand the role of the second order effect of the form: 
\begin{align}
&[  {V}_1  \overline  {V}_2]  + [  {V}_2  \overline  {V}_1] + [  {V}_1  \overline  {V}_1]_{\pm} + [  {V}_2  \overline  {V}_2]_{\pm}     
\end{align}
As we know from resurgence theory, $ [  {V}_a  \overline  {V}_a]_{\pm}$  type effects are are two-fold ambiguous, and the  ambiguous part  cancels the renormalon ambiguity in perturbation theory. Their non-ambiguous part leads to something extremely interesting. It leads to an anomalous dimension for various important operators. 

We would like to determine the amplitude of these events, importantly, their signs compared to vertex operators   ${V}_a$. 
Consider $\R^2$ as  
$\R^+ \times S^1$. In the presence of $\Omega_F$ background, we have  ${V}_a$ and $\bar {V}_a$ events. 
To write the correlated amplitude, we write an integral over the quasi-zero mode Lefschetz thimble. 
This gives us: 
\begin{align}
[  {V}_1  \overline  {V}_2]  &= \Big( -\gamma- \log \Big( \frac{A}{g} \Big)  \Big)  [  {V}_1]  [\overline {V}_2]  \cr
[  {V}_1  \overline  {V}_1]_{\pm}  &=  \Big( -\gamma- \log\Big( \frac{A}{g}\Big)   \pm \im  \pi  \Big)  [  {V}_1]  [\overline {V}_1] 
\label{ambiguity-2}
\end{align}
Again, by resurgent structure \cite{Dunne:2012ae,   Behtash:2018voa}, ambiguous imaginary parts cancel between Borel resummed perturbation theory and non-perturbative effects. Now, we can write combinations of non-perturbative effects as
\begin{align}
    2 \Upsilon & \left[  (\bar \psi_\rmL \psi_\rmR)^2  +   (\bar \psi_\rmR \psi_\rmL)^2  + 2    (\bar \psi_\rmL \psi_\rmR) (\bar \psi_\rmR \psi_\rmL) \right],   \qquad 
    {\rm 
    where} 
     \qquad  \cr 
    2 \Upsilon  &= K^2  \left( -\gamma- \log\left( \frac{A}{g}\right)  \right)    \rme^{- S_I}       < 0
\end{align}
The crucial point is that there is a sign difference between the leading order fractional  instanton effects and second order correlated fractional  instanton effects.\footnote{This is similar to double-well quantum mechanics. $E_0$ is shifted down by $e^{-S_I}$  and $E_1$ is shifted up by instanton effects, but $E_0 + E_1$ remains the same at leading order in semi-classics. However, at second order, $[I \bar I] \sim e^{-2 S_I}  $ correlated events shifts    both  $E_0$  and  $E_1$ up  by the same amount.   
 In particle on a circle with unique minimum at $\theta=0$,  instantons shift the ground state energy down, but
 both  $[I \bar I]$ and  $[I I]$ shift ground state energy up.  So, there is a relative sign between instanton effects and correlated instanton effect in purely bosonic theories. 
 In general, however, 
  the amplitude of the two-events  is a subtle  and deep issue.  If we were to consider $\mathbb {CP}^1$ with  $\mathsf n$ Dirac  fermions ($\mathsf n=1$ is supersymmetric theory),  it is possible to show that 
$ {\rm Fugacity}[  {V}_1  \overline  {V}_1]_{\pm} =   {\rm Fugacity} [  {V}_1  \overline  {V}_2]  \rme^{ \pm \im \mathsf n  \pi } $
 where  $\rme^{\im \mathsf n  \pi }$  
  is called hidden topological angle \cite{Behtash:2015kna} that arise from Lefschetz thimble integration. This extra phase is super-important, and is  responsible for the vanishing of the ground state energy in supersymmetric theories,  see    \cite{Behtash:2015kna} for details.  }
Therefore, we can write the combinations of bion contributions as: 
$2  \Upsilon \left[  (\bar \psi_\rmL \psi_\rmR)^2  +   (\bar \psi_\rmR \psi_\rmL)^2 - \frac{1}{2}   j_{\mu}^2   \right] $
where  $ j_{\mu} = \bar \psi  \gamma_{\mu} \psi $, and $\Upsilon < 0$.
After understanding  these non-trivial facts that arise from resurgence theory, we can proceed with  abelian bosonization   similar to 
  \eqref{Coleman2}. Following almost identical steps with Fradkin's book \cite{fradkin_2013}, 
  we obtain 
 \begin{align}
 L &=   \frac{1}{2} (\diff  \widetilde \phi)^2  +  \frac{\Upsilon }{\pi} (\diff  \widetilde \phi)^2 - (m')^2(\theta)  \cos (\sqrt {2\pi}  \widetilde \phi)-
   4 \Upsilon  \cos ( 2 \sqrt {2\pi}  \widetilde \phi) 
  \end{align}
 where the first part of the kinetic  term is the image of fermion kinetic term, the second part is the image of $j_{\mu}^2$, which is sourced  by  neutral bions $ [  {V}_1  \overline  {V}_1] + [  {V}_2  \overline  {V}_2]  $, and interaction term arise from charged bions $[  {V}_1  \overline  {V}_2]  + [  {V}_2  \overline  {V}_1]$.   Note that neutral bion contribution produce a non-linearity  which effects the dimensions of the charged bion operator, and this effect will be important.

 Setting $\theta=\pi$, and 
 defining     $\varphi  = \left({1 + \frac{2 \Upsilon }{\pi}} \right)^{1/2} \widetilde \phi$ turns the kinetic term to canonical form.  
 Using   $\beta^2 = \frac{2\pi}{1 + \frac{2 \Upsilon }{\pi}}$,   we can write the action as
  \begin{align}
 L =  \frac{1}{2}   (\diff  \varphi)^2 -  4  {\Upsilon}\cos ( 2 \beta \varphi) 
  \end{align}
  Since 
  \begin{align}
  \Delta \Big[\cos ( 2   \beta \varphi) \Big] =  \frac{2}{1 + \frac{2 \Upsilon }{\pi}}
  \label{anomalous}
  \end{align}
  the charged bion  (doubly charged vertex  operator),  which was   exactly marginal at leading order in semi-classical expansion where $\Upsilon=0$, 
is driven towards irrelevance at second order in semi-classics  where    $\Upsilon <0$. As a result, the theory at  $\theta=\pi$ is gapless. 

\subsection{Why  is this working? (Improving Polyakov's  optimism with RG) }
A source of quantitative error in the formalism is following. 
The instanton, as we know, has a size moduli, $\rho$.  In the classical theory, which is scale invariant, to determine 
the single instanton contribution,  
we end up with the integral 
\begin{align}
Z_1 \sim V_2  \int \frac{d \rho}{\rho^3} \rme^{-\frac{4 \pi}{g^2(\rho)}}  \sim 
V_2  \int \frac{d \rho}{\rho^3} \;  (\Lambda \rho)^N =  V_2 \Lambda^N   \int d \rho \rho^{N-3}
 \label{naive-size}
\end{align}
which is IR divergent for all $N$. Furthermore, one should  not use the  one-loop result for the strong scale for $\rho>  \rho_{\rm c} \sim \Lambda^{-1} $ since $g^2(\rho)$ becomes large, but  at least, at large-$N$ limit, the one-loop result 
for $\beta$ function becomes  exact. 

\begin{figure}[t]
\vspace{-1cm}
\begin{center}
\includegraphics[width = 0.7\textwidth]{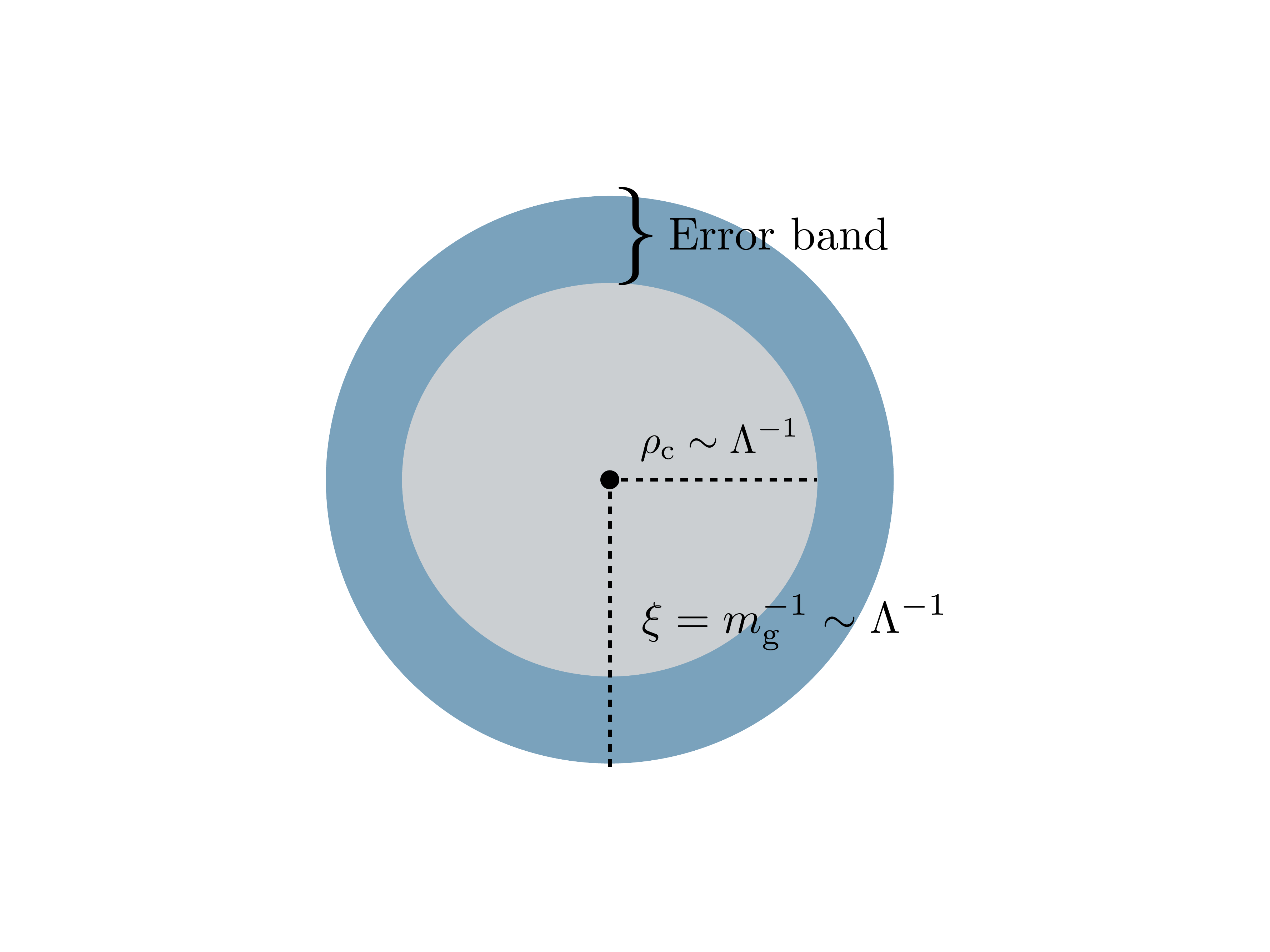}
\vspace{-1.5cm}
\caption{This diagram depicts a source of quantitative error.  We use one-loop result for strong scale, which should be reliable for 
$\tilde l \lesssim \rho_c \sim \Lambda^{-1}$ and unreliable  for $\tilde l \gtrsim \rho_c$, where it leads to an IR divergence.   At the same time,  the proliferation of fractional vortex instantons with fugacity  $\rme^{-S_I/N}$ generates a finite correlation length $\xi \sim \Lambda^{-1}$. Because of the Debye screening, this cuts-off the interaction  between  fractional vortex instantons at distances larger than  $\xi \sim \rho_c$, essentially removing the bad zone from integration. Depending on the numerical values of   $\xi \sim \rho_c$, there may be a band which induce quantitative error.  
We then use RG arguments to find relevant operators at $\theta=0$. Since RG is independent of $\theta$, we can  extrapolate our effective theory to arbitrary $\theta$.  In  $\mathbb {CP}^{1}$, this gives us a  gapless  theory at $\theta=\pi$.  Despite the fact that ``naive" error band becomes infinite,   this result is reliable because the determiner of error is $\theta=0$.  Therefore,  both gap and gaplessness are robust qualitative predictions of the formalism.  
 }
\label{fig:error}
\end{center}
\end{figure}

If we go ahead and do a summation over all fractional vortex instanton  effects,   and construct a grand canonical ensemble of 
topological defect configurations,  we learn that the system induce  a finite correlation  length $\xi=m^{-1} \sim  \Lambda^{-1} $.   
This immediately tells us that one  should  not extrapolate \eqref{naive-size} naively. Instead, we  interpret the  Debye length as a cut-off,   replacing  
\begin{align}
   \int d \rho \rho^{N-3}  \mapsto  \int d \rho \rho^{N-3} \rme^{-m \rho}  .
   \label{cut-off}
   \end{align}
This removes  the  IR divergent bad domain $\rho > \xi$ from integration, see Fig.\ref{fig:error}.   Since $\rho_c \sim \xi$,   it is impossible to do qualitative error  in this way, and  the infrared  theory  for $\mathbb {CP}^{N-1}$ which captures low lying modes   is  described by the  mass deformed $SU(N)_1$ WZW \eqref{eq:general_bosonization2}, where mass deformation is governed by $m\sim \mu \rme^{-S_I/N} = \Lambda$.

The logic   of the previous  paragraph  follows Polyakov,  \cite{Polyakov:1987ez}, Section 6.1.  However, in the important case of $N=2$, the  rationale of Polyakov   breaks down around   $\theta\sim \pi$ where correlation length diverges as  $\xi(\theta) = \Lambda^{-1} \frac{1}{|\cos (\theta/2)|^{2/3} } $. 

Should we not trust this  result as the  bad   domain 
$\rho \in [\Lambda^{-1}, \infty) $   resurfaces in the integration \eqref{cut-off}? Is this prediction unreliable? 
  Our claim is otherwise. We should trust it as much as we trust the $\theta=0$ results.  We will prove this by improving Polyakov's logic with renormalization group.

%
 
  The  $\theta=0$ and $\theta \neq 0$ are not independent theories, we are just adding a topological theta angle to the theory. 
 If we did  everything right at $\theta=0$ and constructed the correct  long distance theory based on  renormalization group of topological defects, why  should  it  fail by turning on a topological angle   at  $\theta=\pi$?   

Indeed,  at $\theta=0$, we showed that the vertex operators $\rme^{\im  l  \beta \varphi}$ have dimensions 
 \begin{align}
\Delta_l \equiv \Delta[\rme^{\im  l  \beta \varphi}]  =\frac{ l^2 \beta^2}{4\pi}  
=  \frac{1}{2 (1 + \frac{2 \Upsilon}{\pi})} l^2, 
 \end{align}
 Therefore, unit charge vertex operator has dimension $ \sim \half$ and is highly relevant, 
 $l\geq 3   $  are  irrelevant, and   $l =2  $ vertex operator is marginally irrelevant. 
This construction is certainly valid at $\theta = 0$ where our EFT is self-consistent.  
 The scaling dimensions  are  {\it independent} of $\theta$ angle.
If we construct  effective theory correctly at $\theta=0$, turning on $\theta$ will not change that fact, and despite the fact that 
 $\xi (\theta) \gg  \rho_c$, we must  obtain the correct results at arbitrary $\theta$.   

 The theory does not become gapless at $\theta=\pi$ because $\rme^{\im   \beta \varphi}$ becomes irrelevant, to the contrary, its is extremely relevant.   The theory becomes gapless because of the destructive topological interference between $ {V}_1  =  \rme^{-\frac{S_I}{2}}    \rme^{\im    \beta \varphi +  \im \frac{\theta}{2}}   $ and $ \bar {V}_2 = 
  \rme^{-\frac{S_I}{2}}    \rme^{\im   \beta \varphi -  \im \frac{\theta}{2}}$, and the fact that the  $l= 2  $ vertex operator is marginally irrelevant.

 
 For general $\mathbb {CP}^{N-1}, N\geq 3$ theories,  we are even safer in our construction. 
  There are  only three  types of relevant operators in  the renormalization group of vertex opetators,   (see \eqref{relevance-stuff} for a full list)
  \begin{align}
   {\rm dim}({V}_a ) = \Big(1-\frac{1}{N} \Big)  \cr
{\rm dim}({V}_a {V}_{b} )= 
    2 \Big(1-\frac{2}{N} \Big), 
\qquad     {\rm dim}({V}_a  \overline {V}_{b} )&= 2    \qquad a \neq b, 
  \end{align} 
  and all the rest is  irrelevant. Since  ${\rm dim} \sim 2$  for   $[{V}_a  \overline {V}_{b} ]$ and $ [{V}_a {V}_{b} ]$ operators, they  do not alter the conclusions of  ${\rm dim} \sim 1$   ${V}_a $ operators, 
   we can ignore their effects accepting quantitative error.  Since we are after the more modest goal of determining the existence of gap, we hope this is acceptable. 
  Therefore,  our demonstration of mass gap for $N \geq 3 $ theories on $\R^2$ is on a robust  footing.    
  
  \vspace{0.5cm}
  \noindent 
{\bf Mass deformed  WZW interpretation:}
  Of course, the above statements translate to non-abelian bosonization language  nicely \eqref{eq:general_bosonization2}. 
The  massive $U$ field  in the mass deformed  $SU(N)_1$ WZW   describe the massive adjoint  $\bar z_a (x) z^b(x) $ field and singlet (slightly heavier)     $\bar z_a (x) \rme^{\im \int_{x}^y a } z^a(y) $ field in the spectrum,  $(N^2 -1) \oplus 1$.  In the large-$N$ limit,   $ \tr U $  mass deformation has dimension   $\Delta \sim 1$.  Double trace operators   $| \tr U |^2,  (\tr U)^2$  have 
  $\Delta \sim 2$. 
 These  describe  the low lying spectrum of the  $\mathbb {CP}^{N-1}, N\geq 3$ model, an adjoint and a singlet whose masses are at  the strong scale of the theory.

 For the 
 $N=2$ case,  we need to pay more attention. Let us write
  \begin{align} 
   U_{ab} \sim 
  \rme^{\im  \sqrt {\frac{4 \pi}{N} }    \widetilde \phi }   g_{ab}
  \end{align}
  in \eqref{eq:general_bosonization2}. 
In this case,  at $\theta=0$, we have a massive triplet and slightly heavier singlet. In fact, as $\theta$ goes from $0$ to $\pi$, the gap between lowest lying triplets and singlet widens, and it becomes fully justified to integrate out the heavy  $ \widetilde \phi$ mode. In this case, 
in the IR,  $\Delta [\tr g]=\half $, but this operator disappears from deformed WZW action  due to topological interference induced by $\theta$ term.  A possible mass term may be generated by the double trace operator,  $|\tr g|^2$. But this operator  has $ \Delta[ |\tr g|^2]  >2 $ \eqref{anomalous}  and is marginally irrelevant. Therefore,  at $\theta=\pi$, we obtain  a {\it gapless triplet} and a massive singlet. (Same conclusion for $N=2$ theory are also  reached by Affleck in Ref.\cite{Affleck:1988wz} by different means.)

\section{What are we summing over in semiclassics in dYM theory on  $\mathbb{R}^3\times S^1$?}
Deformed Yang-Mills  theory is center-symmetry preserving  double-trace deformation of the Yang-Mills action on small  $\mathbb{R}^3\times S^1$ 
\cite{Unsal:2008ch, Shifman:2008ja}.
In the sense of gauge invariant order parameters, it is  continuously  connected to pure YM on $\R^4$.  The deformation 
 allows a  {\it calculable (semi-classical)}   regime in which one can study non-perturbative properties such as confinement, $\theta$ angle dependence, mass gap generation.  See \cite{Unsal:2007jx,Unsal:2007vu,Unsal:2008ch, Poppitz:2011wy, Poppitz:2012sw,Poppitz:2012nz, Argyres:2012vv,Argyres:2012ka, Anber:2011gn, Anber:2015wha,  Misumi:2014raa, 
   Cherman:2014ofa, Poppitz:2008hr,Aitken:2017ayq, Thomas:2011ee, Zhitnitsky:2013hs,   Anber:2017rch, Itou:2019gtg, Anber:2013sga, Kitano:2017jng, Kan:2019rsz}.\footnote{Modeling of Yang-Mills vacuum  
     outside its region of validity  of semi-classics  is  described in \cite{Liu:2015jsa, Liu:2015ufa,Larsen:2018crg}.   
     The construction in this paper  provides some retrospective  rationale  for  this study.  In particular, the coupling of YM to $\Z_N$ TQFT  proves that fractional topological charge $W= 1/N$ configurations is there even at strong coupling of pure YM theory.}
     There is by now compelling evidence that deformed YM, QCD(F) with $\Omega_F$ twist, , QCD(adj) and a number of other  theories  on  small  $\mathbb{R}^3\times S^1_L$ regime   are continuously connected  to the corresponding theories on $\mathbb{R}^4$. Recently,  \cite{Bonati:2018rfg, Bonati:2019kmf} showed  by numerical lattice simulations that topological susceptibility at small $\R^3 \times S^1_L$ dYM theory is identical to the one of pure YM theory on large $T^4 \sim \R^4$. 
    It is important to recall that   in these theories, the $S^1_L$ circle  never has a thermal interpretation in the standard sense  \cite{Unsal:2007jx}. 
  
It should be noted that there also had been   important body of work on $T^3 \times \R $ by using 't Hooft twisted boundary conditions, pioneered by 
Gonzalez-Arroyo,  Garcia-Perez and van Baal \cite{ GarciaPerez:1993jw, GonzalezArroyo:1995zy,  
GonzalezArroyo:1995ex, GonzalezArroyo:1996jp, Zhitnitsky:1991eg, GarciaPerez:1992fj, Bruckmann:2003ag, 
GarciaPerez:1999bc, GarciaPerez:2007ne, vanBaal:2000zc, Perez:2010jx, GarciaPerez:1993ab, Lebedev:1988wd, 
GarciaPerez:1999hs}. (See also \cite{Witten:1982df, Witten:2000nv,  Cohen:1983fd} for other uses of twisted boundary conditions.)
 These works,  in the small   $T^3$ regime, aimed to construct a semi-classical description of the vacuum and confinement mechanism 
  based on 't Hooft's fractional topological charge  $W=1/N$ and action 
$S=S_I/N$ configurations.   As we emphasized multiple times, 't Hooft's original solutions on $T^4$ are constant (space-time independent) and to a certain degree, (apart from its high  value  in showing that  action  $S=S_I/N$ configuration do exist),  are fairly uninteresting. 
The biggest stumbling block in this program  had been   the lack of exact analytic  non-trivial (spacetime dependent) solution with $S=S_I/N$. However, these 
authors were able to  demonstrate the existence of non-trivial  BPS $S=S_I/N$  (time dependent) configurations  by using lattice techniques, leaving no doubt for their existence, though these works are not sufficiently   appreciated. 
 Some  semi-classical dynamics due to  fractional instantons is   understood.  (See \cite{GonzalezArroyo:1995zy} and  also \cite{vanBaal:2000zc} for a review.)
   It should be noted that these works are prior to the  explicit analytical solutions of $W=1$ non-trivial holonomy calorons and their fractionization to $W=1/N$ constituents. 
And possible connections between fractional instantons and monopole instantons are partly discussed in \cite{GarciaPerez:1999hs,GarciaPerez:1999bc}, but it is not concluded that monopole-instantons live in $PSU(N)$ bundle, a missing link between the two that we filled up  in the present work. 

Our analysis of Yang-Mills theory on $M_4$ can be viewed in some sense putting the analysis of gauge theories on  $T^3 \times \R$ and 
$\R^3 \times S^1$ in complete agreement. In particular, we are able to do so by proving that the monopole-instanton  are actually exact solutions  in the $PSU(N)$ bundle.   This fulfills a   (numerically substantiated)  key assumption in the analysis of Gonzalez-Arroyo et. al.  Then, we show that by coupling a TQFT to gauge theory, that  the configurations with action $1/N $ are also present in the strong coupling domain.

Returning back to $\R^3 \times S^1$ studies,   despite the fact that  the long distance theory of the theories on $\R^3 \times S^1$  
 reduce to 3d theory, the microscopic theory is 4d.   The deformed YM and QCD theories  on  $\R^3 \times S^1_L$ are closer in all respect to their 
 decompactifcation limit on $\R^4$ than the theories on $\R^3$ limit.\footnote{In fact, even at  small-$L$, where $L \Lambda   \sim O(N^0)$, taking $N \rightarrow \infty$, these theories are equivalent to the theories on $\R^4$.  In certain sense, effective circle size is $L_{\rm eff}=NL$ and at $N=\infty$,  for the neutral sector observables (this is the sector singlet under zero-form part of center symmetry), it is fair to interpret this as if there is no compactification.  This is called large-$N$ volume independence. 
  }
 For example, 
 these theories have  theta angle parameters,  multi-branch structure, mass gap determined by 4d strong scale,  global ABJ anomalies.  
Very often,  they also have the same global and mixed anomalies as the 4d theories \cite{Tanizaki:2017qhf, Cherman:2017dwt,Sulejmanpasic:2020zfs}.

Below, we would like to determine  in detail  the configurations that contribute to partition function of deformed Yang-Mills in the semi-classical domain 
 in the same spirit as our simple quantum mechanical  $T_N$ example.  But  before that, we would like to provide a quick review of the deformed theory and set the notation. 
 
 \vspace{0.5cm}
\noindent
{\bf Quick overview of deformed Yang-Mills: Remembrance of things past   }
\begin{itemize} 
\item In pure Yang-Mills theory, center symmetry is broken at small circle size (high-temperature) on $\R^3 \times S^1$ \cite{Gross:1980br}. 
Center-symmetry can be  stabilized at small-$L$ by the  addition of double-trace operator  $ \Delta {\cal L} = \sum_{k=1}^{ \lfloor  {N \over 2}  \rfloor} 
|\tr U_3^k|^2$ where $U_3$ denotes Polyakov loop along compact $x_3 \sim x_3+ L$ circle. 
\footnote{ We refer to non-compact directions as $x_1, x_2, x_4$. This convention is more useful when we describe the same set-up with the insertion of 't Hooft flux.} 
 Despite the fact that this term looks non-local, it can arise from local action, e.g. Yang-Mills + (massless or massive) adjoint fermions at small $S^1$ 
 with periodic boundary conditions   induce the center stabilization dynamically.  See \cite{Unsal:2007jx, Unsal:2008ch, Shifman:2008ja} for details. 

\item The minimum of the holonomy potential is given by $\rme^{\im \bm{\phi}_\star}= (1,\omega,\omega^2,\ldots, \omega^{N-1})$ up to Weyl permutations, where $(\rme^{\im \bm{\phi}_\star})_i= (U_3)_{ii} $.
This forces the dynamics to abelianize at small-$L$, and at long distances, the theory is described by maximal abelian sub-group  $SU(N)\to U(1)^{N-1}$, just in terms of photons. Off-diagonal gluons gets  masses  $\ge {2\pi/NL}$.  

\item There are multiple useful basis to describe the  IR physics.  The diagonal components of gauge fields and compact scalar can be written as 
$\bm{a}=\sum_{i=1}^{N-1}  a_i \bm{\alpha}_i$  where $a_i$ are canonically normalized gauge fields and 
$\bm{\phi}=\sum_{i=1}^{N-1}\phi_i \bm{\alpha}_i=(\phi_1,\phi_2-\phi_1,\ldots, -\phi_{N-1})$. 
 The periodicity of $\bm{\phi}$ is governed by the root lattice, 
$\bm{\phi} \sim \bm{\phi} + 2 \pi \bm{\alpha}_i,   \bm{\alpha}_i  \in \Gamma_{r}$, 
\begin{align}
\bm \phi \in \frac{\R^{N-1}}{2 \pi \Gamma_{r}}
\label{phi-cell}
 \end{align}
hence,  $\phi_i$ are periodic variable with period $2 \pi$.  $\bm{\phi}_\star$ is the center of Weyl chamber where dynamics completely abelianize. 

\item We can perform abelian duality  \cite{Polyakov:1987ez} and express the free 3d action as 
\begin{align}
S_{\mathrm{eff}}&=\int \left({1\over 2g^2 L}\Bigl|\diff \bm{\phi} \Bigr|^2+{g^2\over 8\pi^2 L}\Bigl|\diff \bm{\sigma}-{\theta\over 2\pi} \diff \bm{\phi} \Bigr|^2\right)  
\end{align}
In most  theories, we can also forget about $ \bm{\phi}$ as it is gapped by holonomy potential, and can be dropped 
in long distance effective field theory. 
The dual photon field  $\bm{\sigma}$   has  periodicity  determined by weight lattice,   $\bm{\sigma}\sim \bm{\sigma}+2\pi \bm{\mu}_i ,  \; 
 \bm{\mu}_i  \in \Gamma_w$. 
\begin{align}
\bm \sigma  \in \frac{\R^{N-1}}{2 \pi \Gamma_w}
\label{s-cell}
 \end{align}
 A useful basis for our purpose is $\bm{\sigma}=\sum_{i=1}^{N-1} \sigma_i \bm{\mu}_i$, hence 
 $\sigma_i   \sim  \sigma_i  + 2\pi$, and 
 $\sigma_i  $  are   $2\pi$-periodic scalars. 

\item There are    $N$ types of the fundamental monopole instantons, and  the monopole operators are given by  
\be
{\cal M}_a(x)=\rme^{-S_I/N} \rme^{\im \bm{\alpha}_a \cdot \bm{\sigma}(x)} \rme^{\im \theta/N}\quad (a=1,\ldots, N). 
\label{mon-op}
\ee 
$N-1$ of these monopoles are the regular ones associated with $SU(N)\to U(1)^{N-1}$ adjoint Higgsing. The monopole associated with the affine root   
$\bm{\alpha}_N= -  \sum_{a=1}^{N-1} \bm{\alpha}_a $ 
 \cite{Lee:1997vp, Lee:1998bb, Kraan:1998kp,Kraan:1998pm,Kraan:1998sn}   has  the same action with the rest of the simple roots due to compactness of the adjoint scalar or equivalently, due to the fact that  Polyakov loop acts as a group valued field at a center-symmetric background.  See also \cite{Collie:2009iz}.

 \item The sum over all monopole instantons  generates  an effective  potential.  The grand canonical ensemble of the monopole gives the partition function: 
 \begin{align}
 Z= \int \Diff \bm \sigma  \;  \exp  \Big[  -\Big(   \int {g^2\over 8\pi^2 L} |\diff \bm{\sigma} |^2 -
 2K  \rme^{-S_I/N}\sum_{a=1}^{N} \cos\left(\bm{\alpha}_a \cdot \bm{\sigma}+{\theta \over N}\right) \Big) \Big]
  \label{master}
 \end{align}
  which leads to a non-perturbative mass gap, finite string tension with N-ality $k \lesssim N/2$,  $N$-branched vacua,  and {\rm CP} breaking at $\theta=\pi$. 
  
  \item The  effective field theory description in 
 in terms  monopole-instanton and  bion  local fields  is based on the parametric scale separation between the dual photon and $W$-boson, and  is valid provided   $(\Lambda LN) \lesssim 1$. However, 
 EFTs  can be used to determine {\bf some}  observables,  but {\bf not  all} observables. For example, despite the fact that  EFT  \eqref{master} can produce mass gap,  string tension with $N$-ality $k \lesssim N/2$  \cite{Unsal:2007jx, Unsal:2008ch,Polyakov:1975rs}, exact value of chiral condensate in ${\cal N}=1$ SYM \cite{Davies:2000nw},   it cannot produce string tensions with  $N$-ality $k \gtrsim  N/2$ correctly.  Reproducing those require the inclusion of $W$-bosons, which are present in microscopic theory, but not in EFT.  We will dwell onto these subtleties a bit more precisely in the conclusion section.

\end{itemize}

 \subsection{Mini-space formalism: Global constraints in monopole-instanton sums}
 \label{mini-space}
A number of issues come to mind with this construction 
and has been raised over the years. 

\vspace{0.3cm}
\noindent
{\bf Configurations contributing to partition function:} 
 One is the appearance of $\theta/N$ in the monopole operator  \eqref{mon-op} and effective action \eqref{master} that arise from a monopole-instanton 
 with topological charge $Q= \frac{1}{N}$.    Yet, if we consider the $SU(N)$  theory on  orientable 4-manifolds,   it is well-known that the topological charge is quantized in integer units $W \in  \Z$.  Therefore, the configurations   that contribute to the partition function must have a theta dependence of the form $\rme^{i W \theta}$. How is this compatible with the appearance of  $\theta/N$ in the monopole operators and effective action?  

If we compactify the theory on small $\R^3 \times S^1_L $ down to quantum mechanics, $ T^2 \times S^1_\beta \times S^1_L$, then,  in the path integral representation of the partition function $Z(\beta) = \tr  \; \exp [{-\beta H_{T^2 \times S^1_L}}]$, we are supposed to sum over fields that obey periodic boundary conditions.   Yet, the presence of a monopole implies that 
the magnetic flux piercing $T^2$ $ \bm \Phi_{\rm mag} (\tau) = \int_{T^2}  \bm B $ changes by 
\begin{align}
\bm \Phi_{\rm mag} (\beta)-  \bm \Phi_{\rm mag} (0)=  \left( \int_{T^2}  {\bm B} \right) \Big|_{\tau=0}^{\tau=\beta}  =  \frac{2 \pi} {g}  \bm \alpha_a,  \qquad a =1, \ldots, N.
\end{align}
Such configurations are present in the theory, but just like our quantum mechanical example, they should not contribute to partition function,  which is a sum   over { \it periodic } paths.  But of course, despite the fact that these contribution are absent in the sum, most of the non-perturbative low energy phenomena  are sourced by them, again just like our QM example.

The image of the issue raised on $T^2 \times S^1_\beta \times S^1_L$ for the infinite volume theory can be simply described as follows. 
The magnetic flux at infinity on $\mathbb{R}^3\times S^1$ is a choice of boundary condition. If we choose the flux at infinity to be zero,  the net flux from all sources must vanish by Gauss's law, 
the only configuration that can contribute to the partition function must satisfy magnetic neutrality at infinity 
$ \int_{S^2_{\infty}} \bm B_{\rm total} = 0  $.  
%
%
%
%
%
%
%

Below, we show that  the master partition function \eqref{master} is aware of these issues and cleverly takes these constraints into account, as mentioned in passing  in \cite{Tanizaki:2019rbk}. 
We  prove in detail that magnetic charge neutrality comes naturally from 
 \eqref{master}. Furthermore, as a bonus,  once magnetic charge neutrality emerges, the topological charge of corresponding configurations become integer quantized.  (Opposite is not true, configurations with integer winding number need not be magnetically neutral.) We also note that the  points about global neutrality  are also emphasized in \cite{Diakonov:2007nv, Diakonov:2010qg}, though we disagree with the naming of monopole-instanton 
 as dyon, since it is not an eigenstate of electric charge operator.  

\noindent
{\bf Minispace formalism:}  The path integration over the fields  $ \int \Diff \bm \sigma \;  \rme^{-S}$ has a zero mode 
 part,   the integration over the space of constant (space-time) independent $\bm \sigma$. In this subspace, the measure reduce to an ordinary  integral over  the fundamental cell of  ${\bm \sigma}$  field  given in \eqref{s-cell}, 
 \begin{align}
 \int_{ \sigma \in \frac{\R^{N-1}}{2 \pi \Gamma_w} }   [d{\bm \sigma}]  \equiv  
\Big( \prod_{i=1}^{N-1}  \int_0^{2\pi} d \sigma_i   \Big) 
 \label{measure}
 \end{align}
The action  over the constant modes reduces to  $s= -2 V_3 K  \rme^{-S_I/N}\sum_{a=1}^{N} \cos\left(\bm{\alpha}_a\cdot \bm{\sigma}+{\theta \over N}\right)$. So, the zero mode integration becomes (set $ \xi= V_3 K  \rme^{-S_I/N}$)
\begin{align}
z(\theta) & = \int_{ \rm cell}  [d{\bm \sigma}]  \; \rme^{-s} \cr
&= 
\Big( \prod_{i=1}^{N-1}  \int_0^{2\pi} d \sigma_i   \Big) 
  \prod_{a=1}^{N}   \rme^{ \xi \rme^{ \im \left(\bm{\alpha}_a \cdot \bm{\sigma}+{\theta \over N}\right)} } \rme^{ \xi \rme^{ - \im \left(\bm{\alpha}_a\cdot \bm{\sigma}+{\theta \over N}\right)}}
\end{align}
We call $s(\bm \sigma) $ mini-space action and $z(\theta)$ mini-space partition function. 
Using the good basis  $\bm \sigma=  \sum_{i=1}^{N-1} \sigma_i {\bm \mu_i}$,  we have 
$\bm{\alpha}_i \cdot \bm{\sigma} = \sigma_i, \;  n=1, \ldots, N-1 $ and  $\bm{\alpha}_N\cdot \bm{\sigma} =  - (\sigma_1+ \ldots + \sigma_{N-1}) $, which allows us to easily perform the integration. 
\begin{align}
z(\theta)= 
\prod_{a=1}^{N} \left( \sum_{n_a=0}^\infty   \sum_{\bar n_a=0}^\infty   \right)    &  \frac{\delta_{n_1 - \bar n_1, n_N -\bar n_N} \ldots \delta_{n_{N-1}  - \bar n_{N-1} , n_N -\bar n_N}       }{n_1! \bar n_1 ! \ldots  n_N! \bar n_N! }  \cr
&  \xi^{n_1+ \ldots + n_N+ \bar n_1+ \ldots + \bar n_N} \rme^{i \frac{\theta}{N} (n_1+ \ldots + n_N -( \bar n_1+ \ldots + \bar n_N))} 
  \label{crude}
\end{align}
where $N-1$ constraints comes from the  integration over the $N-1$ independent $\sigma_i$ field.  The solution of the constraints is 
\begin{align}
n_1-\bar n_1=  n_2-\bar n_2=  \ldots = n_N-\bar n_N=  W 
\label{constraint-5}
\end{align}
where  $W \in \Z$. 
This enforces magnetic neutrality, i.e, magnetic flux at infinity is zero.  This constraint tells us that $n_a - \bar n_a$ is independent of $a$. Clearly, if $n_a - \bar n_a$ is zero, both magnetic neutrality and topological neutrality are guaranteed.  But what if 
$n_a - \bar n_a = W,   \; \forall a $, an $a$-independent  excess of monopoles over anti-monopoles? In that case, 
magnetic neutrality is guaranteed thanks to the fact that  $\sum_{a=1}^N \bm{\alpha}_a=0 $.  Furthermore, when this condition is satisfied, the   corresponding configuration has an integer topological charge given by $W \in \Z$!
Therefore, the  mini-space partition function can be written as 
\begin{align}
z(\theta)= \sum_{W \in \mathbb Z} \left( \sum_{\bar n_1=0}^{\infty} \ldots \sum_{\bar n_N=0}^{\infty} \right) \frac{1}{(\bar n_1+ W)! \bar n_1!  \ldots  ( \bar n_N+ W)!  \bar n_N! }  \xi^{2 \bar n_1 + \ldots + 2 \bar n_N + N W} \rme^{ i W \theta} 
\label{GeneralizedB}
\end{align}
 This elegant formula  tells us many interesting things: 
 \begin{itemize}
 \item The configurations that contribute to the 
partition  function have fractional  actions and integer topological charges:
\begin{align}
S&= \frac{S_I}{N} (2 \bar n_1 + \ldots + 2 \bar n_N) + S_I |W| \in   S_I \left( \frac{2}{N} |k| +  |W| \right), \qquad  k, W \in  \Z  \cr
 Q&= W \in \Z
\end{align}
just like our simple quantum mechanical $T_N$ model. 
 
\item Despite the fact that the monopole instantons have fractional topological charge, only  their magnetically neutral combinations   contribute to the sum.  Magnetic neutrality   enforces the  quantization of topological charge as well.  This satisfies the boundary condition at infinity on  $S^2_{\infty}$.

\item This construction also holds for  any sufficiently large $M_3$ times small $S^1_L$. We can think $\R^3$ as the infinite volume limit of  $M_3$. \end{itemize}

We can decompose mini-space partition function into its Fourier modes $z(\theta) = \sum_{W \in \Z} z_W  \rme^{\im  W \theta }$, and comparing with  \eqref{GeneralizedB}; we learn that 
\begin{align}
z_W&=  \left( \sum_{\bar n_1=0}^{\infty} \ldots \sum_{\bar n_N=0}^{\infty} \right) \frac{1}{(\bar n_1+ W)! \bar n_1!  \ldots  (n_N+ W)!  n_N! }  \xi^{2 \bar n_1 + \ldots + 2 \bar n_N + N W}  \cr 
&= [I_W(2 \xi)]^N
\label{GeneralizedB2}
\end{align}
where $I_W(2 \xi)$ is modified Bessel function. $z_W$ carries the information of all configurations (within the dilute monopole gas approximation) contributing to the topological sector with charge $W \in \Z$.    

 Tunneling events  that contribute to the partition  function  are the transitions between pure gauge configurations which differ in the  winding number  $W$ only, just like regular instantons on $\R^4$. Net magnetic flux is  zero at infinity, but this does not mean that the corresponding configurations are boring. Instead, each contributing configuration has sub-structure,  $n_a - \bar n_a = W, \;  \forall a $ monopole-instantons.

\subsection{Vacuum energy density and  multi-branch structure}
 A final remark is on the $N$-branched vacuum structure of deformed Yang-Mills theory. Monopole induced potential has $N$ extrema within the fundamental domain of $\bm \sigma$  field given in \eqref{s-cell}. These are:
\be
\bm{\sigma}_k={2\pi\over N}k \bm{\rho} \equiv  {2\pi\over N}k \sum_{i=1}^{N-1}\bm{\mu}_i, 
\ee
with $k=0,1,\ldots ,N-1$ and $ \bm{\rho} $ is  the Weyl vector.  Approximately $\frac {N}{2}$ of these branches are meta-stable for a given value of $\theta$ \cite{Aitken:2018mbb}. As $\theta$ is varied, the set of meta-stable vacua changes. From this, we can deduce the 
 vacuum energy densities as
\be
\mathcal{E}_k=- 2 N  K \rme^{-S_I/N} \cos\left({\theta+2\pi k \over N}\right). 
\ee
As a result, we can write the partition function for the theory defined on a  $3$-manifold as 
 \be
Z(\theta) = \sum_{k=0}^{N-1}  \rme^{ 2 N V_3  K  \rme^{-S_I/N} \cos\left({\theta+2\pi k \over N}\right)} \equiv \sum_{k=0}^{N-1} \rme^{ 2 
 N \xi  \cos\left({\theta+2\pi k \over N}\right)} 
 \label{vacen}
\ee
which can be converted into a fractional instanton sum:
\begin{align}
Z(\theta)  &=  N \sum_{ W \in \Z}  \sum_{n=0}^{\infty}   \sum_{\bar n =0}^{\infty}   \frac{1}{n!} \frac{1}{\bar n !}  \left(  N \xi   \rme^{ \im  \frac{ \theta} {N}}  \right)^{n }     \left(N \xi \rme^{ -  \im \frac{\theta} {N}}  \right)^{\bar n }     
 \delta_{n-\bar n  - W N,0}  \cr
& =  N \sum_{ W \in \Z}    \sum_{\bar n =0}^{\infty}   \frac{1}{(n + W N)!} \frac{1}{\bar n !}   (N\xi)^{2 \bar n + W N} 
   \rme^{ \im  W   \theta} \cr     
  & =  N \sum_{ W \in \Z}   I_{W N}(2 \xi N)   \rme^{ \im  W   \theta} 
  \label{partsum2}
   \end{align}
   In  \eqref{partsum2}, quantization of topological charge emerges as a result of the constraint,  $\delta_{n-\bar n  - W N,0} $. 
In this formula, there is no information about magnetic charges of the individual monopole instantons.  That information is washed away by the choice of the vacuum branch in \eqref{vacen}. At the $k^{\rm th}$ branch, all monopole operators with non-zero adjoint charge (also called GNO charge)  but zero  't Hooft  charge   acquire the same vev, $\langle   \rme^{  \im \bm{\alpha}_a \cdot \bm{\sigma} } \rangle  = \rme^{ \im 2\pi k \over N} $ \cite{Aitken:2018mbb}   and the information about individual magnetic charges  $\bm{\alpha}_a$  of $N$ types of monopoles is lost.  
In this sense, \eqref{GeneralizedB} has more information than \eqref{partsum2}. It knows that $n_a -\bar n_a = W \;\;   \forall a$. This naturally implies  $\sum_{a=1}^N (n_a - \bar n_a) \equiv n-\bar n = W N $.

\section{Coupling Yang-Mills theory  to $\Z_N$ TQFT }
The Lagrangian  of  $SU(N)$ Yang-Mills theory is
\be
{\cal L}_{\mathrm{YM}}={1\over 2g^2_{\mathrm{YM}}}\int \tr[F\wedge \star F]+{\im\, \theta_{\mathrm{YM}}\over 8\pi^2}\int \tr[F\wedge F] 
\label{SU}
\ee
where the second term is properly  quantized topological term:
\begin{align}
Q=\frac{1}{8\pi^2}\int \tr[F\wedge F]  \in \Z
\label{quantize-3}
 \end{align}
 $SU(N)$ YM theory possess a  $\Z_N^{[1]}$ electric one-form symmetry, which acts on  Wilson line operators. 
 
 To probe the topological configurations in YM theory, we follow  the same method as in  our $T_N$ quantum mechanics example.  
 We turn on a  background gauge  field  for the  $\Z_N^{[1]}$  1-form symmetry \cite{Kapustin:2014gua, Gaiotto:2014kfa},   introducing    pair of 
  $U(1)$ 2-form and 1-form  gauge fields $(B^{(2)}, B^{(1)}) $  satisfying 
  \begin{align}
N B^{(2)}= \diff B^{(1)},  \qquad N \int B^{(2)}=  \int \diff B^{(1)} = 2 \pi \Z
\label{ZN-YM}
\end{align}
 The  action corresponding to the  $\Z_N$ topological gauge theory is given by 
 \begin{align}
 Z_{{\rm top}, p} &= \int  \Diff   B^{(2)} \Diff B^{(1)}    \Diff  C^{(2)} \;  \rme^{ \im \int C^{(2)}  \wedge ( N B^{(2)}-\diff B^{(1)}) + \im p {N\over 4\pi}\int B^{(2)}\wedge B^{(2)}   }  
 \label{top-3}
 \end{align}
 where $C^{(2)}$ is Lagrange multiplier, and $p$ is the discrete theta angle. 
 
To couple the $SU(N)$ YM theory to the background gauge field $B^{(2)}$, we  promote the $SU(N)$ gauge field $a$ into a $U(N)$ gauge field  $\tilde a$. The $U(N)$ gauge field is related to dynamical $SU(N)$ gauge field $a$ locally as 
\be
\widetilde{a}=a+{1\over N}B^{(1)}. 
\ee
To gauge $\mathbb{Z}_{N}^{[1]}$, we introduce gauge invariance under the one-form gauge transformation, 
\be
B^{(2)}\mapsto B^{(2)}+\diff \Lambda^{(1)},  \qquad B^{(1)}\mapsto B^{(1)}+N\Lambda^{(1)},  
\ee
and hence, 
\be
\widetilde{a}\mapsto \widetilde{a}+\Lambda^{(1)},   \qquad  \widetilde{F} \mapsto  \widetilde{F} + \diff \Lambda^{(1)} 
\ee
In writing $SU(N)$ theory   in  a  $\Z_N$  TQFT  (equivalently  't Hooft flux) background,  
we  have to replace $SU(N)$ field strength with the 
gauge-invariant combination of the $U(N)$ field strength $\widetilde{F}=F(\widetilde{a})$ and $B^{(2)}$:
\be
F(a)\Rightarrow \widetilde{F}-B^{(2)}. 
\label{replace}
\ee
The action of the $SU(N)$ theory in the $B^{(2)}$ background can be written as:
\be
{S}[ B^{(2)}, B^{(1)},  \widetilde a]={1\over 2g^2_{\mathrm{YM}}}\int \tr[  (\widetilde{F}-B^{(2)})\wedge \star  (\widetilde{F}-B^{(2)})]+{\im\, \theta_{\mathrm{YM}}\over 8\pi^2}\int \tr[  (\widetilde{F}-B^{(2)}) \wedge  (\widetilde{F}-B^{(2)})] 
\label{SUflux}
\ee
If we further  wish to obtain the partition function of the  $(SU(N)/\Z_N)_p$ gauge theory, then, we have to  sum over all  $B^{(2)}$, $\Z_N$ valued 2-form gauge fields. Namely, 
 \begin{align}
 Z_{(SU(N)/\Z_N)_p} =  \int  \Diff   B^{(2)} \Diff B^{(1)}    \Diff  C^{(2)}  \Diff  \widetilde a \;   \rme^{ \im \int C^{(2)}  \wedge ( N B^{(2)}-\diff B^{(1)}) + \im p {N\over 4\pi}\int B^{(2)}\wedge B^{(2)}   }   \rme^{-S[ B^{(2)}, B^{(1)},  \widetilde a] }
 \label{top-5}
 \end{align}
where   
\begin{align}
{N\over 8\pi^2}\int B^{(2)}\wedge B^{(2)}\in  \frac {1}{N} \mathbb{Z}
\label{flux-5}
\end{align}
is the $\Z_N$ 't Hooft flux. 
 This partition function is the counter-part of 
 our simple quantum mechanical example \eqref{tpf2}. For other background fields that one can turn on in Yang-Mills theory, see \cite{Wan:2019oyr}.

In $SU(N)$  theory, topological charge  is quantized in integer units  \eqref{quantize-3}.  When we turn on $B^{(2)}$ background, the topological term  becomes 
\begin{align}
Q&= {1\over 8\pi^2}\int \tr[  (\widetilde{F}-B^{(2)}) \wedge  (\widetilde{F}-B^{(2)})]   \cr 
&= \underbrace{{1\over 8\pi^2}\int \tr[  (\widetilde{F} \wedge \widetilde{F}  )}_{\in \Z}  -  \underbrace{\frac{N}{8 \pi^2} \int  B^{(2)} \wedge B^{(2)})]  }_{\in \frac{1}{N} \Z}  \in  \frac {1}{N} \mathbb{Z}.
\label{quanPSU}
\end{align}
fractionally quantized, just like our simple quantum mechanics and $\mathbb {CP}^{N-1}$ examples. 
In the second line, the first term is integer quantized because $ \widetilde{F} $   is in the $U(N)$ bundle.  
 The second term corresponds to $\Z_N$ 't Hooft flux background.

Let us now determine the saddles in the  theory with  non-trivial $B^{(2)}$ background, and their actions. The action functional for $\theta=0$ can be manipulated by using Bogomolny factorization:
\begin{align}
S_{}&={1\over g^2_{\mathrm{YM}}} \int \tr[(\widetilde F - B^{(2)})\wedge \star  (\widetilde F -B^{(2)})]  \cr
&= {1\over 2 g^2_{\mathrm{YM}}} \int \tr \left[ \left( (\widetilde F - B^{(2)}) \pm   \star  (\widetilde F -B^{(2)})\right)   \wedge \star \left( (\widetilde F - B^{(2)}) \pm   \star  (\widetilde F -B^{(2)})\right)  \right  ]  
\cr
& \mp   \frac{8 \pi^2}{g^2}  \frac{1}{8 \pi^2}  \int   \tr[ (\widetilde F -B^{(2)}) \wedge  (\widetilde F -B^{(2)}) ]
\end{align}
The BPS bound is saturated by modified version of the instanton equation in the $B^{(2)}$ background:  
\begin{align}
(\widetilde F-B^{(2)})=  \mp \star( \widetilde F-B^{(2)})
\label{Mod-inst}
\end{align}
This equation is the counterpart of \eqref{ins1}  in QM and \eqref{ins3} in $\mathbb{CP}^{N-1}$ in appropriate $\Z_N$ backgrounds.  
Using  \eqref{quanPSU}, we see that the action is bounded from below by 
\be
S  =  \mp \frac{8 \pi^2}{g^2}   \frac{1}{8 \pi^2}  \int   \tr[  (\widetilde F -B^{(2)}) \wedge  (\widetilde F -B^{(2)}) ]
 =  \frac{S_I}{N}
 \label{Mod-inst-ac}
\ee

When we turn on a $ B^{(2)} $ flux, we are {\it still} in $SU(N)$ theory, just like in our QM example, when we consider 
the $\tr [\mathsf U^\ell  e^{-\beta H}]$, we are still in the  $T_N$ model.  If we sum over all backgrounds, 
  we move to  $(SU(N)/\Z_N)_p$  gauge theory just like we moved to  $(T_N/\Z_N)_p$ model in QM. 

Formally,  in quantum mechanics,   it takes the insertion of $\mathsf U^\ell$ into the state sum to see explicitly the existence of fractional topological charge  
saddles. (Intuitively, just looking to the Figures, this is of course obvious.)  This  enforces the following interpretation.  
We have to accept that  in the   $SU(N)$  theory, saddles with fractional charge  $\frac{1}{N} \Z$ and action $\frac{S_I}{N}= \frac{8 \pi^2}{g^2N} $ exist,  but they are configuration in the $SU(N)/\Z_N$ bundle. As such, all  it takes is the insertion of a 't Hooft flux to see these configurations  in the $PSU(N)$ bundle in the  original $SU(N)$ theory. Then, we can build configurations in the 
$SU(N)$ theory which live in the $SU(N)$ bundle, which are  non-BPS composites of objects in the $PSU(N)$ bundle. 
Therefore, in exact analogy with our quantum mechanical example, there are  topological configurations with fractional action,  $\frac{2|n|}{N}S_I  + |W| S_I$ and integer topological charge.  This strongly suggests that the  non-perturbative expansion parameter in the theory is $  \rme^{-  S_I/N} = \exp[- \frac{  8 \pi^2 }{g^2(\mu)N }]$, which is   exponentially more important than BPST instanton \cite{Belavin:1975fg}: 
 \begin{align}
\rme^{- \frac{  S_I}{N} } \gg   \rme^{- 2 \frac{  S_I}{N} }\gg   \rme^{- 3 \frac{  S_I}{N} } \gg \ldots \gg   \underbrace{\rme^{-  {  S_I} }}_{\rm instanton}  
\label{frac-instanton-3}
\end{align}
 Remarkably, the picture that emerges in the large 4-manifold $M_4$ is almost identical to our construction of dilute gas of monopole instantons on deformed YM  and 
 $\N=1$ SYM theory on $\R^3 \times S^1$.

\subsection{$T^3 \times S^1_L$  with 't Hooft flux and monopole-instantons} 
We can use the idea of double-trace deformation both in $SU(N)$ and $PSU(N)$ theory, because the double-trace 
operator can be viewed as a  trace in adjoint representation which belongs to both groups. Therefore, regardless of the existence of center symmetry, in both case, holonomy potential has a minimum at $\rme^{\im \bm{\phi}_\star}=(1,\omega,\omega^2,\ldots, \omega^{N-1})$ at the center of Weyl chamber. Hence, in both case, dynamics abelianize to maximal abelian subgroup. 
 
 The magnetic monopoles that participate to non-perturbative dynamics  have magnetic charges in root lattice,  $ {\bm \alpha} \in  \Gamma_{r}$, 
 both for $SU(N)$ and $PSU(N)$, corresponding to  monopole operators ${\cal M}_a(x) \sim \rme^{\im \bm{\alpha}_a \cdot \bm{\sigma}(x)}$  \eqref{mon-op} in the effective field theory. 
   The distinction between the two is that  $PSU(N)$  admits test magnetic charges ${\bm \mu} \in \Gamma_w$, which are forbidden in $SU(N)$, corresponding to probe operators  $\rme^{\im \bm{\mu}_a \cdot \bm{\sigma}(x)}$ where  $\bm{\mu}_a$ are fundamental weights. 
The magnetically  charged operators  which  are non-trivial under $\Gamma_w/\Gamma_r$
 are immediately in $PSU(N)$ bundle, but how about dynamical monopoles $ {\bm \alpha} \in  \Gamma_{r}$  which carry 
 $Q=1/N$?    What is the  precise role they play? 
    
    \vspace{0.5cm}
\noindent
 {\footnotesize
{ \bf Reminder about monopoles:}  Consider a monopole with charge $\bm \mu \in  \Gamma_w $. Its  magnetic field at asymptotic large distances  is:
  \begin{align}
 \mathsf {\bm B} = \frac{\bm {\hat r}}{r^2}  (\bm \mu\cdot \bm H)
\label{mag-flux}
\end{align} 
where   $\bm H = (H^1, \ldots, H^{N-1} )$ denote the Cartan generators of $SU(N)$ algebra. 
For  $\bm \mu$ which is a weight of  fundamental representation,   $(\bm \mu\cdot \bm H)= \frac{T}{2N}$ up to permutations, 
where  $T= {\rm diag}(1,\ldots, 1, -(N-1))$.
 If $\bm \mu$    is in $N$-index representation or adjoint representation, 
$ (\bm \mu_N \cdot \bm H) \sim  (\bm \alpha \cdot \bm H)  \sim \frac{T}{2}$.

 Now, consider a sphere $S^2$ surrounding the monopole. 
The non-abelian gauge connection on the north and south hemisphere  patches, corresponding to the above magnetic flux, is given by:  
  \begin{align}
A^N_{\phi} &=  \frac{1}{r}  \frac{T}{2N} \frac{ (1-\cos \theta)}{\sin  \theta} \cr
A^S_{\phi} &=  \frac{1}{r}   \frac{T}{2N}   \frac{ (-1-\cos \theta)}{\sin  \theta}
\end{align} 
The two gauge connections are same up to gauge transformation. On a small strip along the equator where the patches overlap, we have 
  \begin{align}
A^N_{\phi}  - A^S_{\phi}  =   \frac{1}{r \sin \theta}     \partial_\phi  T (\phi) \cr
\end{align} 
where 
\begin{eqnarray}
T(\phi)   = \frac{\phi}{N} T \qquad {\rm and} \;\;  \rme^{\im T(\phi+ 2\pi)}= \rme^{\im T(\phi)} \rme^{-\im \frac{2\pi}{N}}
\end{eqnarray}
i.e. charge $\bm \mu$  configuration      lives in  $PSU(n)$ bundle. 
 Concerning charge $ {\bm \alpha}$ monopole, which is dynamical in the theory, we have the transition matrix  $(\rme^{\im T(\phi)} )^N$, which is an element of   $PSU(N)$ bundle which can lift to  $SU(N)$ bundle.  But still, its topological charge is $1/N$.  We discuss its importance below. 
}   

\begin{figure}[t]
\vspace{-2cm}
\begin{center}
\includegraphics[width = 1\textwidth]{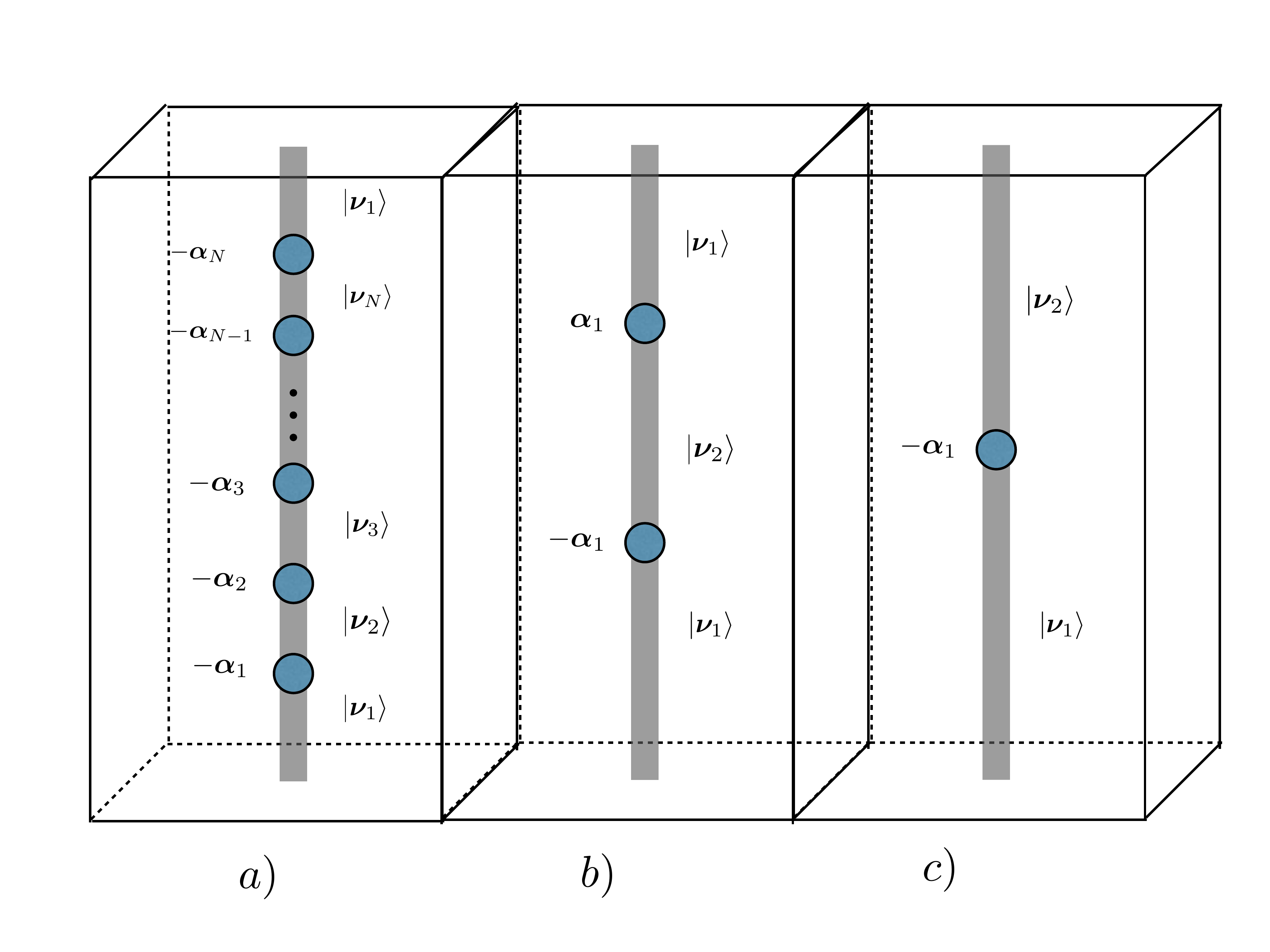}
\vspace{-1.5cm}
\caption{Consider YM in the background of    $\frac{N}{2\pi}  \int B^{(2)}_{12} = \ell_{12}= 1$  unit of   't Hooft flux, $Z_{\ell_{12}}=   \tr [ \rme^{-\beta H_{ \ell_{12}}} ]$.  The existence of  $\ell_{12}=1$,  in finite volume, leads to  $N$-perturbatively  degenerate minima, just like $T_N$ QM model.    a)  A non-trivial  periodic (in Euclidean time) configuration with topological charge $W=-1$ contributing to  $Z_{\ell_{12}}$     is a collection of $N$  dynamical monopole instantons, each of   which  have zero  
 't Hooft charge, but non-zero magnetic charge.    b) A periodic configuration with topological charge $W=0$ contributing to  $Z_{\ell_{12}}$. 
 c) This configuration does not contribute to $Z_{\ell_{12}}$   as it is not periodic in Euclidean time, but it contributes to 
  the sum over transition amplitudes, 
    $  Z_{\ell_{12}\ell_{34}} =   \tr [ \rme^{-\beta H_{ \ell_{12}}} (\mathsf U_{\rmc})^{\ell_{34}} ],  \; \ell_{34}=1 $. 
  The aperiodicity in the euclidean time direction can also be undone, and replaced 
 with an $\frac{N}{2\pi}  \int B^{(2)}_{34} = \ell_{34}= 1$ background.  }
\label{fig:monopole}
\end{center}
\end{figure}

%
%

\vspace{0.5cm}
\noindent
{\bf Monopole-instantons in the    't Hooft flux background:} In this part, we would like to see the role of monopole-instantons in the non-trivial  't Hooft flux background  which is in the $PSU(N)$ bundle, and determine an  analytic solution to \eqref{Mod-inst} when the theory is compactified on large $T^3$ (serving as regularization of  $\R^3$) times a small $S^1_L$. First of all, following our construction in quantum mechanics and 
$\mathbb {CP}^{N-1}$ model, we know that we can undo the background and substitute it with boundary conditions.  

Then, we can check if our monopole-instantons are sensible solutions within $PSU(N)$,   fitting  with the boundary conditions. We will come to the conclusion that our monopole-instantons with topological charge $Q_a=1/N$,  action $S_a= S_I/N$  and magnetic charge $ {\bm  \alpha}_a$  have an interpretation as tunneling events between configurations in the $PSU(N)$ bundle in the semi-classical domain of the theory. 

We can think turning on $k$ unit of   't Hooft magnetic flux in 12-plane,  (in $3$-direction), in multiple  ways. The simplest way is to start with twisted boundary conditions on $T^2 \times \R^2$, and undo the twist in favor of   background $B^{(2)}_{12}$  fields. Gray band in Fig.  \ref{fig:monopole} is a  depiction of such classical background.  This flux background  exists without any reference to magnetic monopoles. On each $T^2$ slice,  we have 
\begin{align}
  \frac{N}{2\pi}  \int B^{(2)}_{12}  = k \equiv \ell_{12}  \;\;\;  {\rm mod}  \;\; N
\end{align}
 flux. A useful  way to think about it is as follows.  
 Such 't Hooft flux lines can be viewed as 
the field line  created between magnetic monopoles with charges ${\bm \mu}_k$  and $-{\bm \mu}_k$, associated with magnetic $N$-ality $k$ and $-k$.   (But recall that such   charges are not present in the $SU(N)$ theory, neither as dynamical object nor as probes.  Even if when we gauge the 1-form center symmetry,  and move  to say  $PSU(N)_0$ theory which is locally equivalent to $SU(N)$ theory,   such monopoles are not present dynamically. They are only allowed as probe charges.) These  magnetic flux lines carry a magnetic flux through 
the 12-surface as well, given by 
\begin{align}
\int_{12}  \bm B = \frac{2 \pi} {g}  \bm \mu_k 
\end{align}
So, we have to think that these flux lines are present and upper and lower plane in \ref{fig:monopole} are identified despite the fact that 
charges are not present in the theory.\footnote{In certain sense, we can think of 
the  't Hooft magnetic flux    $\ell_{12} =k$  sourced by ${\bm \mu}_k$  and $-{\bm \mu}_k$ pair  
similar to the   mythical symbol ouroboros, a snake swallowing its own tail. We can forget about source and sink, and just take the flux lines into consideration. }   
Another useful way to think about the  $\ell_{12} =k$  flux line is a non-dynamical center-vortex, see \cite{Greensite:2003bk} about center-vortices. Indeed, a center-vortex    passing through a fundamental Wilson loop  in 12-plane contributes to 
$W(C) = \rme^{\im \oint a} $ as 
$\rme^{\im \int_{12}  \bm B} =  \rme^{\im  \frac{2 \pi k}{N}}$. All these perspectives are useful once we think about tunneling events in flux backgrounds.

%
%
%

%
   
Now, we can give an interpretation of the tunneling events  corresponding to dynamical monopole-instantons with charge  
${\bm  \alpha}_a \in \Gamma_r$, in the  $SU(N)$  theory on $T^2 \times S^1_\beta \times S^1_L$ with $\ell_{12}=1$ unit of magnetic 't Hooft flux background, see  Fig. \ref{fig:monopole}. 
First, note that  the configurations in the  $\ell_{12}=1$ can be attached different  magnetic flux, corresponding to any representative of the 
fundamental (defining) representation:
\begin{align}
{\bm \nu}_1\equiv {\bm  \mu}_1,  \qquad {\bm \nu}_2 \equiv {\bm  \mu}_1 -   {\bm  \alpha}_1, \qquad {\bm \nu}_3 \equiv {\bm  \mu}_1 -   {\bm  \alpha}_1 -    {\bm  \alpha}_2,   \qquad \ldots  \qquad 
     {\bm \nu}_N \equiv {\bm  \mu}_1 -    \sum_{a=1}^{N-1} {\bm  \alpha}_a, 
     \label{magneticflux-2}
\end{align}
The  magnetic flux passing through the $T^2$ for these configurations are: 
\begin{align}
\bm \Phi =    \int_{T^2}  {\bm B}=  \frac{2 \pi} {g} {\bm \nu}_a   ,  \qquad a =1, \ldots, N.
\end{align}
Assume that the flux is uniform through the surface of torus $T^2$.\footnote{One does not need to make this assumption, but it makes the discussion a bit simpler, without sacrificing correctness.   The 't Hooft flux insertion cost an energy, which depends only on $\ell_{12}=k$. For example, 
for    $\ell_{12}=1$, it does not matter what magnetic flux out of the list \eqref{magneticflux-2}  is associated with it.}
  Then for these $N$-configurations,    the energy of the corresponding states are exactly degenerate {\it even at finite volume.}  If ${\bm B}_a=  \frac{1}{A} \frac{2 \pi} {g} {\bm \nu}_a$,  their  energies are 
  \begin{align}
E_a & =   \half  \int_{T^2}  {\bm B_a}^2=  \frac{1}{2A}  \Big(\frac{2 \pi} {g} \Big)^2 {\bm \nu}_a^2 =    \frac{1}{2A}   \Big(\frac{2 \pi} {g} \Big)^2  \Big(1-\frac{1}{N} \Big) ,  \qquad a =1, \ldots, N. 
\label{degenerate}
\end{align}
The tunneling between the  flux configurations   $| {\bm \nu}_a\rangle$   and $| {\bm \nu}_{a+1}\rangle$  
changes magnetic flux  by simple and affine rooot ${\bm \alpha}_a, \;  a=1, \ldots, N $,  
 but does not change 't Hooft flux, which is fixed,  $\ell_{12}=1$.  
\begin{align}
 \Delta &    \int_{T^2}  {\bm B} =     \frac{2 \pi} {g}   \left( {\bm \nu}_a  - {\bm \nu}_{a+1} \right) = -   \frac{2 \pi} {g} {\bm \alpha}_a, 
  \qquad a =1, \ldots, N.  \cr
\Delta  & \left(  \frac{N}{2\pi}  \int B^{(2)}_{12}  \right) =  0   \;\;  
\end{align}
This change in the magnetic flux is associated with  the magnetic charge of the  dynamical monopole instantons present in the theory. 

\subsection{Two types of monopole events, Born-Oppenheimer and $T_N$ QM}
One needs to be careful concerning two distinct class of tunneling events in compactified set-up on $T^2 \times S^1_L \times S^1$.  These are between 
\begin{itemize}
\item[$a)$] Monopole tunneling events between exactly degenerate minima  as described above, and  
\item[$b)$]  Tunnelings that become degenerate only in ${\rm Area}(T^2) \rightarrow \infty  $ limit. 
\end{itemize}
The discussion above implies the existence of  is  different from the discussion of monopole-instantons in the absence of  't Hooft flux background, for example, in the  Polyakov model, see page 226 of Ref.\cite{Banks:2008tpa} (which is one of the rare textbooks which discusses Polyakov model and Hamiltonian interpretation of the tunneling.)      A monopole-instanton in the case of Polyakov model 
   {\bf always} changes the energy of vacuum state at finite   ${{\rm Area}({T^2})}$.    If $\bm \Phi =    \int_{T^2}  {\bm B} =   \frac{2 \pi} {g} {\bm \alpha}_a n_a   $ is magnetic  flux (no summation over $a$), then the change in energy at finite volume  between the zero-magnetic flux state and   $\bm \Phi$ flux state is:
\begin{align}
\Delta E =   \int_{T^2}  \half  {\bm B}^2 = \half \frac{     \left( \frac{2 \pi} {g} \right)^2 n_a^2   }{ {\rm Area}({T^2})}   >0   
\end{align}
Therefore, at finite volume, these states are not degenerate.  Only when the  area tends to infinity,  the cost of energy becomes vanishingly small.  
\begin{align}
\lim_{ {\rm Area}({T^2}) \rightarrow \infty }  \Delta E  = 0  
\end{align}
and these states become degenerate with the zero magnetic flux state. 
These are the perturbative   vacua, which only emerge in the 
${\rm Area}({T^2})  \rightarrow  \infty$ limit. The tunneling events between them  is the Hamiltonian interpretation   of monopole instantons  in Polyakov model.  

In  the presence of   't Hooft flux background, say  $\ell_{12}=1$, 
there is a genuine $N$-fold  vacuum degeneracy at finite  ${{\rm Area}({T^2})}$ associated with monopoles in the affine  root system. 
The classical degenerate vacua at finite volume are the states   $| {\bm \nu}_a\rangle, a=1, \ldots, N$.  These are the counterpart of our 
perturbative $|j \rangle, j=1, \ldots, N$ in our quantum mechanical $T_N$ model.   Therefore, in some respect, there are two distinct class of tunneling events in the case of  $\ell_{12}=1$ background.  For example, starting with the state  $| {\bm \nu}_1\rangle$, tunneling with  $- {\bm \alpha}_1$  and   $+{\bm \alpha}_1$    takes us to  states
 $|  \bm \nu_1 -  \bm  \alpha_1  \rangle    = | {\bm \nu}_2 \rangle$,  and 
$| {\bm \nu}_1 + \bm \alpha_1 \rangle  = | 2 {\bm \nu}_1   -  {\bm \nu}_2  \rangle$. 
\begin{align}
& E_{| {\bm \nu}_1 - \bm \alpha_1 \rangle}   -  E_{| {\bm \nu}_1 \rangle    } =0 
\cr 
& E_{| {\bm \nu}_1 + \bm  \alpha_1 \rangle}   -  E_{| {\bm \nu}_1 \rangle    }     = \half \frac{     \left( \frac{2 \pi} {g} \right)^2 4   }{ {\rm Area}({T^2})}   >0    \underbrace {\longrightarrow}_{  {\rm Area}({T^2}) \rightarrow \infty } 0
\end{align}
where the first type  is associated with the exact $N$-fold perturbative  degeneracy at finite volume  and the latter is associated with the emergent degeneracy as in the Polyakov model.

Since all magnetic flux states except for  $| {\bm \nu}_a\rangle, \;  a=1, \ldots, N$ have high  energy in the small $T^2 \times \R \times S^1_L$ limit, within  Born-Oppenheimer approximation,  the Yang-Mills theory with $\ell_{12}=1$ flux actually reduces to our simple $T_N$ model.  The role of the fractional instantons  $\I_a$ in the QM system is played by monopole-instantons with charge $-\bm \alpha_a$ in the Yang-Mills quantum mechanics. 
This is a concrete and nice correspondence with quantum mechanics and Yang-Mills theory. For example, it allows us to derive the  vacuum energy densities  for the $N$-branched  vacua \eqref{vacen} and  fractional theta angle dependence  by simple quantum mechanics. However,  we are more interested in the large-$T^4$  limit in general.

\subsection{What does $\Z_N$ TQFT background has anything to do with monopoles?}
Now, we can consider 
the partition function just in the background of  't Hooft flux $\ell_{12}$. 
 \begin{align}
 Z_{\ell_{12}} & =   \tr [ \rme^{-\beta H_{ \ell_{12}}}  ]   = \int_{\rm pbc} \Diff a \;  \rme^{-S(a, B^{(2)}_{12})}
  \label{tpf-10}
 \end{align}
This is the exact counter-part of  the regular partition function in our $T_N$ model with $N$-degenerate minima.   
The existence of potential in  the QM model is a substitute for 
the effect of ${ \ell_{12}}$ flux in Yang-Mills theory.  
To determine \eqref{tpf-10}, 
we have  to sum over all periodic  configurations,  in particular,  
\begin{align}
\Phi(\beta)= \Phi(0) 
\end{align}
Therefore, exactly the same constraint that we obtained in the zero flux sector \eqref{constraint-5} is still operative,  
$n_a-\bar n_a =  W, a=1, \ldots, N$ and independent of $a$ for dynamical monopoles of type $\bm \alpha_a$. 
The configurations contributing to the sum must satisfy magnetic neutrality, which automatically impose quantization of the 
topological charge $W \in \Z$.  

Exactly as in our QM example, where in order to see
what contributes to the state sum, we need to look to a transition matrix element, between the states $ | {\bm \nu}_a\rangle $  and $ | {\bm \nu}_{a+1}\rangle $. This is what is called an electric 't Hooft flux insertion.  Let us make this more explicit. 

The $\Z_N^{[1]}$ form center symmetry of the theory on $\R^4$, upon compactification  on   
$ T^4 \equiv T^2 \times  S^1_L \times S^1_\beta $, becomes  $(\Z_N^{[0]})^4$.  The states $| {\bm \nu}_a\rangle$  described above are related to each other 
by a center-transformation, $\mathsf U_{\rmc}  | {\bm \nu}_a\rangle = | {\bm \nu}_{a+1}\rangle$.  
 Insertion of this operator in  partition function corresponds to $\ell_{34}=1$ units of  electric 't Hooft flux.
 \begin{align} 
 Z_{\ell_{12}\ell_{34}}  & =   \tr [ \rme^{-\beta H_{ \ell_{12}}} (\mathsf U_{\rmc})^{\ell_{34}} ]   \cr
& =    
  \sum_{a=1}^{N} \;\;    \langle \bm \nu_a | \rme^{-\beta H_{ \ell_{12}}} (\mathsf U_{\rmc})^{\ell_{34}}   | {\bm \nu}_a\rangle  \cr
 & =    
  \sum_{a=1}^{N} \;\;    \langle \bm \nu_a | \rme^{-\beta H_{ \ell_{12}}}  | {\bm \nu}_{a+ \ell_{34}} \rangle  \cr 
 &=   \int_{\rm \Phi(\beta)=  \Phi(0) +  \frac{2\pi }{g} \bm \alpha_{a, a+ \ell_{34}}  } \Diff a \;  \rme^{-S(a, B^{(2)}_{12})}  \cr
  &=   \int_{\rm pbc } \Diff a \;  \rme^{-S(a, B^{(2)}_{12}, B^{(2)}_{34} )}  
  \label{tpf-9}
 \end{align}
 In the fourth step,  we have the boundary conditions corresponding to a transition amplitude: 
 \begin{align}
\Phi(\beta)= \Phi(0) +   \frac{2\pi }{g} \bm \alpha_{a, a+ \ell_{34}}  \equiv  \Phi(0) +   \frac{2\pi }{g} ( \bm \nu_a - \bm \nu_{a + \ell_{34}})
\label{tunneling-SUN}
\end{align}
which is the change in the magnetic flux, corresponding to a dynamically allowed  monopole event  in the root lattice $\Gamma_r$. 
 This is the twisted boundary condition in terms of magnetic flux description. 
 In the last line of \eqref{tpf-9},  we  undo the twisted boundary condition into a periodic boundary condition,  just like in the simple QM example, and transmute the effect to the background $B^{(2)}_{34}$ flux.   Note that we can also think twisted boundary condition in different ways. In particular, we could have written it more conventionally as $U_3(\beta=L_4)= \rme^{\im \frac{2 \pi}{N} {\ell_{34}} }U_3(0)$. This is equivalent to above. This is because the center transformation acts on the center-symmetric  holonomy field  
 $\rme^{\im \bm{\phi}_\star}=(1,\omega,\omega^2,\ldots, \omega^{N-1})$ 
as $\rme^{\im \bm{\phi}_\star}  \rightarrow   \rme^{\im  \frac{2 \pi}{N} } \rme^{\im \bm{\phi}_\star}$. The action is invariant under this transformation if 
we act on the  fields by appropriate cyclic permutation,  e.g, it takes $\sigma_k \rightarrow \sigma_{k-1}$  \cite{Cherman:2016jtu}.  This  cyclically permutes  background magnetic flux, and corresponds to the  twisted boundary condition on fourth line.  \footnote{The gauge invariant definition of the $U(1)^{N}$ photons are 
\begin{align}
F_{\mu \nu, k}= \frac{1}{N} \sum_{p=0}^{N-1} \;  \rme^{-\im \frac{2 \pi  k p}{N}} \;  \tr( U_3^p  F_{\mu \nu})
\end{align} 
where   $F_{\mu \nu}$ inside trace is the non-abelian gauge field strength. Crucially, counter-part of the adjoint Higgs field is the Polyakov loop.  Under a zero-form center transformation, Polyakov loop 
transforms as   $ U_3^p  \rightarrow  \rme^{\im \frac{2 \pi   p}{N}}  U_3^p$. Hence, the   photon in the 3d theory,  {\it transforms} under the center transformation as $ F_{\mu \nu, k} \rightarrow   F_{\mu \nu, k-1}$. The same is also true for the  dual photon. Clearly, the monopole operator $\rme^{\im  \bm{\alpha}_k \cdot \bm{\sigma}  }    \rightarrow   \rme^{\im  \bm{\alpha}_{k-1} \cdot \bm{\sigma}  } $ and the flux states are cyclically shifted, $ |\bm \nu_k \rangle \rightarrow  |\bm \nu_{k-1} \rangle$. In other words, the zero-form part of center-symmetry acts like  $\Z_N$ discrete translation  symmetry  in the $T_N$   model.   }


\vspace{0.5cm}
 \noindent
{\bf The oddity of this result:}  On $\R^3 \times S^1$, the action and topological charge of the monopole-instantons with magnetic charge $Q_m= \frac{2 \pi}{g} \bm \alpha_a $ is determined by center-symmetric  gauge holonomy: 
\begin{align}
{\rm Diag} (U_3) = \rme^{\im \bm{\phi}_\star}=(1,\omega,\omega^2,\ldots, \omega^{N-1}) 
\label{center-sym}
\end{align}
This background is quantum mechanically stable in QCD(adj) and can be made stable in Yang-Mills by using the double trace deformation. Minimal  action and topological charge in this background is given 
by 
 \begin{align} 
 S_a= \frac{4 \pi }{g^2} (\bm \alpha_a. \bm{\phi}_\star) = \frac{8 \pi^2}{g^2N},  \qquad  \qquad
Q &= \frac{1}{2\pi} (\bm \alpha_a. \bm{\phi}_\star) = \frac{1}{N} 
\label{frac5}
 \end{align}

On the other hand, it is also well-known that in the non-vanishing   't Hooft flux  background, or equivalently, by coupling YM to $\Z_N$ TQFT, minimal action and topological charge are also $\frac{1}{N}$  quantized. 
 \begin{align}
 S  =    \frac{8 \pi^2}{g^2}   \frac{1}{8 \pi^2}  \int  B^{(2)} \wedge  B^{(2)}  
 =  \frac{8\pi^2}{g^2N}, \qquad 
Q= {N\over 8\pi^2}\int B^{(2)}\wedge B^{(2)}  = \frac{\ell_{12} \ell_{34}}{N}  = \frac{1}{N}
\label{flux-7}
\end{align}
Clearly, except for the results, these two constructions do not look anything alike. One is the action of a dynamical monopole instanton on the zero 't Hooft flux background and the other seems to be a property of the classical  $B^{(2)}$   background associated with 1-form $\Z_N^{[1]}$ center symmetry. In the latter, there seems to be naively no data of the monopole-instanton, and it is quite tempting that, despite the agreement of the final results, these two construction has nothing to do with one another.

 But again, truth is subtler. Let us recap  what we did in this section again because it is interesting. 
  On $T^2 \times S^1_L \times S^1_\beta$, assume first that center-symmetric background \eqref{center-sym} is stabilized. Turning on  $\ell_{12}=1$ on $T^2$ part gives us $N$- exactly degenerate minima at the classical level. There are $N$-configurations which carry 1-unit of 't Hooft flux, but they  also carry an orientation, that we labeled by  
  $\bm \nu_a, a=1, 2, \ldots$ in   \eqref{magneticflux-2}.   $\ell_{34}=1$ background is equivalent to 
  a twisted boundary condition $U_3(\beta=L_4)= \rme^{\im \frac{2 \pi}{N} {\ell_{34}} }U_3(0)$.  But as we explained above, this amounts to the transition amplitude between  $\bm \nu_a$  and   $\bm \nu_{a+1}$, meaning that the difference of the magnetic flux on the $\tau= \beta$ and  $\tau= 0$ is just the charge 
  $ Q_m= \Delta     \int_{T^2}  {\bm B} =     \frac{2 \pi} {g}   \left( {\bm \nu}_a  - {\bm \nu}_{a+1} \right) = -   \frac{2 \pi} {g} {\bm \alpha}_a $. This  is the charge of our dynamical monopole-instanton event.  \footnote{
  If the minima for $U_3$ is classically center-broken configurations $U_3 = \rme^{\im  \frac{2 \pi}{N} j} {\mathbbm 1} $ given in    \eqref{centerbroken},  then, the twisted boundary conditions 
   $U_3(\beta=L_4)= \rme^{\im \frac{2 \pi}{N} {\ell_{34}} }U_3(0)$ takes $|j\rangle$ to $|j+\ell_{34}\rangle$. This gives a dynamics which is  different from above. See Appendix~\ref{sec:TEK}. It would be interesting to understand the connection between these two constructions more precisely.}

 Now, we realized our promise. To find  the 
relevant saddles of $SU(N)$ gauge  theories,  we should first consider either  $SU(N)$ with  background gauge  field $(B^{(2)},B^{(1)}) $  for the  $\Z_N^{[1]}$  1-form symmetry or  $(SU(N)/\Z_N)_p$ theory where $\Z_N^{[1]}$ is gauged. 
After finding the  configurations in the   $(SU(N)/\Z_N)_p$,   we can  patch  them up   to find the ones that can be lifted to $SU(N)$ theory.  These are  fractional action  (e.g. $2S_I/N$)   configurations that contribute to the partition function of  $SU(N)$ theory. Indeed, the mini-space formalism that we derived in Sec.\ref{mini-space} reveals this fact. 
The construction and effective field theory based on monopole-instantons and bions 
 works in semi-classical weak coupling domain. The fact that  non-perturbative effects are controlled by 
 action $\frac{S_I}{N}$ and topological charge $W=1/N$ defects is true both in  strong coupling domain and weak coupling domain. 

 We also note that  $\mathbb {CP}^{N-1}$ model with an $\Omega_F$ and  $\ell_{12}$ flux background   and Yang-Mills theory on the center-symmetric background  with  't Hooft fluxes $\ell_{12}, \ell_{34}$ maps to identical problems from  
 the point of view of mixed anomalies \cite{Yamazaki:2017ulc,Yamazaki:2017dra}.  For a 
  general  discussion of background field coupling to  detect  't
Hooft anomalies of 4d Yang-Mills,  and their connection to   2d $ \mathbb {CP}^{N?1}$,  see also \cite{Wan:2018zql}. 
   It is not a coincidence that the combination of  \eqref{frac5} and \eqref{flux-7}  in $d=4$  is same as   \eqref{frac8} in  $d=2$.  
 
 
 \vspace{0.5cm}
 \noindent
{\bf Monopole-instantons  in $B^{(2)}$ background:} In order to determine the fractional instantons on arbitrary $M_4$, 
we are supposed  to solve in 4d the BPS  equation \eqref{Mod-inst}  in the $B^{(2)}$ background. 
If one of the directions is compactified, in one-lower dimension,  the background field decomposes as 
$B^{(2)}_{\rm 4d} = B^{(2)}_{\rm 3d} + B^{(1)}_{\rm 3d} \wedge \frac{ \diff x_3}{L_3} $  and 
the monopole instanton equations become
  \begin{align}
(\widetilde F-B^{(2)}_{\rm  3d})=  \mp \star  ( \widetilde {\diff_a a_3} -B^{(1)}_{\rm  3d})   
\label{Mod-inst-comp}
\end{align}
  However, as we did in quantum mechanics (compare \eqref{ins1} with   \eqref{ins2}) and 
  ${\mathbb CP}^{N-1} $ model,  we can undo the background and convert it to boundary conditions that the monopole instanton has to satisfy. This  is nothing but the regular monopole instanton equations \cite{tHooft:1974kcl, Polyakov:1974ek}
\begin{align}
F =  \mp \star  {\diff_a a_3} 
\end{align}
Our construction tells us that these well-known solutions fits with the  't Hooft flux boundary conditions and are indeed non-trivial solutions in $PSU(N)$ bundle.  

\vspace{0.5cm}
{\footnotesize
\noindent
{\bf Remarks:}
Although we phrased the fact that monopole instantons with topological charge $Q=1/N$ and action 
$\frac{8 \pi^2}{g^2N}$ fits perfectly with the $SU(N)$ with twisted boundary conditions or $PSU(N)$ bundle,
 the exact solutions within $PSU(N)$ bundle  had historically been a difficult endeavor.  
In 1981, 't Hooft showed that some {\it constant} abelian  solutions that  had topological charge $1/N$ 
and were easy  to find, but     had total action which never descends below $ 8 \pi^2(N-1) /(g^2N)$.   Then, he found some non-abelian {\it constant}  solutions that has action $8 \pi^2 /(g^2N)$,  a  desired property. Historically, however, it was not easy to 
determine the  time dependent  or space-time dependent solutions. 
   Gonzalez-Arroyo et. al.  found by simulations on latticized $T^3 \times \R$ that  time-dependent fractional instanton solutions with action ${1}/{N}$ exist  in the presence of 't Hooft flux
\cite{GarciaPerez:1989gt, GarciaPerez:1992fj, Montero:2000mv,GonzalezArroyo:1997uj}, see also  \cite{Gonzalez-Arroyo:2019wpu} for recent work.
One of our aim in this work was  to determine these solutions analytically,  and  we were able to interpret the well-known  monopole-instanton solutions 
   on   $\R^3 \times S^1_L$ as  non-trivial solutions  in the flux background.   }

\subsection{Fractionalization in $\N=1$  $SU(N)$ theory on $M_4$}

The evidence we gather suggest that the fundamentally important configurations in $SU(N)$ gauge theory 
are fractional  topological charge and  fractional action instantons.  
 The fact that such configurations exist is a mathematical statement about classification of 
the classical configurations,  determined by the properties by the  $SU(N)/\Z_N$ bundle.  However, one may ask, 
why such a statement, is not in apparent contradiction with the global chiral symmetry,  ABJ anomaly  or quantization of 
topological charge in $SU(N)$ theories on orientable 4-manifolds $M_4$?

Let us consider $\N=1$ supersymmetric  Yang-Mills theory with $G=SU(N)$ gauge group. 
This model has a $\Z_{2N}$ chiral symmetry which is broken dynamically to  $\Z_{2}$ by the formation of  fermion-bilinear condensate. 
 \begin{equation}
  \label{cc2}
 \langle  k|   \tr \lambda \lambda  |k \rangle  = N \Lambda^3\, \rme^{\im \frac{2 \pi k + \theta}{N}}, \qquad  k=0,1, \ldots, N-1 
 \end{equation}
  leading to $N$ isolated  supersymmetric vacua \cite{Witten:1982df}.
 However, we claim that the classical configurations in this theory should be determined by using 
 $SU(N)/\Z_N$ bundle. But in a  non-trivial flux background \eqref{flux-5},   the chiral symmetry is reduced to  $\Z_2$, instead of  $\Z_{2N}$. Each of these configurations has two exact fermion zero modes.  One may ask why the existence of objects with just  $2$  fermion zero modes   is not in contradiction  with the ABJ anomaly \cite{Bell:1969ts} in the $SU(N)$ theory (with no background turned on).    
 In particular, we expect
 \begin{align}
 {\cal I}_a \sim \rme^{- \frac{  S_I}{N} }   \tr \lambda \lambda
 \label{notin}
\end{align}
to be present in the Euclidean vacuum, even when  the description of  the vacuum is non-semiclassical. 
Of course, the answer is tied with the fact that 
only integer topological charge configurations   are in the $SU(N)$ bundle,  hence, only such configurations are part of the $SU(N)$ theory. However, the key point is 
we can construct $SU(N)$  configurations out of $PSU(N)$ bundle in different ways. 
For example, 
 \begin{align}
  I_{W=0}&=   {\cal I}_a (x_a)   \overline {\cal I}_a (  \overline x_a) \cr   
 I_{W=1}&=   \prod_{a=1}^{N} {\cal I}_a (x_a) \sim  \rme^{- S_I  + i \theta}  \prod_{a=1}^{N}  (\tr \lambda \lambda)(x_a)
 \label{list}
\end{align}
 are both in the  $SU(N)$ bundle, but not   $ {\cal I}_a (x_a)$ itself.   As long as one fractional instanton here (at this spacetime point) has a friend  fractional anti-instanton on the dark side of the Moon, the combination lives  happily in the $SU(N)$ bundle. The right hand side  in \eqref{list} is the first configuration to reduce the chiral symmetry down to $\Z_{2N}$ in $SU(N)$ gauge theory. 
    The proliferation of fractional instantons with the global constraint \eqref{constraint-5} is also   capable of  breaking the discrete chiral symmetry \cite{Davies:2000nw, Unsal:2007jx} at weak coupling.

  To see the presence of the fractional events on $\R^4$, let us reverse engineer  the fractional instanton sum. 
Consider a soft mass deformation, $\Delta {\cal L}_m =  \frac{m}{g^2}   \tr \lambda \lambda + {\rm h.c.} $ of $\N=1$ SYM.   In this case,  the vacuum energy density  is modified into 
${\cal E}_k =  -\frac{m}{g^2}   \langle  k|   \tr \lambda \lambda  |k \rangle     + {\rm c.c.}$ at leading order in $m$. 
The partition function on a four manifold can be written as:
\begin{align}
Z(\theta)= \sum_{k =0}^{N-1}  \rme^{ 2 m  N^2 \Lambda^3 V_4  \cos  \frac{ \theta+ 2 \pi k} {N} }
\label{multi}
\end{align}
where $\Lambda^3 = \mu^3 \rme^{-S_I/N} =  \mu^3 \rme^{- \frac{8 \pi^2}{g^2(\mu)N}} $ is the strong scale and $V_4$ is the volume of the 4-manifold that the theory is defined.  Based on our quantum mechanical example, we can rewrite the partition function as 
\begin{align}
Z(\theta)&= N \sum_{ W \in \Z}  \sum_{n=0}^{\infty}   \sum_{\bar n =0}^{\infty}   \frac{1}{n!} \frac{1}{\bar n !}  \left( m
N^2  \mu^3 V_4   \rme^{-\frac{S_I}{N} + \im  \frac{ \theta} {N}}  \right)^{n }     \left( m N^2  \mu^3 V_4   \rme^{- \frac{S_I}{N} -  \im \frac{\theta} {N}}  \right)^{\bar n }     
 \delta_{n-\bar n  - W N,0} \cr
  &= N \sum_{W \in \Z}   \left[ I_{NW} (2 m N^2 \mu^3 V_4  \rme^{-\frac{S_I}{N}} ) \right]  \rme^{\im  W \theta }    \qquad ({\rm strong\; coupling, \;no\;}  \; {\rm 't \;  Hooft } {\; \rm  flux}) \qquad 
  \label{constraint-3}
\end{align}
A few remarks are in order:
\begin{itemize}
\item  Individual  terms in the sum are   sourced by the solution of self-duality equation \eqref{Mod-inst} in the $SU(N)/\Z_N$ bundle. 
 Minimal configurations have action $S_I/N$ and topological charge $\frac{1}{N}$, and they do not contribute to the partition function, but they contribute to physical observables. 

\item The constraint  $\delta_{n-\bar n  - W N , 0}$ guarantees that the sum  is over integer topological charge configurations  $W \in \Z$ which belong to $SU(N)$.  This is ultimately the reason why  the proliferation of fractional instantons with the global constraint \eqref{constraint-3} is in agreement with ABJ anomaly,  and integer quantization of topological charge on 4-manifolds, and is also  capable of  breaking  chiral symmetry dynamically.

\item The  solutions  in  $PSU(N)$ bundle has  4 bosonic zero modes, which may perhaps interpreted as the position moduli. $V_4$ may be viewed as the volume of the bosonic moduli.   $\mu$ us Pauli-Villars renormalization scale. It appears with the
combination  
$\mu^{n_b -  n_f/2 } = \mu^3$ where $n_b=4, n_f=2$ are the numbers of bosonic and fermionic zero modes.  The moduli space of an instanton 
in $SU(N)$ theory, which has $N$ constituents,   can be parametrized in terms of $N$ 4-position, $ a_i \in \R^4, \;  i=1, \ldots, N$.
\end{itemize}

\section{Prospects and  comments}

{\footnotesize
 \begin{center}
\begin{tabular}{|c|c|c|c|c|} 
 \hline Theory   &
   \shortstack{  $N$-fold    classical    \\  degeneracy}
    &    \shortstack{Partially twisted \\ partition function} &   \shortstack{ Twisted \\partition function}  &   
   \shortstack{Minimal topological charge  \\ in  TQFT  background} \\ 
 \hline 
$T_N$ &      $V(Nq) $ & $Z= \tr [ \rme^{-\beta H}]$  &  $Z_\ell= \tr [ \rme^{-\beta H} \mathsf U^\ell]$ &$ \frac{1}{2\pi} \int A^{(1)}=  \frac{ 1 }{N}  $   \\  
 \hline
$\mathbb {CP}^{N-1}$  &  $\Omega_F$-twist  &  $Z_{\Omega_F}= \tr [ \rme^{-\beta H_{\Omega_F}}]$ & 
  $Z_{\Omega_F, \ell }= \tr [ \rme^{-\beta H_{\Omega_F}}   \mathsf U^\ell ]$    &$ \frac{ 1}{2 \pi}  \int B^{(2)}=  \frac{ 1 }{N} $ 
  \\ 
 \hline
$SU(N)$ YM &    $\ell_{12}$  flux  &   $Z_{\ell_{12}}= \tr [ \rme^{-\beta H_{\ell_{12}}}]$    &
$Z_{\ell_{12} \ell_{34}}= \tr [ \rme^{-\beta H_{\ell_{12}}}   \mathsf U^{\ell_{34}}]$   
 &   $  \frac{N}{8 \pi^2} \int B^{(2)} \wedge B^{(2)}  =  \frac{ 1 }{N}  $ 
 \\ 
 \hline
\end{tabular}
\end{center}
}

We developed three parallel constructions, in $d=1,2,4$  QM and QFTs.  
The analogous quantities are listed above. 
The classical $N$-fold degenerate minima in QM is a consequence of the  potential    $V(Nq) $.
 In $\mathbb {CP}^{N-1}$ it is induced by turning on $\Omega_F$ background for a global symmetry, and  in the YM theory, it is induced by turning on $B^{(2)}= \ell_{12} $ background  in the 12-plane. The simplest way to realize that these $N$-fold  classical degeneracy is not a fiction is to recall that in   $\mathbb {CP}^{N-1}$ and YM theory, the vacua at the quantum level  is $N$-branched.  The classical (as well as all orders perturbative minima) that emerges in the second column  is the precursor of that well-know fact. 
   In Yang-Mills, in the 3-direction associated with circle $S^1_L$, we may or may not have a center-symmetric holonomy field $U_3$.    
   Our main application is in center-symmetric background, but the application in the Appendix~\ref{sec:TEK} is in classically center-broken background.  The $N$-fold perturbative  degeneracy is independent of that. 
    The generalized partition function in third column is periodic in $\beta$ and it receives contribution only from $W \in \Z$ integer  topological charge configurations.  
The important point is that  these configurations can and do have fractional action, which starts with $\frac{2S_I}{N}$, in the $W=0$ sector. Observables are controlled by $\frac{S_I}{N}$. To see these configurations more explicitly, we inspect the fully twisted partition functions, also involving a twist associated with  $S^1_\beta$ circle (say, the 't Hooft flux  in 34-plane in gauge theory and 12-plane in  $\mathbb {CP}^{N-1}$).   We are free to put  all twists as  boundary conditions. Alternatively,  we can   use periodic boundary conditions everywhere,  and  turn on classical  background fields.   This is equivalent to coupling the QFT to $\Z_N$ TQFT, and reveals the true nature of non-perturbative configurations in the original theory without any backgrounds.

\vspace{0.5cm}
 \noindent 
{\bf Pure Yang-Mills theory on $\R^4$:} Coupling YM theory to a $\Z_N$ TQFT tells us that we need to use $PSU(N)$ bundle to build up 
$W \in \Z$  configurations in the $SU(N)$, which generically possess fractional action. 
The construction  instructs us that  non-perturbative expansion parameter in pure $SU(N)$ Yang-Mills theory is 
 \begin{align}
 \rme^{- \frac{S_I}{N} + \im \frac{\theta}{N} } =  \rme^{- \frac{ 8 \pi^2}{g^2(\mu) N} + \im \frac{\theta}{N} }  
 \end{align}
This is  exactly like the monopole-instanton factor in deformed Yang-Mills theory on $\R^3 \times S^1$  \cite{Unsal:2008ch}  where $\mu$ is Pauli-Villars  scale.   
This parameter is exponentially more important than the 4d instanton amplitude \cite{Belavin:1975fg}.   

\vspace{0.5cm}
\noindent 
{\bf QCD and general $SU(N)$ gauge theory, NP expansion parameter:} In his original works,   't Hooft argues  that the  twisted boundary conditions cannot be applied to QCD with fundamental quarks as there is no longer a center symmetry.  
Though the absence of center symmetry is a correct statement, his negative  conclusion can 
 can be avoided in two different ways. 
\begin{itemize}
\item 
If  ${\rm gcd}(N_f, N)  \neq 1$, we   can impose 't Hooft twisted boundary conditions   by turning on first an $SU(N_f)$  background.  This becomes most efficient for $N_f=N$,  where we can show that expansion parameter becomes  $\exp \left[ - {S_I}/{N} + \im {\theta}/{N} \right] $. 
\item If we turn on $U(1)_V/\Z_N$ background \cite{Roberge:1986mm} (which is part of genuine vector-like symmetry of the theory),  then,  we can  impose twisted boundary  conditions even for $N_f=1$.  In fact, our construction in $\mathbb {CP}^{N-1}$ is the dual of this idea, where we had $U(1)$ gauge structure, and $SU(N)/\Z_N$ global symmetry,  see Sec. \ref{flux-cp}. In 1-flavor QCD, we have $SU(N)$ gauge structure and $U(1)/\Z_N $ global symmetry. In either case, we can impose the same twists. 
As a 
result, we were able to turn on 't Hooft flux in a theory with quarks. Therefore, the expansion parameter is always  
$\exp [- { 8 \pi^2}/{g^2 N}]$ regardless of matter content of $SU(N)$ gauge theory. 
\end{itemize}

\vspace{0.5cm}
\noindent 
{\bf Renormalization group for  topological defects:} 
If   $\exp [- { 8 \pi^2}/{g^2 N}]$  is expansion parameter in all $SU(N)$ QCD-like theories, what distinguishes an asymptotically free IR-CFT  from a confining theory which exhibit chiral symmetry breaking and confinement?   We need to formulate a renormalization group, similar to \cite{Kosterlitz:1974sm, fradkin_2013}, which tells us relevance vs. irrelevance of  defect operators.  Indeed,  we used such a method at $\theta=\pi$ of $\mathbb {CP}^{1}$ model to exhibit the irrelevance of doubly-charged vertex operators and show  conformality,  and to prove the relevance of  vertex operators  at $\theta=0$, and mass gap.

\vspace{0.5cm}
 \noindent 
{\bf Rethinking  deformed Yang-Mills theory on $\R^3 \times S^1$:} 
On $\R^3 \times S^1$, a monopole-instanton  possesses four bosonic zero mode. This  introduces the Pauli-Villars  $\mu$ dependence of the form  $\mu^4$. The fluctuation operator over the non-zero modes  in the background of a monopole gives  
$[{\rm Det} (-D^2)]^{-1}=  (\mu  r_{\rm mon})^{-1/3} $ where  $ r_{\rm mon}= m_W^{-1} = \frac{LN}{2\pi}$ is the characteristic size of monopole-instantons, which is the inverse $W$-boson mass,  determined by the adjoint Higgsing.  
  The  combination of the zero and non-zero modes just gives the expected result, $\mu^{4-(1/3)}= \mu^{11/3}$ which combines with the exponential to give the 1-loop renormalization group invariant scale,  
  \begin{align}
  \mu^{11/3}    \rme^{- \frac{ 8 \pi^2}{g^2(\mu) N} + \im \frac{\theta}{N} } = \Lambda^{11/3}  \rme^{ \im \frac{\theta}{N} } 
  \label{accident?}
\end{align}
The mass gap on semi-classical domain on $\R^3 \times S^1$ is given by: 
   \begin{align}
   m^2(\theta)  = \Lambda^2  (\Lambda LN)^{5/3}  \;  \Big[{\rm Max}_k  \cos\Big({\theta+2\pi k \over N}\Big) \Big],  
   \qquad LN\Lambda \ll 1,  \; {\rm semiclassical \;domain}  
\end{align}
This result is based on a justified  effective field theory based on proliferation of monopole-instantons.  The merit of this formula is that 
it captures the $\theta$ angle dependence and existence of mass gap  correctly. 
 As mentioned above, $LN$ is the monopole-instanton size $r_{\rm mon}$, and semi-classics is reliable  provided  $r_{\rm mon} \Lambda \lesssim 1$.

 One of the new insights of this work is  to recognize that the monopole-instantons can also be interpreted as tunneling events in the $PSU(N)$ bundle,   in a fixed 't Hooft flux sector. See Fig.\eqref{fig:monopole}.   This may seem a trivial observation, 
but   this interpretation is important, because 't Hooft himself,  in 1981,    was only capable of finding {\it constant} solutions on $T^4$ on $PSU(N)$ bundle    and these solutions required certain ratio of scales in $T^4$ geometry to achieve BPS bound.  (He says in his paper, ``Considering the difficulty we had in finding them  it looked worth-while to publish the result." \cite{tHooft:1981nnx}). 
 After this  work,  the only non-trivial solutions that has been obtained were via numerical lattice simulations. See \cite{Gonzalez-Arroyo:2019wpu} for an up to date report.

 The fact that monopole-instantons are in $PSU(N)$ bundle makes one wonder if they may be weak coupling reincarnation of configurations with action 
 $S_I/N$ on arbitrarily large $T^4_{\rm large} \sim \R^4$   down to  $T^3_{\rm large} \times S^1_{\rm small} \sim \R^3 \times S^1_{\rm small}  $.  
  It is tempting to speculate that on $\R^4$, the fractional instantons can produce a gap of the form 
  $
m^2(\theta)  = \Lambda^{11/6}  (\rho )^{5/6}  \left[{\rm Max}_k  f\left({\theta+2\pi k \over N}\right) \right]^{p} + \ldots $ 
where  $\rho$ is a size moduli  of instanton, and $p$  can be a real number (not necessarily one as in semi-classical domain) of order one.   (Recall that in  $\mathbb {CP}^{1}$,  we showed that mass gap is of the form   
$m(\theta)  = \Lambda |\cos(\theta/2)|^{2/3}$.)
Very likely, similar to $\mathbb {CP}^{N-1}$ where Debye length  provides a self-consistent cut-off over the  instanton size moduli \eqref{cut-off},  there may be a similar mechanism on  $\R^4$ where $\rho$ is cut-off at order $\Lambda^{-1}$. 

\vspace{0.5cm}
 \noindent 
{\bf k-strings and center-vortices:} In pure Yang-Mills, there is only one fundamental string, and $k$-string tensions are  determined by $N$-ality of representation of sources  in the strong coupling domain.  However, 
in Polyakov model on $\R^3$ \cite{Polyakov:1987ez}  and Seiberg-Witten theory on $\R^4$ \cite{Seiberg:1994rs}, there are $N-1$ types of fundamental strings,  with $N-1$ different string tensions, astray from the pure Yang-Mills theory (but an unavoidable property of these theories, see e.g.  \cite{Douglas:1995nw}). 
In deformed YM, ${\cal N}=1$ SYM or QCD(adj)  on $\R^3 \times S^1$, there is again only one fundamental  string tension, just like pure Yang-Mills, thanks to unbroken $\Z_N$ center symmetry and  the absence of elementary adjoint Higgs field \cite{Poppitz:2017ivi}. 

Despite this  remarkable fact, it is important to  notice  that the infrared physics of  deformed Yang-Mills theory (as well as Polyakov model  \cite{Polyakov:1975rs}) {\bf cannot} be described by just writing an EFT based on proliferation of 
 monopole-instantons, one should keep in mind the existence of $W$-bosons as emphasized  \cite{Greensite:2016pfc, Greensite:2003bk}. If one forgets  $W$-bosons, 
this   would lead to non-zero tension for   adjoint or $k=N$  string,  which is incorrect,  as well as quantitatively  incorrect string tensions for   $k  >N/2$. 

  But fortunately,  dynamical $W$-bosons are in the full microscopic theory, and they can easily screen  $N$-ality zero  sources.   
There is no doubt  that: 
\begin{itemize}
\item [{\it i)}]  
In the weak coupling  semiclassical regime on  $\R^3 \times S^1$ center-symmetric background, 
there exists an EFT description in terms  monopole-instanton and  bion  local fields (ie, fields which create local excitations with the quantum numbers of the fractional-instanton along with their  fermionic  zero modes, and local bion fields).  A number of low-energy observables, mass gap, 
topological susceptibilities,  chiral condensates, and   
 $N$-ality   $k   \lesssim N/2$  (single winding) Wilson loop expectation values  are accurately (and sometimes exactly)  described by  the  local EFT based on these topological defects in the semi-classical domain.  This is as explicit as it can ever be \cite{Aharony:1997bx, Unsal:2007jx, Unsal:2008ch,Davies:2000nw}, and there is no doubt of its validity either in supersymmetric or non-supersymmetric theories.

\item [{\it ii)}]  However, to  screen the adjoint probes, and to get the right string tensions for  $k  \gtrsim N/2$, 
 one must keep $W$-bosons in the description, (which we should not call EFT anymore),  otherwise 
 these observables will come out to be wrong.  In other words, Wilson loops with $k  \gtrsim N/2$ is {\bf not} in the set of low-enegy observable  that EFT can address correctly. 

\item[ {\it iii)} ]     We do not know a    semi-classical effective field  theory describing the role of center-vortices, even when a large subset of  non-perturbative observables 
  are   completely  describable in terms of  semi-classics and EFT,  as in   $ {\cal N}=1$ SYM or QCD(adj) on $\R^3 \times S^1$,  or Polyakov model on $\R^3$.  
\end{itemize}
%
 
\noindent
The  EFT for deformed YM on  small  $\R^3 \times S^1$ center-symmetric regime is based on the parametric separation of scales between the non-perturbative dual photon mass $m_\gamma \sim  \Lambda (\Lambda LN)^{5/6}$ and the $W$-boson mass, $\frac{2 \pi}{LN} \sim\Lambda  (\Lambda LN)^{-1}   $, and EFT is valid provided 
\begin{align}
\frac{m_\gamma}{m_W}\sim  (\Lambda LN)^{11/6} \lesssim 1
\end{align}
But as we emphasized above,  the  EFT can be used to determine {\bf some}  observables,  but {\bf not  all} observables. For example, it is not capable of producing 
$k  \gtrsim N/2$ string tensions or multi-winding loops for some $k  \lesssim N/2$ probes. 
 To produce those correctly, we must incorporate into EFT  $W$-bosons, the heavy degrees of freedom, which are not part of EFT. Once this is done, one cannot call the combined  theory as EFT, better call it microscopic theory. But it has an  EFT subsector. 

Perhaps, a reasonable proposal is to aim to demonstrate that 
$\{$monopole-instanton + bions + $W$ bosons$\}$ system  is equivalent to center-vortices in weak coupling domain.  In some ways, it seems like one needs to incorporate $W$-bosons to the description and then, integrate it out. But then, very likely, there will not be a local  description. In other words, we suspect that there may not be a local  semi-classical EFT description for center-vortices at all.

But we still think that center vortices is a major part of the story, very likely not more or less important than fractional instantons, and the two are possibly  intimately connected. 
 There are  some  other interesting facts  about the center-vortices that support this perspective. 
 't Hooft magnetic flux background itself is actually a (non-dynamical) center vortex, and dynamical monopole-instanton tunneling events can be viewed as junction on them (See Fig.\eqref{fig:monopole}) and do not change the charge of  't Hooft  background 
 as it has zero 't Hooft charge.    
 On $T^4$, the intersection  point of center-vortices acts as sourcing topological charge $W=1/N$  \cite{Engelhardt:2010ft},  which naturally lives in $PSU(N)$ bundle, and smoothed  center-vortex configurations are related to instantons \cite{Trewartha:2015ida}, which we proved  to be 
 composites of fractional instantons with $W=1/N$ even on large $T^4$ strong coupling domain.
 
   Finally, probably  most importantly,  when we couple $SU(N)$ Wilson lattice gauge theory  \cite{Wilson:1974sk} to $\Z_N$ TQFT,  
   and gauge it,    
   vortices become local minima of the action for any $N$,    
    obviating various  restrictions concerning stability of center-vortex configurations  \cite{Bachas:1982ep, Greensite:2016pfc}.   However, one should be careful with the fact that   
    there are multiple lattice  $PSU(N)_0$ theories   (with or without magnetic matter),  and the one which is locally same as $SU(N)$ theory should  not have magnetic matter charged under magnetic center $N$-ality.   
     We are currently investigating  $PSU(N)_0$  lattice gauge theories  (with or without magnetic matter)    in order to reveal these connections precisely.

\vspace{0.5cm}
 \noindent 
{\bf The importance of reparametrization of instanton moduli space: }
The instanton in $SU(N)$ gauge theory  on $\R^4$ has  $4N$ bosonic zero modes and $q$-instantons have $4Nq$ moduli parameters.   The analogy with the analysis of  Refs.\cite{Berg:1979uq, Fateev:1979dc}   suggest that perhaps,  we should parametrize the moduli space of instanton as  
\begin{align}
\{a^{m}_j \},  \qquad m=1, \ldots q,    \; j=1, \ldots, N 
\end{align} 
These are $Nq$  $4$-positions (or quaternions)  that we may  think as fractional instanton positions.    

An intriguing  fact  that we learned in the course of this work is that the monopole-instantons on $\R^3 \times S^1$ are in $PSU(N)$  bundle  and come exactly as   
$ \mu^{11/3} \exp[- \frac{ 8 \pi^2}{g^2 N} + \im \frac{\theta}{N} ]$ as we already knew from \cite{Unsal:2008ch}.  
\begin{itemize}
\item
 The non-perturbative  expansion parameter in {\bf all}  center symmetric theories  in weak coupling domain on $\R^3 \times S^1$  is this factor.   
 \item
The non-perturbative expansion parameter in the strongly coupled domain  on  $\R^4$  for {\bf all}  theories that admit a coupling to $\Z_N$ TQFT is the same  non-perturbative  factor. 
 
 \end{itemize} 
   It is  hard to believe that this is an accident.

The picture we obtain from $\R^2$ and  $\R \times S^1$  suggests something deeper,  compliant with the previous paragraph.  The instanton  solutions of \cite{Berg:1979uq, Fateev:1979dc} on $\R^2$ do {\bf not} physically break an instanton into pieces. (Action density and topological density are single lumps for one instanton). 
 It is a reparametrization of the usual solution. The brilliant thing  is that the fluctuation determinant in the background of parametrization   \eqref{moduli} admits an interpretation {\it as if} we are summing over fractional vortex instantons, which carry fractional charge and fractional action, as it appears in   \eqref{sum-vortex}.  To see the fractional instantons explicitly, we need to do an extra work, for example, couple the theory to $\Omega_F$ background (see \cite{Dunne:2016nmc} for a review), or 't Hooft flux background,  or consider it in non-trivial holonomy background \cite{Lee:1997vp, Lee:1998bb, Kraan:1998kp, Kraan:1998pm, Kraan:1998sn}, and then,  dial the moduli parameters. For small moduli, one still does not see 
 physical separation, and only for sufficiently large moduli parameters, one sees fractionalization   explicitly (see Fig. \ref{fig:fraction}). May be,  BPST instanton is far more clever than us, and  it hides its deepest secrets inside in its moduli parametrization. 
 

\vspace{0.5cm}
 \noindent 
{\bf  QCD(F) on $\R^4$ vs. $\R^3 \times S^1$:}   Think of QCD with fundamental quarks, on $M_4$. Turn on a  background 
for the global symmetry $U(1)_V/\Z_N$, upon which 
 we can couple the system to $\Z_N$ TQFT.   Our claim is that this system should admit topological excitations that  are controlled by the action  $S_I/N$, and these are relevant to real QCD on $\R^4$.  

Consider,   for example, QCD with $N_f$ flavors on $\R^3 \times S^1$.  For concreteness, let us take  $N_f \leq N$, but $N_f$ can be made arbitrary.  
Then, in a center-symmetric background for gauge holonomy,  we can also turn on an $\Omega_F \in SU(N_f)$ as in \eqref{flavor-twist}. 
Then,  we have $N$ monopoles with action  $S_I/N$.   $N-N_f$ of the monopoles do not have a fermi zero mode (call this set   ${\cal S}_1$) and  $N_f$  of them have   2 fermi zero modes each (call this  set  ${\cal S}_2$).  This  is the set-up introduced in \cite{Cherman:2016hcd, Cherman:2017dwt} to explain 
the transmutation of $\frac{\theta + 2 \pi k }{N_f}$ dependence  in $N_f$ flavor QCD  to the  $\frac{\theta + 2 \pi k }{N_c}$ in pure Yang-Mills as the fermions are decoupled by increasing their masses.  

Monopole instantons on   ${\cal S}_1$ each has 4  bosonic zero modes, 3 position and one angular.  Perhaps, in the light of our intentions, it is wise to denote 
$\diff^3a \diff \phi \equiv L^{-1} \diff^4a$. 
Monopole instantons on   ${\cal S}_2$ each has 4  bosonic zero modes, 3 position and one angular, and 2 fermi zero modes. 

Then, the measure, in the background of the monopole instantons, which amounts to the calculation of pre-exponential factor, takes the form:
\begin{align}
{{\rm d} \mu_{{\rm mon}, j}} \sim \left\{ \begin{array}{ll}
  \mu^{4}   d^4a \;       \rme^{- \frac{ 8 \pi^2}{g^2 N} +  \im \frac{\theta}{N} }   \;  \underbrace{ \left({\det}^{'} [ -D^2] (\mu) \right)^{-1}}_{ (r_{\rm mon} \mu)^{-1/3}},   &  \cr 
  =  \mu^{4-\frac{1}{3} }       \rme^{- \frac{ 8 \pi^2}{g^2 N}  +  \im \frac{\theta}{N} }   d^4a  \ldots  & {j \in {\cal S}_1}   \cr  \cr
 \mu^{4 }   d^4a \;     \mu^{-1}  d^2 \xi_f    \rme^{- \frac{ 8 \pi^2}{g^2 N} +  \im \frac{\theta}{N}  }  \;  \underbrace{ \left({\det}^{'} [ -D^2] (\mu) \right)^{-1}}_{ (r_{\rm mon} \mu)^{-1/3}}  
 \underbrace{ \left({\det}^{'}_{\cal R} [ \gamma_{\mu} D_{\mu}] (\mu) \right)}_{ (r_{\rm mon} \mu)^{1/3}}, &    \cr 
 =  \mu^{4-\frac{1}{3} }      \mu^{-1+ \frac{1}{3}}    \rme^{- \frac{ 8 \pi^2}{g^2 N} +  \im \frac{\theta}{N}  }      d^4a \;  d^2 \xi_f      \ldots &   {j \in {\cal S}_2}
\end{array} \right. 
\end{align} 
where $ {\det}^{'}$ is the fluctuation operator over non-zero modes, which also induces $\mu$-scale dependence. 
 \begin{align} 
\I_{\rm BPST} \sim  \prod_{{j \in {\cal S}_1} } {\cal M}_j  \prod_{{j \in {\cal S}_2} } {\cal M}_j  & \sim   \left(\mu^{4-\frac{1}{3} }    \rme^{- \frac{ 8 \pi^2}{g^2 N}  +  \im \frac{\theta}{N} }    \right)^{N-N_f}  
\left( \mu^{4-\frac{1}{3} }      \mu^{-1+ \frac{1}{3}}    \rme^{- \frac{ 8 \pi^2}{g^2 N}  +  \im \frac{\theta}{N}  }  \right)^{N_f}  \cr
&=  \mu^{\frac{11}{3}N - \frac{2}{3}N_f }    \rme^{- \frac{ 8 \pi^2}{g^2} +  \im \theta  }
\end{align} 
The part of this expression  involving Pauli-Villars scales and how it combines  to produce renormalization group invariant strong scale  by combining $N$ constituent looks like an elegant conspiracy. 
The standard  calculations  of  the  pre-exponential factor for a  BPST  instanton (in the  standard parametrization)  can be found in NSVZ review \cite{Vainshtein:1981wh}.

It is very tempting that the structure above arises in $\R^4$ when QCD(F) is considered on $U(1)_V/\Z_N$ background and the theory is coupled 
to a $\Z_N$ TQFT. We suspect that in the decompactification to   $\R^4$,  
$(r_{\rm mon} \mu)$  type  factors  on $\R^3 \times S^1$ will be replaced by  combinations such as 
$(|a^m_j -a^n_k| \mu)$,  $(|\bar a^m_j -\bar a^n_k| \mu)$, or  $(|a^m_j - \bar a^n_k| \mu)$. 
 Once we consider   original QCD on $\R^4$, we should probably inspect proliferation of the fractional instantons which should appear in a way similar to  monopole-instantons do on  $\R^3 \times S^1$.  They must  
 obey the global constraints  similar to monopole-instantons, and  we should see if one can obtain a self-consistent construction of QCD vacuum for general $N_f$.

\acknowledgments
I am deeply grateful to Aleksey Cherman, Yuya Tanizaki,   Tin Sulejmanpasic  for many discussions at   odd hours. 
I am also thankful to Margarita  Garcia Perez,  Thomas Sch\"afer, Patrick Draper, Mohamed Anber, Gerald Dunne, Tony Gonzalez-Arroyo, Mikhail Shifman, David Gross,   Philip Argyres, Rob Pisarski,  Pierre van Baal for past discussions.   Although I have not met with Polyakov in person,  I am thankful to him for his  unpretentious beautiful  book.  
My work  is supported by the U.S. Department of Energy, Office of Science, Division of Nuclear Physics under Award DE-SC0013036.


 \appendix

\section{Yang-Mills  with flux  on $1^3 \times S^1_\beta$ lattice   vs. $T_N$ model}
\label{sec:TEK}
Let us consider a 4d lattice gauge theory  with discrete space and continuous time, $N_s^3 \times \R$. 
Then, to reduce the model to simple quantum mechanics, we  take  $N_s = 1$ in the same spirit with the twisted Eguchi-Kawai model  \cite{GonzalezArroyo:1982ub, GonzalezArroyo:1982hz, 
GonzalezArroyo:2010ss}, by employing twisted boundary conditions, but  we do not take the large-$N$ limit. 

We consider $1$-site theory with Lagrangian
\begin{align}
g^2 {\cal L} = \tr |\diff \widetilde U_i|^2 + \tr |[ \widetilde U_i,  \widetilde U_j]|^2
\label{mat1}
\end{align}
in the  $A_0=0$ gauge. In path integral, we assume that $\widetilde U_i $ obey  't Hooft  twisted boundary conditions on the $1^3$ lattice. 
 \begin{align}
  \widetilde U_{i, {\bf n} + N_j {\bf e_j}} & =  \Omega_{j }   \widetilde U_{i, {\bf n}}  \Omega^{\dagger}_{j} 
      \label{link-trans} 
 \end{align}
 Consistency at, e.g.   $N_1 {\bf e_1} + N_2 {\bf e_2}$ requires (recall $N_1=N_2=1$):
\begin{align}
 \Omega_{1}    \Omega_{2}        =  \Omega_{2  }    \Omega_{1}   \rme^{ \im \frac{2 \pi }{N} \ell_{12}}
   \end{align}
where $ \ell_{12} $ is $B^{(2)}$ flux mod $N$.  
Since lattice is only 1-site, this forces the transitions matrices to be independent of space coordinates. 
 This actually leads to subtle and important differences between an arbitrarily small continuum  $T^3 \times S^1_\beta$ theory and   $1^3 \times S^1_\beta$ theory. In the former,  the transition matrices are allowed to depend on spacetime,  but not in the latter, and this has consequences. 

Undoing the twisted boundary conditions in favor of a classical background field, we end up with the flux $\ell_{12}$ in the $12$ face and the matrix model becomes 
\begin{align}
g^2 {\cal L} = \tr |\diff  U_i|^2 + \tr |[U_1 U_2 - \rme^{\im \frac{2 \pi}{N} \ell_{12}} U_2 U_1  ]|^2+ \tr |[U_1, U_3 ]|^2 + \tr |[U_2, U_3 ]|^2
\label{mat2}
\end{align}
where path integral over the $U_i (\tau)$ field is periodic both in time and  space direction (which is reduced to 1-site). 

Let us describe the classical minima of this system.  The minimum of the modified commutator (take $\ell_{12}=1$) is  
$U_1=C$, $U_2= S$, where $C$ and $S$ are clock and shift matrices. The minima of the action correspond to the values of 
$U_3$ which commutes both with  $C$ and $S$. Since $C$ and $S$ are generators of $GL(N, \mathbb C)$ algebra, the only matrix that  commutes with both is proportional to identity. Since $U_3$ lives in $SU(N)$,  the minima are 
\begin{align}
U_3 = \rme^{\im  \frac{2 \pi}{N} j} {\mathbbm 1}, \qquad  j=0, \ldots, N-1
\label{centerbroken}
\end{align}
Therefore, the classical system  for the $\ell_{12}=1$ sector  has $N$-minima, corresponding to   $N$ center broken configurations.  This is analogous to our discussion on large- $T^3 \times S^1_\beta$ where we showed that the $\ell_{12}=1$ sector  was also $N$-fold degenerate, see \eqref{degenerate}.
  Of course, in quantum mechanics, center symmetry cannot break spontaneously due to tunneling effects  at any finite-$N$.

How do we describe the tunneling between these minima?  
Let us take $U_3(\tau) $ of the form 
\begin{align}
U_3= {\rm Diag}(\rme^{\im \Theta_1},\ldots, 
\rme^{\im \Theta_N} ), \qquad  {\rm where} \qquad \Theta_N= -( \Theta_1 + \ldots +  \Theta_{N-1})
\end{align}
 Then, the  (classical) potential can be expressed as 
\be
V(\Theta_i) = - \sum_{i=1}^{N} \cos( \Theta_i -  \Theta_{i+1} )
\label{classicalpot}
\ee
Despite the fact that this potential looks like  Gross-Pisarski-Yaffe (GPY) type  potentials for gauge holonomy \cite{Gross:1980br},   this is  superficial and they are  fundamentally different. The GPY  type-potentials is a  {\it quantum}   effect, starting  at  1-loop order as  $O(g^2)$. 
  However,  \eqref{classicalpot}  is  {\it classical}, not quantum. It is induced by the presence of 
classical 't Hooft flux background. As a result, for example, if one examines instantons in the theory with flux, their action is 
$\sim \frac{1}{g^2}$ since they arise from classical action. In contrast, if one examines tunneling on small $T^3 \times \R$ without flux, the action is  of the form $\sim \frac{1}{g}$,   because these configurations  arise from the balancing of classical actions against quantum induced potential, see eg.  \cite{vanBaal:1986ag,Luscher:1982ma}.


To describe these  tunnelings, one can  take  simple    ansatz   for  $U_3(\tau)$
\be
U_3(\tau)= {\rm Diag}(\rme^{\im \Theta}, \ldots, \rme^{\im \Theta}, \rme^{-\im (N-1) \Theta} )
\ee 
 Then, the Lagrangian becomes 
 \begin{align}
g^2 {\cal L} =  (N-1)N  (\diff \Theta)^2 -  \cos( N \Theta)
\label{mat3}
\end{align}
where $ \Theta \sim  \Theta + 2\pi$. 
This is nothing but our $T_N$ model, with the distinction that the mass of the particle, instead of being set to one, is now
 $(N-1)N$. 

The remnant of the center-symmetry (which becomes $(\Z_N^{[0]})^4$ 0-form symmetry on  $T^3 \times S^1_\beta$) is just    $U_3 \rightarrow   \rme^{\im  \frac{2 \pi}{N} }U_3$ or  equivalently, acting on 
coordinate   
$\Theta$ as a shift symmetry   $\Theta \rightarrow \Theta  +   \frac{2 \pi}{N}$, i.e., the part of center-symmetry which acts on $U_3$ non-trivially becomes the $\Z_N$ shift symmetry of the $T_N$ model.  
   To reveal the fractional instantons in the background of 't Hooft flux, we can proceed exactly as  in our QM example.

  We can build the fractional instanton events by considering the transition amplitudes between the minima 
  which are ${\ell_{34}}$ units apart. 
Hamiltonian now is the one corresponding to \eqref{mat3} or more microscopically, \eqref{mat2}, where $H$ already has the magnetic flux data, so we call it $H_{ \ell_{12}}$. So, the quantity we are calculating is 
\begin{align}
 Z_{\ell_{12}\ell_{34}}  & =   \tr [ \rme^{-\beta H_{ \ell_{12}}} (\mathsf U_{\rmc})^{\ell_{34}} ]   
  \label{tpf-3}
 \end{align}
 which forces twisted boundary condition in $U_3$ by $\ell_{34}$  unit,   $U_3(\beta=L_4)= \rme^{\im \frac{2 \pi}{N} {\ell_{34}} }U_3(0)$. 
 
 These fractional instantons are the ones in the theory with 't Hooft flux, $({\ell_{12}, \ell_{34}}) $. They describe   
tunneling  events in our matrix model.  Roughly, 
 $\dot \theta \sim E_3(\tau)$ and  
  $\cos( N \theta (\tau)) \sim B_3 (\tau)$.  Despite the fact that these are reasonable fractional instanton events in 
  the 1-site model, they do {\bf not} provide a description of the fractional instantons in  continuum theory on small $T^3 \times \R$ in the presence of  't Hooft flux.   For once, their action is $S_a \sim \frac{\sqrt{(N(N-1)}}{g^2 N} $. In fact, at this stage,  we are facing a 
  problem that   't Hooft faced in a time independent continuum version of this problem \cite{tHooft:1981nnx}
  where the simple solutions he found did have an action  which did not descend below $S_a = \frac{(N-1)}{N} \frac{8 \pi^2} {g^2 } $.  (The problem in  \cite{tHooft:1981nnx} is  fixed by introducing space-time dependent transition matrices.) In our case as well, 
this is an artifact of 1-site model, and the fact that we are restricting to space-independent transition functions.   Our continuum formulations tells us that 
even at  arbitrarily small volume $T^3 \times S^1_\beta$, there exist configurations which saturate BPS bound, and this is consistent with lattice simulations on $N_s^3 \times  S^1_\beta$ where   $N_s \geq 2$ \cite{Gonzalez-Arroyo:2019wpu, Montero:2000mv}.

\section{Aside:  How to fix merons in $\mathbb {CP}^1$?}
\label{sec:meron}
Both  instanton  \cite{Polyakov:1975yp}  and  smooth fractional instanton   solutions   (in the $\Omega_F$ or $B^{(2)}$  flux   background)   \cite{Bruckmann:2007zh,  Dunne:2012ae}    have aspects  in sharp contrast with 
meron configurations  \cite{Gross:1977wu}.  But   clearly they look  quite alike as shown in Figure below, where horizontal axis is the angular coordinate  on $S^1_{\infty}$ in Fig.\ref{fig:mapping}. 
   Let us try to make   some sense out of this situation for $N=2$.

\begin{figure}[t]
\vspace{-0.3cm}
\begin{center}
\includegraphics[width = 0.8 \textwidth]{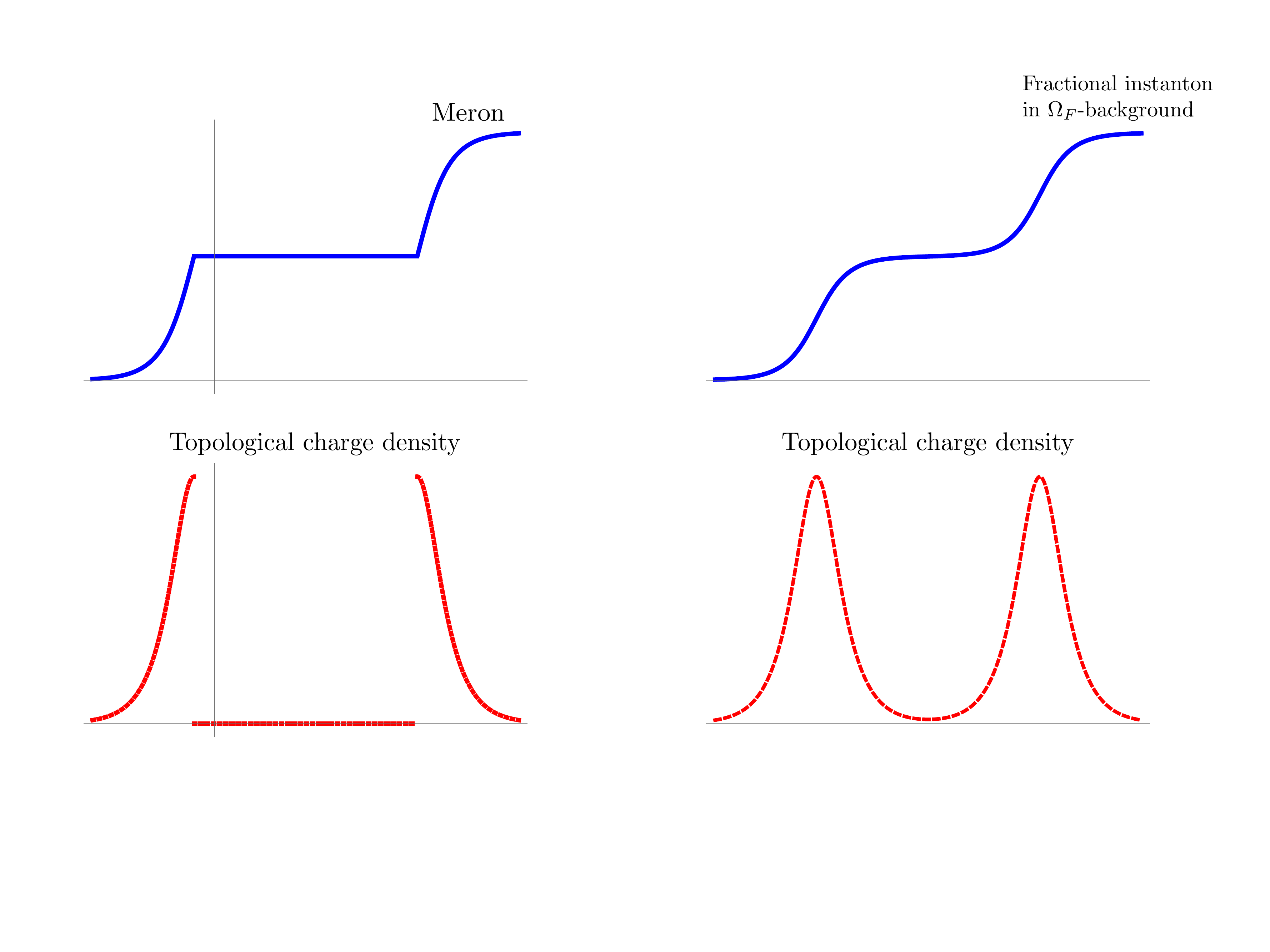}
\vspace{-1cm}
\caption{ 
\footnotesize 
 Horizontal axis is the angular coordinate  on $S^1_{\infty}$ in Fig.\ref{fig:mapping}.
{\bf a)} Meron data  is piecewise continuous or  discontinuous. 
Action is asserted to be log divergent  \cite{Gross:1977wu},  but that needs a  
re-interpretation, see text.  
 {\bf b)} Instanton solution in the $\Omega_F$ background.  A fractional instanton pair is a critical point at infinity. Above configuration is on the Lefschetz thimble of this exact critical point. 
Insertion of $B^{(2)}$ 't Hooft  flux, one can isolate each smooth configuration with topological charge $\half$.
 %
 }
\label{fig:meron}
\vspace{-0cm}
\end{center}
\end{figure}

{\bf 1)} For merons,  configurations,  action densities,  and topological charge densities are either {\it piece-wise continuous} or 
 {\it discontinuous},  and they   possess  logarithmically   divergent action   \cite{Gross:1977wu,Callan:1977gz,  Callan:1977qs}.

{\bf 2)}  Instantons 
 are smooth single lump solutions \cite{Polyakov:1975yp}.   On the other hand,  the     parametrization of  Refs.\cite{Berg:1979uq, Fateev:1979dc}, 
 the  determinant of fluctuation operator  gives a precise mathematical description as if these single lumps ought to be interpreted as composites 
 of fractional vortex instantons, despite the fact that  an instanton does not physically fractionate to two pieces. 
  Gross   uses this parametrization even earlier  and shows that  the interaction between the instanton  and anti-instanton 
 is a dipole-dipole interaction in $2d$  in  \cite{Gross:1977wu}. The two are consistent.

{\bf 3)} The solutions  in \cite{Bruckmann:2007zh, 
 Dunne:2012ae}  are  
generalization of  Ref. \cite{Polyakov:1975yp}  to  the $\Omega_F$  background, and  it  physically  fractionates an instanton  into pieces.  If we want to single out a fractional instanton,  we need to insert a $B^{(2)}$  flux,  and this will guarantee a $W=1/N$ configuration with $S=S_I/N$.   These configurations are smooth and finite action, unlike  merons.   


If we superficially smoothen  the meron,   as suggested  by Gross \cite{Gross:1977wu}, 
 that is  close enough  to  exact analytic solutions in the   $\Omega_F$  background   \cite{Bruckmann:2007zh, Dunne:2012ae}.  
 What is called the  logarithmically   divergent action of meron pair  is the logarithmic Coulomb interaction 
between the  fractional vortex instantons. This also  arise from {\it exact} computation of the determinant of the fluctuation operator \cite{Berg:1979uq, Fateev:1979dc}.

Meron idea is not 
generalized correctly to 
$SU(N)$ gauge theories and   $\mathbb {CP}^{N-1}$  with $N\geq 3$   
\cite{Gross:1977wu,Callan:1977gz,  Callan:1977qs}.  The difficulty is in the motto   ``meron= half-instanton". 
 For a given  $N$,
  the fractionalization   (in center-symmetric holonomy  background, or $\Omega_F$ or  $\Z_N$ TQFT  background) is always  into $N$ constituents, {\bf not more, not less}! Given this fact, it seems very worthwhile to revisit the idea. 

 't Hooft long time ago came up with  a $PSU(N)$ bundle construction, to figure out  configurations with topological charge and action $\frac{1}{N}$.  He was only able to find some exact, but  uninteresting  {\it constant} configurations on $T^4$. 
To be  fair to him, one should state that  no one was able to find non-trivial  (space-time dependent) solutions in $PSU(N)$ bundle  up to date analytically!  Numerically, they are proven to exist in    \cite{GarciaPerez:1989gt, GarciaPerez:1992fj, Montero:2000mv}.  In this work, we interpreted  monopole-instanton solutions as configurations in the $PSU(N)$ bundle with action $S_I/N$ on $\R^3 \times S^1$ and this is a concrete realization of (another)  long sought dream of 't Hooft.

 't Hooft asserts that  his idea of configurations  with $\Z_N$ topological charge are important for the confinement  and mass gap problem,     but not merons or instantons   \cite{tHooft:1977nqb}.  But at least for $SU(2)$ or  $\mathbb {CP}^{1}$, 
  these two ideas (with enough cosmetic changes)  are actually the  same. (I suspect both of them would think otherwise, hopefully up until today.)     The correct part of the meron idea  in $\mathbb {CP}^{1}$   is   the  ``constituents"  that interact with each other via    long range  logarithmic interactions.   
%
 
 Probably, the strangest  insight that our  work brings is following.  Despite the naming ``fractionalization", a physical fractionalization of action density 
 may or may not take place. Physical fractionalization of a  lump 
  is only achieved by turning on appropriate background field.    Yet, the clever parametrization of moduli space   \eqref{moduli}
tells us that mathematically, the moduli parameters acts as if they are the positions of fractional vortex instantons.  This is in some sense, can be viewed as  fractionalization without   fractionalization.

    Sometimes, it takes long time to  figure  out  truth in a wild dream \cite{Gross:1977wu},  and  utility of  a formal construction \cite{tHooft:1977nqb}, perhaps, because the truth lives somewhere in between.   

\bibliographystyle{utphys}
\bibliography{QFT-Mithat} 


\end{document}